\documentclass[11pt,twoside,fleqn,a4paper,notitlepage]{report}

\usepackage{graphicx}
\usepackage[labelsep=quad,labelfont=bf,format=hang]{caption}
\usepackage{subcaption}
\usepackage[percent]{overpic}
\usepackage{amssymb}
\usepackage{textcomp}
\usepackage{units}
\usepackage{exscale}
\usepackage{url}
\usepackage[intlimits]{amsmath}
\usepackage{listings}
\usepackage{fancyhdr}

\allowdisplaybreaks
\newcommand{\E}{\mathcal{E}}
\newcommand{\F}{\mathcal{F}}

\begin{document}

\title{Signal generation in CdTe X-ray sensors\\[1ex]{\small Report ETHZ-IPP-2020-01}}
\author{Oliver Grimm\footnote{E-mail \texttt{oliver.grimm@phys.ethz.ch}}\\[1ex]ETH Z\"{u}rich, Institute for Particle Physics and Astrophysics\\Otto-Stern-Weg 5, 8093 Z\"{u}rich, Switzerland}
\maketitle
\vspace{-5ex}
\begin{center}
\line(1,0){350}
\end{center}
\vspace{4ex}
\noindent\textbf{Abstract}\hspace{1ex} This write-up explains the signal generation mechanism in CdTe semiconductor sensors. Derivations are mostly carried out explicitly, starting with basic semiconductor relations. The analysis is largely applicable to any semiconductor, with the focus being on the Schottky-type CdTe:Cl sensors that are employed in the Spectrometer/Telescope for Imaging X-rays (STIX) instrument on-board the ESA Solar Orbiter mission.

\begin{center}
\vspace{1ex}
\includegraphics[width=0.85\textwidth]{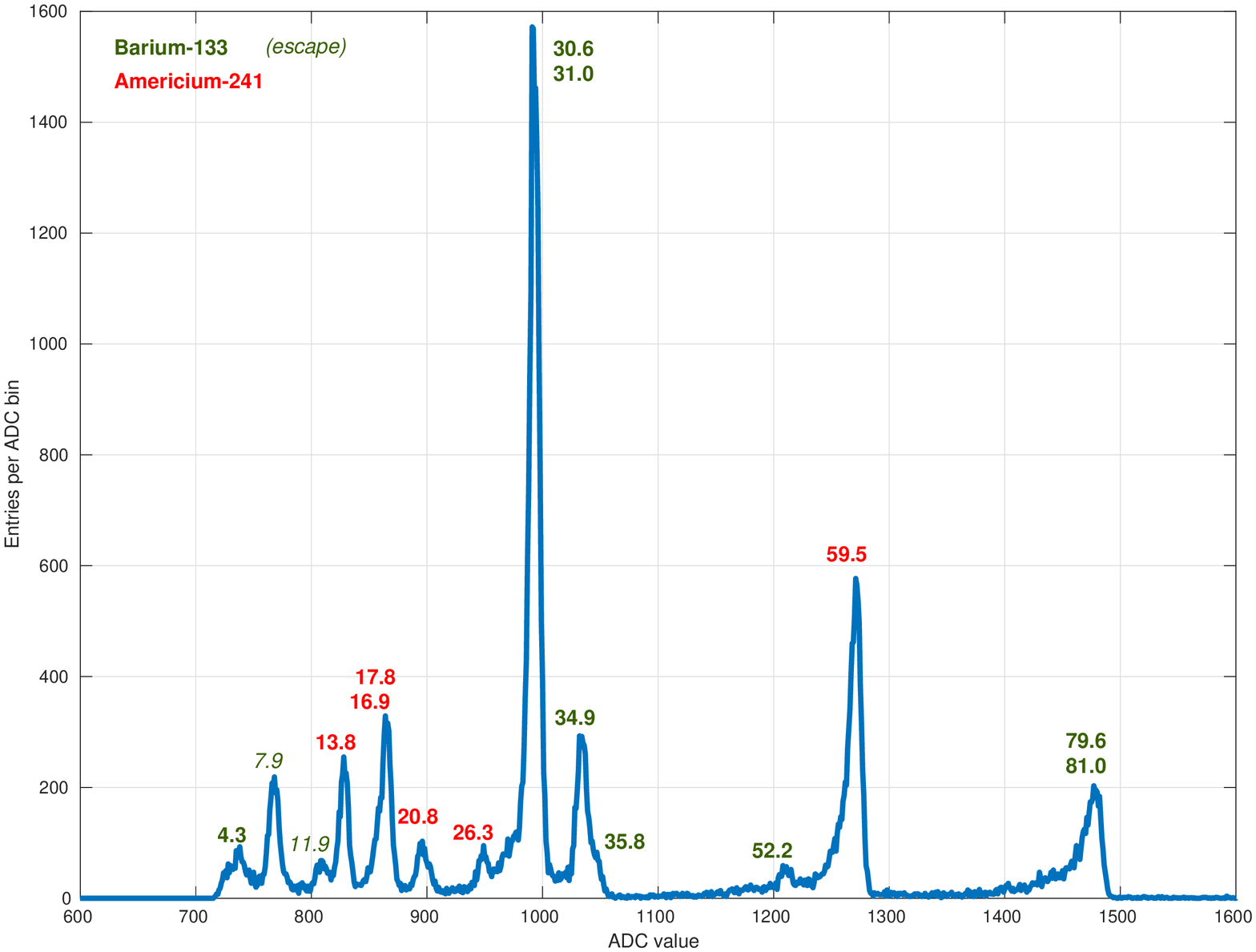}\\
Measured spectrum resulting from exposing a CdTe crystal to $^{133}$Ba and $^{241}$Am X-rays.\\[1.5ex]
Line energies in keV are indicated. A comparison to modelling of the same configuration,\\
implementing the principles described in this report, is shown on page \pageref{Fig:Comparison}.
\end{center}

\tableofcontents

\chapter*{}
\section*{Introduction}
\addcontentsline{toc}{section}{Introduction}

This report explains the electrical signal generation mechanism in the CdTe sensors that are employed in STIX (Spectrometer/Telescope for Imaging X-rays) \cite{Benz12}), an instrument on-board the ESA Solar Orbiter mission \cite{ESA09}. The treatment starts with a localized energy deposition in the sensor and carries this through to the electrical signal that can be measured with an analogue-to-digital converter at the output of the Caliste-SO front-end read-out hybrid \cite{Meu12}.

The formation of a measurable electrical signal after a particle interaction in a semiconductor involves the following steps:
\begin{itemize}
	\item Generation of a localized electron/hole pair cloud in proportion to the deposited energy and to the pair creation energy
	\item Separation and drift of electrons and holes by the applied electric field
	\item Charge induction on the electrodes due to the carrier motion
	\item Diffusion of the carrier clouds due to interactions of the drifting carriers with the crystal lattice
	\item Loss or temporary immobilization of carriers due to traps and recombination
	\item Amplification and conversion of the induced charge signal into a voltage, involving a signal shaping circuit
\end{itemize}
All these steps are covered in this report, except for the initial particle or photon interaction processes and possible fluorescence of the cadmium or tellurium atoms. Comprehensive simulation packages for interactions exist, like for example Geant4. The interface to a signal simulation program based on the principles described here is the energy deposition that results in an initially localized charge cloud.

The basic properties of CdTe sensors and charge transport are covered in Chapter~\ref{Sect:CdTeSensors}. The electronic band structure, helping to visualize the behaviour of the sensors, is derived both with the simple depletion approximation and exactly in Chapter~\ref{Sect:Band-Structure}, which also covers the polarization effect. Generation of the electrical output signal by the drifting charges is explained in Chapter~\ref{Sect:SignalInduction}, for which the Shockley-Ramo theorem is derived. Considerations of some basic effects of the read-out ASIC on the detected signal are made in Chapter~\ref{Sect:ASIC}. The inclusion of the described effects into a numerical signal simulation is outlined in Chapter~\ref{Sect:Implementation}. Finally, Chapter~\ref{Sect:Experimental-Results} shows a number of plots illustrating the behaviour of the STIX CdTe sensors.

For reference, the main symbols used in this report are: $q$ the electric charge, with $q=-e$ for the electron, $k$ the Boltzmann constant, $E(x)$ the electric field at position $x$, $U(x)$ the electric potential, $\E$ denotes electric potential energies, $\F$ Fermi levels, $\psi_k$ the weighting potential of electrode $k$, $\rho$ the charge density as used in Poisson's equation, $\Phi$ the carrier concentration as used in diffusion calculations, $\mu$ the mobility, $D$ the diffusion constant. $\varepsilon_0$ is the permittivity of free space, $\varepsilon_{\text{r}}$ is the relative permittivity of a material. $V$ is the applied bias voltage, $V_\text{d}$ the depletion voltage. $T$ is an absolute temperature, $t$ the time. Bold symbols are used for vector quantities.

\chapter{CdTe properties, carrier transport}
\label{Sect:CdTeSensors}

X-rays are detected in the STIX Detector/Electronics Module by cadmium telluride (CdTe) semiconductor sensors. Bias voltages of several hundred volts are applied to deplete the sensor and to obtain acceptable charge collection. A blocking Schottky contact on the anode limits the leakage current. The sensors are produced by Acrorad, Japan, and the pixel structure on the anode is applied at the Paul Scherrer Institute, Switzerland \cite{Gri15}. The crystals are integrated into the Caliste-SO read-out hybrids by CEA Saclay, France.

\section{Bulk material properties}
\label{Sect:Bulk-Properties}

\begin{figure}
\centering
\includegraphics[width=0.8\textwidth]{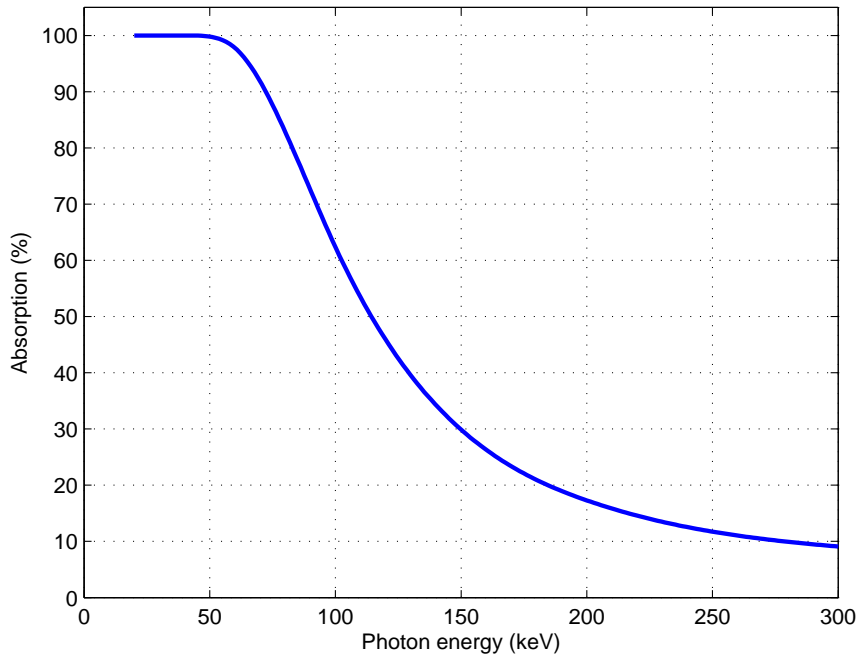}
\caption{Absorption probability of X-ray photons in \unit[1]{mm} thick CdTe}
\label{Fig:CdTeAbsorption}
\end{figure}

The CdTe semiconductor material is well suited for spectroscopic detection of X-rays with high energy due to its high average atomic number of 50, thus high photoelectric absorption coefficient, and high density of \unit[5.85]{g/cm$^{3}$}. A thickness of \unit[1]{mm} absorbs 63\% of all photons at \unit[100]{keV}, see Fig.\,\ref{Fig:CdTeAbsorption}\footnote{The mass attenuation coefficient is taken from a National Institute of Standards and Technology database, accessible at \url{www.nist.gov/pml/data/xraycoef/}. This describes the attenuation of a narrow pencil beam.\label{Footnote:One}}. The relative dielectric constant $\varepsilon_{\text{r}}$ is about 11.

The compound nature of the material results in a crystal quality worse than obtainable with silicon or germanium, thus in relatively poor carrier transport properties. Especially the hole lifetime is not long compared to the drift time over millimetre lengths and significant charge loss occurs. Typical room temperature values for mobility\footnote{The drift speed $v$ of carriers with mobility $\mu$ at electric field $E$ is $v=\mu E$.}, lifetime, and drift time for \unit[1]{mm} distance at \unit[200]{V} bias are as follows.\vspace{2ex}

\begin{tabular}{cccc}
	Carrier type & Mobility (cm$^2$/(Vs)) & Lifetime (\textmu s) & Drift time for \unit[1]{mm} (ns)\\[0.5ex]
	\hline
	Electron & 1100 & 3 & 45  \\
	Hole & 100 & 1 & 500
\end{tabular}\\[0.5ex]
The mobility scales with absolute temperature $T$ approximately as $T^{-3/2}$ for the temperatures of interest here.

Undoped CdTe is typically strongly p-type due to the presence of a large number of acceptor-like cadmium vacancies in the crystal lattice.\footnote{One reason for these vacancies is the higher vapour pressure of cadmium compared to tellurium. This makes it difficult to maintain the correct stoichiometric proportions of the two components for intrinsic material during crystal fabrication.} To compensate the acceptor vacancies, and thus to increase the resistivity and to improve the charge transport, the material is doped with chlorine that acts as a donor. The resulting crystal remains slightly p-type after this procedure \cite{Bis12,Kras13}.

Moderate cooling of the material is sufficient to achieve low leakage current due to the large band gap.\footnote{The band gap is about of \unit[1.6]{eV} at \unit[4]{K} and decreases to approximately \unit[1.5]{eV} at room temperature.} The bulk resistivity of the best obtainable CdTe crystals is about \unit[$2\times10^9$]{$\Omega$\,cm} at room temperature. A \unit[1]{mm} thick piece of \unit[10]{mm$^2$} area, corresponding to the dimensions of one large pixel in STIX, then has a resistance of \unit[2]{G$\Omega$}, or a leakage current of \unit[50]{nA} at \unit[100]{V}. At -20\textdegree C this will be reduced to a few nanoamperes due to the exponential dependence of the resistivity on temperature.

The energy needed to generate one electron/hole pair is \unit[4.43]{eV}, so for \unit[4]{keV} 900 pairs are generated. This would result in a Poisson statistical fluctuations of 3\%. Because the fluctuations are limited by the total energy available, and because very many, small energy deposition processes are to phonons\footnote{Therefore, the energy required to create an electron/hole pair is substantially larger than the band gap, despite CdTe being a direct band gap material.}, the statistical fluctuations are much smaller than this, which is empirically described with the \emph{Fano factor} $F$. The observed fluctuation $\sigma_\text{observed}$ is then related to the Poisson fluctuation $\sigma_\text{Poisson}$ by $\sigma_\text{observed}=\sqrt{F}\sigma_\text{Poisson}$. Typical values of $F$ for CdTe are in the range 0.1 to 0.15, so the statistical fluctuations at \unit[4]{keV} will be about 1\%. This is small compared to the best electronic noise contribution of about \unit[420]{eV} and is thus usually ignored.

\section{Electrodes}

\begin{figure}
\centering
\includegraphics[width=0.8\textwidth]{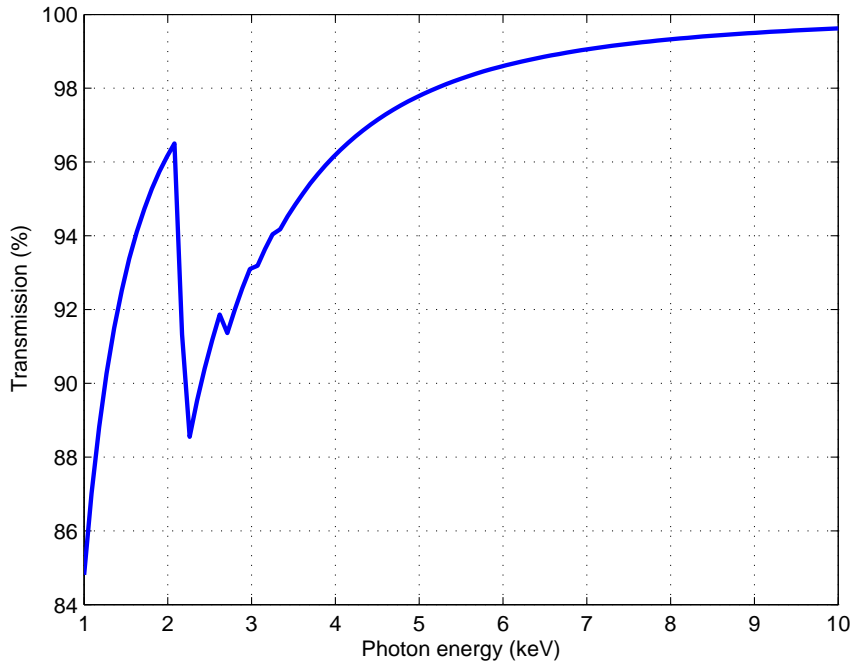}
\caption{Transmission probability of X-ray photons through \unit[16]{nm} of platinum}
\label{Fig:PtTransmission}
\end{figure}

The cathode of the \unit[10x10]{mm$^2$} area CdTe crystals is a monolithic platinum electrode. As the platinum work function of \unit[5.4]{eV} is similar to that of slightly p-type CdTe ($\approx$\unit[5.6]{eV}), a nearly Ohmic contact is formed, passing majority carriers easily. This is the entrance electrode for X-rays, thus its transmission characteristic is of importance. The thickness was measured using Rutherford backscattering spectrometry at ETH Z\"{u}rich at three positions for one crystal to be \unit[16]{nm} with \unit[1]{nm} spread. The X-ray transmission as function of photon energy for this thickness is shown in Fig.\,\ref{Fig:PtTransmission}.\footnote{See footnote \ref{Footnote:One} for the data source.}

The anode is a three layer structure of gold, titanium and aluminium. The innermost aluminium layer forms a Schottky contact (the work function of Aluminium is about \unit[4.2]{eV}, lower than that of CdTe), blocking the hole majority carriers. The gold layer is applied for electrode protection, and the intermediate titanium layer for better adhesion of the two other metals.

\begin{figure}
\centering{}
\begin{overpic}[width=0.55\textwidth,bb=90 230 460 580,clip]{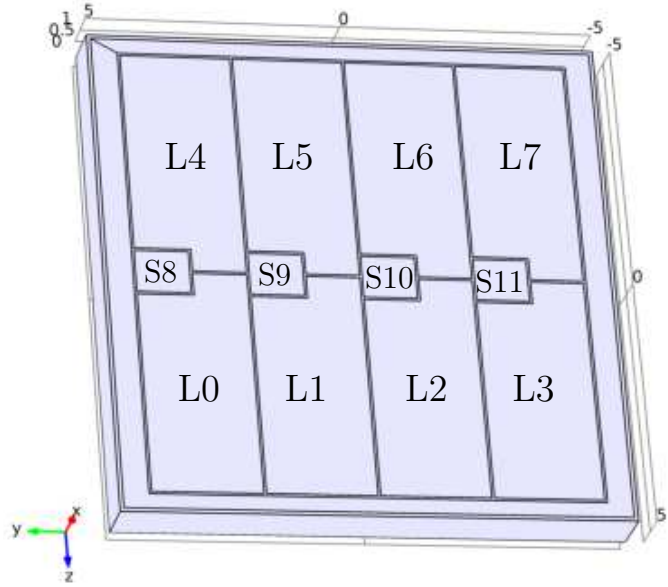}
\put(24,65){\Large L4}
\put(40,65){\Large L5}
\put(58,65){\Large L6}
\put(74,65){\Large L7}
\put(26,30){\Large L0}
\put(42,30){\Large L1}
\put(60,30){\Large L2}
\put(76,30){\Large L3}
\put(21,48){\large S8}
\put(38,47.5){\large S9}
\put(54,47){\large S10}
\put(70.5,46.5){\large S11}
\end{overpic}
\caption[Pixel and guard ring pattern of the CdTe anode]{Pixel and guard ring pattern of the CdTe anode. The outside dimension of the crystal is $10\times 10\times 1$ mm$^3$. The coordinate system and pixel numbering used in this report is shown. The x axis points from the cathode to the anode, thus out of the image plane. The origin of the coordinate system is centred on the cathode.}
\label{Fig:PixelPattern}
\end{figure}

The originally monolithic anode is etched into a pattern of 8 large and 4 small pixels and a surrounding guard ring, as shown in Fig.\,\ref{Fig:PixelPattern}.\footnote{The arrangement of the pixels in four stripes is required for the detection of the large-scale Moire structure that is the basis of imaging in STIX. The additional use of two large and one small pixel per stripe allows to adjust the effective area depending on the incoming photon rate to limit pulse pile-up and dead-time by disabling pixels. Smaller pixels also have less capacitance, lowering the electronic noise of the amplifier. The guard ring prevents surface leakage current along the edges of the crystal to influence the pixels.} The etching process is critical as all metal layers should be reliably removed to avoid shorts between the pixels, but the CdTe bulk should not be damaged excessively by overetching.

The Schottky barrier limits the dark current to thermionically generated majority carriers. In practice, currents below \unit[40]{pA} for a large pixel have been obtained at -20\textdegree C and \unit[300]{V} bias. Values below \unit[60]{pA} are required to not generate excess noise in the read-out amplifier.

\section{Electric field}
\label{Sect:Electric-Field}

The electric potential $U(\mathbf r)$ within some volume can be found by solving the Poisson equation $\nabla^2 U(\mathbf r) = -\rho(\mathbf r)/(\varepsilon_{\text{r}} \varepsilon_{0})$ for a given charge density distribution $\rho(\mathbf r)$ and given boundary conditions. The electric field $\mathbf E(\mathbf r)$ follows from $\mathbf E(\mathbf r)=-\nabla U(\mathbf r)$.\footnote{The nabla operator is $\nabla = (\frac{\partial}{\partial x}, \frac{\partial}{\partial y}, \frac{\partial}{\partial z})$. A linear relationship between electric and displacement field is assumed, and $\varepsilon_\text{r}$ to be constant within the volume of interest.} One dimensional considerations will be sufficient in the following, so the relevant equation is
\begin{equation}
\frac{\partial^2 U(x)}{\partial x^2} = -\frac{\rho(x)}{\varepsilon_{\text{r}}\varepsilon_{0}}.
\label{Eq:Poisson-Equation}
\end{equation}

The electrode-free gaps between the 12 pixels and on the inside of the guard ring in Fig.\,\ref{Fig:PixelPattern} are \unit[50]{\textmu m} wide, and a \unit[75]{\textmu m} wide area outside of the guard ring is also without metallization. Since the thickness of the crystal is \unit[1]{mm}, the field direction is very nearly perpendicular to the large crystal faces everywhere except close to the outer edge and within about \unit[50]{\textmu m} of a pixel boundary. The resulting change in charge transport direction near the pixel boundaries has only a marginal effect on the induced electric signal, as most signal results from movement in the bulk of the crystal. Therefore, the field direction is assumed to be exactly along the x coordinate. Also edge effects are neglected, as they would affect mainly the guard ring, which is of no interest as the guard ring will not be allowed to trigger or be read-out in STIX.\footnote{The reason it is connected to a powered channel of the ASIC is to guarantee that is has exactly the same electrical potential as the pixels. This optimizes the protection against edge-related leakage current.}

In the following, the orientation of the \emph{DEM Unit Reference Frame} (URF) coordinate system is used, see Fig.\,\ref{Fig:PixelPattern}. The origin is at the centre of the platinum cathode, with the x axis extending towards the anode (incoming X-rays propagate along this axis).

\begin{figure}
\centering
\includegraphics[width=0.8\textwidth]{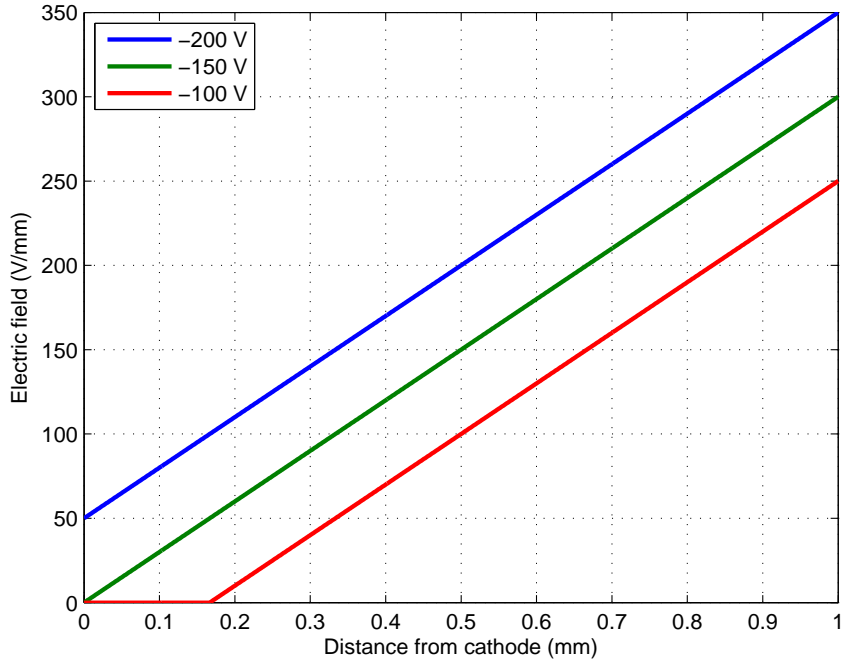}
\caption[Electric field for three applied voltages]{Electric field for an assumed depletion voltage of $V_\text{d}$=\unit[-150]{V} and for three applied bias voltages.}
\label{Fig:ElectricField}
\end{figure}

To determine the electric potential, the space-charge $\rho(\mathbf r)$ in the bulk has to be known. This in turn depends on the electric field. First, the case of an applied voltage $V$ is considered that is sufficiently high that all mobile charges are swept out (full depletion). Then the effective doping concentration determines the space-charge. Since the CdTe material is p-type, the space-charge $\rho(x)=\rho$ is negative. It is also constant throughout the volume, so from (\ref{Eq:Poisson-Equation})
\begin{equation}
	U(x) = -\frac{\rho}{2\varepsilon_{\text{r}}\varepsilon_{0}} x^2 + C_1 x + C_2.
\end{equation}
The integration constants follow from the boundary condition $U(0)=V$ and $U(d)=0$, describing a voltage $V$ applied to the cathode with respect to the anode. The potential distribution inside the crystals is thus
\begin{equation}
	U(x) = -V_\text{d}\left(\frac{x}{d}\right)^2 + \left( V_\text{d} - V \right) \frac{x}{d} + V, \qquad\qquad \text{with}\quad V_\text{d} = \frac{\rho d^2}{2\varepsilon_{\text{r}}\varepsilon_{0}},
\label{Eq:Potential-DepletionApproximation}
\end{equation}
and the electric field resulting from this potential
\begin{equation}
	E(x) = -\frac{\partial U(x)}{\partial x} = -\frac{2V_\text{d}}{d^2} x + \frac{V_\text{d} - V}{d}.
	\label{Eq:ElectricField}
\end{equation}
The condition of full depletion is $V/V_\text{d}>1$. At smaller voltages part of the CdTe volume will remain undepleted and thus charge neutral with $\rho(x)=0$. $E(x)$ will vanish in this region, as seen in Fig.\,\ref{Fig:ElectricField} for distances smaller than $x_\text{d}=d\cdot\left(V_\text{d}-V\right)/(2V_\text{d})$ from the cathode. The detector will lose sensitivity in particular for low-energy X-rays that all interact close to the cathode.

More complete considerations of the electric potential and field distributions are made in the next chapter.

\section{Charge drift}

Carriers with mobility $\mu$ will move with drift speed $v(x)=\mu E(x)$ in an electric field\footnote{Charge transport also occurs in regions with vanishing field due to diffusion. The actual driving force for charge transport in a semiconductor results not from electric potential differences alone, but from electro-chemical potential differences.}. The mobility is positive for holes (positive charges) and negative for electrons.\footnote{This reflects the convention that electrons move against the direction of the electric field.} From this equation the time as function of distance for a carrier moving from a start location $x_0$ can be calculated:
\begin{multline}
	t(x) = \int_{x_0}^x\frac{\text{d}t}{\text{d}x'} \text{d}x' = \int_{x_0}^x\frac{1}{\mu E(x')} \text{d}x' =
	\frac{d}{\mu} \int_{x_0}^x\frac{1}{V-V_\text{d}+2 V_\text{d}x'/d} \text{d}x' \\
	= \frac{d^2}{2 \mu V_\text{d}} \ln\frac{\displaystyle V-V_\text{d}+2 V_\text{d} \frac{x}{d}}{\displaystyle V-V_\text{d}+2 V_\text{d} \frac{x_0}{d}}.
	\label{Eq:TimeDistanceRelation}
\end{multline}
For electrons $x>x_0$ and for holes $x<x_0$, therefore $t(x)$ is always positive. The drift time $t_0$ through the whole crystal is
\begin{equation}
	t_0 = \frac{\pm d^2}{2 \mu V_\text{d}} \ln\frac{V-V_\text{d}}{V+V_\text{d}},
	\label{Eq:FullDriftTime}
\end{equation}
with + applying to holes and - to electrons. For $V=\unit[-200]{V}$, $V_\text{d}=\unit[-150]{V}$ and $d=\unit[1]{mm}$ this yields \unit[60]{ns} for electrons and about ten times longer for holes.

\section{Diffusion}

The expansion of a charge cloud with density distribution $\Phi(\mathbf{r},t)$ due to diffusion is governed by Fick's second law
\begin{displaymath}
	\frac{\partial\Phi(\mathbf{r},t)}{\partial t} = D \nabla^2 \Phi(\mathbf{r},t),
\end{displaymath}
with the diffusion constant $D$. Inserting a spherically symmetric, normalized Gaussian distribution
\begin{equation}
	\Phi(\mathbf{r},t) = \frac{1}{\left(\sqrt{2\pi}\sigma(t)\right)^3} \exp\left(-\frac{\mathbf{r}^2}{2 \sigma(t)^2}\right)
	\label{Eq:SphericalGaussian}
\end{equation}
into this differential equation, $\sigma(t)$ has to obey\footnote{This result holds also in the one- and two-dimension cases. The prefactor in (\ref{Eq:SphericalGaussian}) for $k$ dimensions has to be written $(\sqrt{2\pi}\sigma(t))^{-k}$, so that integration over the $k$ dimensional volume yields unity.}
\begin{displaymath}
	\sigma(t) \frac{\partial\sigma(t)}{\partial t} - D = 0 \Longrightarrow\quad \boxed{\sigma(t) = \sqrt{2Dt}.}
\end{displaymath}

Carriers obtain a constant speed for a given electric field instead of accelerating because of repeated collisions with the crystal lattice, transferring energy from the carriers to phonons. Diffusion occurs for the same reason, therefore the constants describing drift (the mobility $\mu$), and diffusion (the diffusion constant $D$) are related. The Einstein relation\footnote{For deriving, assume some electrical potential $U(x)$ that allows for a stationary distribution $\Phi(x)$ in the presence of diffusion (a potential well). Then the particle current density by drift in one direction, $\mu\,E(x)\,\Phi(x)$, is balanced by an opposing diffusion current density $D\,\partial\Phi(x)/\partial x$. Boltzmann statistics gives in thermal equilibrium $\Phi(x) \propto \exp(-eU(x)/(kT))$. With $E(x)=-\partial U(x)/\partial x$ the Einstein relation follows.} states
\begin{equation}
    D = \frac{\mu\,k\,T}{q},
    \label{Eq:EinsteinRelation}
\end{equation}
where $k$ is the Boltzmann constant, $T$ the absolute temperature, and $q$ the charge. Since $\mu$ and $q$ have the same sign, $D$ is always positive.

The spread after the maximum drift time $t_0$ is
\begin{equation}
	\sigma(t_0) = \sqrt{\frac{kTd^2}{e\,V_\text{d}} \ln\frac{V-V_\text{d}}{V+V_\text{d}}}.
	\label{Eq:MaximumSpread}
\end{equation}
For $V=\unit[-200]{V}$, $V_\text{d}=\unit[-150]{V}$, $d=\unit[1]{mm}$ this gives for room temperature \unit[18]{\textmu m}, smaller than the gaps between pixels.

\begin{figure}
\centering
\includegraphics[width=0.8\textwidth]{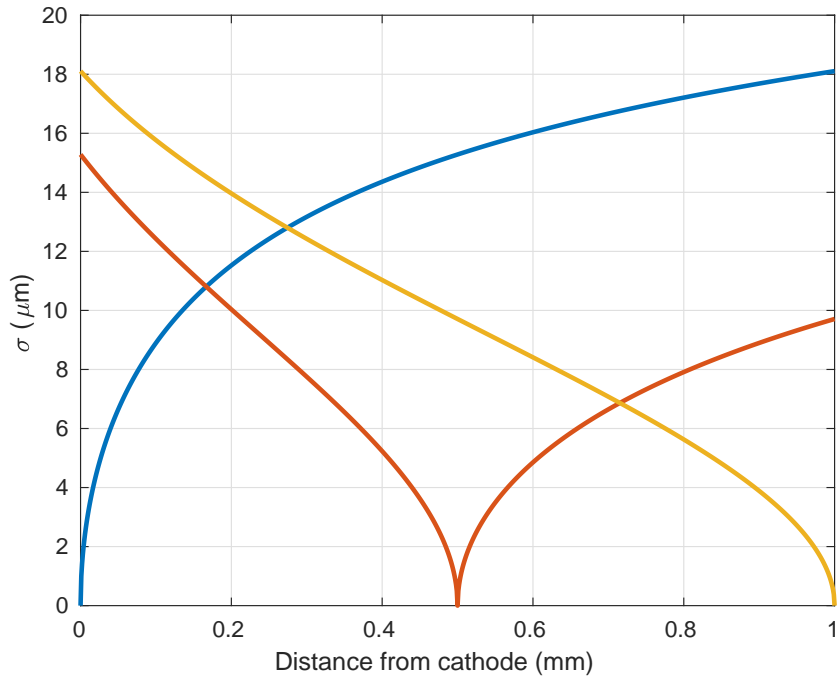}
\caption[Diffusion width]{Diffusion width (\ref{Eq:DiffusionSigma}) for $x_0$ at 0, \unit[0.5]{mm}, and \unit[1]{mm}. Holes drift towards the left, electrons towards the right of the primary interaction. Room temperature, $V_\text{d}$=\unit[-150]{V}, and $V$=\unit[-200]{V}.}
\label{Fig:Diffusion}
\end{figure}

The final expression for the diffusion width as function of distance travelled is
\begin{equation}
    \boxed{\sigma(x) = d\sqrt{\frac{k\,T}{q\,V_\text{d}}\ln\frac{V-V_\text{d}+2 V_\text{d} \frac{x}{d}}{V-V_\text{d}+2 V_\text{d} \frac{x_0}{d}}}.}
    \label{Eq:DiffusionSigma}
\end{equation}
Although this is independent of $\mu$, the effect on electrons and holes is not identical because of the electric field shape (see Fig.\,\ref{Fig:ElectricField}). This is illustrated in Fig.\,\ref{Fig:Diffusion}, giving the evolution of $\sigma(x)$ for three primary interaction depths. The holes drift slower near the cathode due to the lower field, and thus the charge cloud has more time to expand. This difference in $\sigma(x)$ for the same drift distance is reduced at higher bias voltages.

Note that these are idealized estimates. The electron/hole pairs are generated by the primary photoelectron along its irregular trajectory, with a range between a few micrometer around \unit[10]{keV} and several tens of micrometer at a \unit[100]{keV}, so the initial charge distribution is not exactly spherical.

\section{Carrier loss}
\label{Sect:CarrierLoss}

Various mechanisms can result in recombination of electron/hole pairs before they delivered the maximum energy to an external circuit (i.e. are lost before reaching their electrodes). Some effects are unavoidable, e.g. radiative recombination, others involving recombination centres depend on the purity of the material.

The random thermal motion of carriers with energy $3kT/2$ involves speeds of typically $\sqrt{3kT/m^*}$, where $m^*$ is the effective mass. For carrier electrons the effective mass is about 10\% and for holes about 35\% of the free electron mass, resulting in thermal speeds of \unit[$3.7\times 10^5$]{m/s} and \unit[$2.0\times 10^5$]{m/s} for electrons and holes, 20 and 120 times larger than their drift speeds. Therefore each carrier samples a large volume for recombination possibilities, and thus the time, not the drift distance, is the relevant parameter to characterize charge loss. With the life time $\tau_\text{e}$ of electrons and $\tau_\text{h}$ of holes, the number of carriers $n_\text{e}(t)$ and $n_\text{h}(t)$ after time $t$ is
\begin{equation}
	n_\text{e}(t) = n_0\,\exp\left(\frac{-t}{\tau_\text{e}}\right), \qquad n_\text{h}(t) = n_0\,\exp\left(\frac{-t}{\tau_\text{h}}\right).
	\label{Eq:CarrierDecay}
\end{equation}
$n_0$ is the number of electrons and holes initially created.

It is observed experimentally that spectral lines of low-energy photons, which interact close to the entrance electrode, show a pronounced tail towards low signal amplitudes. This is due to a damage layer close to the surface from crystal cutting and/or electrode deposition, resulting in an additional charge loss depending on distance from the cathode. For modelling, this is included by replacing $n_0$ by $n_0 \cdot (1-\exp(-x/\lambda_\text{entrance}))$. Good agreement with measurements is obtained for $\lambda_\text{entrance} \approx \unit[6.5]{\text{\textmu}m}$ \cite{Gri19}.

Electron loss is becoming much more severe after the sensors have been exposed to energetic protons \cite{Gri20}.

\chapter{Band structure}
\label{Sect:Band-Structure}

The electronic band structure is a useful concept to understand the behaviour of a semiconductor, especially near the electrodes. The account given here for the particular case of CdTe will follow closely the general derivations in \cite{Pier96}. A very comprehensive coverage of the subject can also be found in \cite{Boer02}. To avoid confusion, the electric field is designated with the letter $E$, whereas calligraphic letters ($\E$, $\F$) are used for energies.

A fundamental relation for the discussion of band structures is the \emph{Fermi-Dirac distribution} $f(\E)$. It gives the probability that an electronic energy level $\E$ is occupied in equilibrium at temperature $T$,
\begin{equation}
f(\E) = \frac{1}{\displaystyle 1+\exp\left(\frac{\E-\F}{kT}\right)},
\label{Eq:FermiDiracDistribution}
\end{equation}
where $\F$ is the \emph{Fermi level}. The probability that a hole energy level is occupied (i.e. that an electron is not present) is $1-f(\E)$. The requirement of equilibrium entails not only constant temperature, but also that no external potentials are applied.

The derivations in \cite{Pier96} generally assume a non-degenerate semiconductor: the Fermi level should lie within the energy gap between conduction and valence band with a distance to the band edges of at least several $kT$. Numerical values will be calculated in the following for the temperature T = \unit[273]{K} (0\textdegree C), where $kT=\unit[24]{meV}$.

For stationary (time-independent) conditions, the Maxwell-Faraday equation states that the rotation of the electric field $\mathbf E(\mathbf r)$ vanishes, thus it can be written as the gradient of a scalar \emph{electrostatic potential} $U(\mathbf r)$,
\begin{equation*}
\mathbf E(\mathbf r) = -\nabla U(\mathbf r).
\end{equation*}
From Gauss's law follows the Poisson equation
\begin{equation*}
\nabla^2 U(\mathbf r) = -\frac{\rho(\mathbf r)}{\varepsilon_{\text{r}} \varepsilon_{0}}
\end{equation*}
for a given charge density $\rho(\mathbf r)$. Note that the force $\mathbf F = q \mathbf E = -q \nabla U(\mathbf r)$ on a charge $q$ pushes electrons ($q=-e$) towards higher potential and holes towards lower potential. The \emph{electrostatic potential energy} is defined as $\E=q U$, thus charges of both signs are pushed towards lower potential energies, as expected.

Note also that the energy band diagrams in this section always show electron energies, also for the valence band where the effective carriers are holes.

\section{Isolated energy levels}

\begin{figure}
\centering
\includegraphics[width=0.9\textwidth]{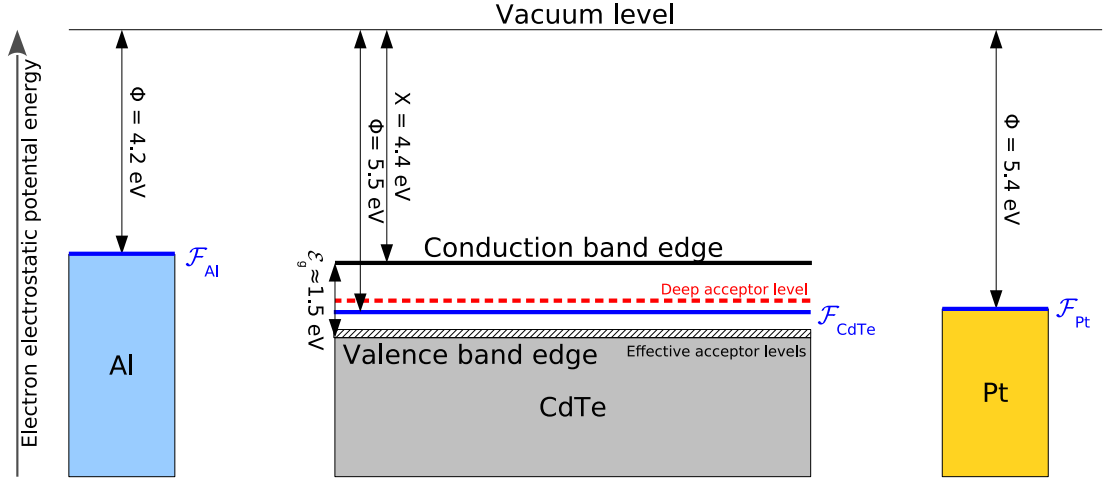}
\caption[Energy levels of isolated slabs of aluminium, CdTe, and platinum]{Electronic energy levels of isolated pieces of aluminium, CdTe, and platinum. Fermi levels are indicated by $\F$, work functions by $\Phi$, and the semiconductor electron affinity by $\chi$. The electrostatic potential energy increases from bottom to top.}
\label{Fig:IsolatedEnergyLevels}
\end{figure}

To construct the band structure for the Al-CdTe-Pt arrangement, first the energy levels of the isolated materials and the Fermi levels are shown in Fig.\,\ref{Fig:IsolatedEnergyLevels}.

The Fermi level $\F$ in (\ref{Eq:FermiDiracDistribution}) is equivalent to the energy needed to add one additional electron from the vacuum, and thus can be identified with the \emph{electrochemical potential} of the electrons.\footnote{The terminology is slightly inconsistent: whereas the electrostatic potential is measured in volt and the electrostatic potential energy in joule, as expected, the electrochemical potential is also measured in joule, although the word 'energy' is not used. Note also that in the context of semiconductor physics, the convention is such that \emph{chemical potential} includes only effects resulting from concentration gradients. The \emph{electrochemical potential} includes electric field effects as well, and this potential appears in (\ref{Eq:FermiDiracDistribution}).} This can be easily seen at zero temperature and for continuous energy bands when all energy levels up to the Fermi level are filled ($f(\E<\F)=1$) and all other are empty. The lowest energy level available for an additional electron is then at $\F$.

The semiconductor Fermi level depends on the hole concentration $p$ as \cite[(2.37)]{Pier96}
\begin{equation}
\F_\text{CdTe} = \F_\text{i} - kT\ln\left(\frac{p}{n_\text{i}}\right),
\label{Eq:Fermi-Level}
\end{equation}
where the Fermi level of the intrinsic (undoped) material $\F_\text{i}$ is given in terms of the conduction and valence band energies $\E_\text{c}$ and $\E_\text{v}$ by
\begin{equation}
\F_\text{i} = \frac{\E_\text{c} + \E_\text{v}}{2}+\frac{3kT}{4}\ln\left(\frac{m^*_\text{v}}{m^*_\text{c}}\right).{}
\label{Eq:Intrinsic-Level}
\end{equation}
$m^*_\text{c}$ and $m^*_\text{v}$ are the effective masses of electrons in the conduction band and holes in the valence band. Typical values for CdTe are $m^*_\text{c} \approx 0.11\,m_\text{e}$ and $m^*_\text{v} \approx 0.35\,m_\text{e}$, with $m_\text{e}$ the free electron mass \cite{Owens12}. $\F_\text{i}$ is only \unit[20]{meV} above mid-gap at 0\textdegree C. $\F_\text{i}$ can be assumed to be exactly at the mid-gap energy for most purposes.

The \emph{intrinsic carrier concentration} $n_\text{i}$ is given as function of the energy gap width $\E_\text{g}=\E_\text{c} - \E_\text{v}$ as \cite[(2.21)]{Pier96}
\begin{equation}
n_\text{i} = \sqrt{N_\text{c} N_\text{v}}\exp\left(\frac{-\E_\text{g}}{2kT}\right).
\label{Eq:IntrinsicConcentration}
\end{equation}
$N_\text{c}$ and $N_\text{v}$ are the \emph{effective number of states} of the conduction and valence band.\footnote{The effective number of states expresses the actual number of states that are available within a given band over all energies, but by referring them to the band edge energy. Thus, for example, the number of electrons $n$ in the conduction band can be calculated simply as $n=f(\E_\text{c}) \,N_\text{c}$. From \cite[(2.13)]{Pier96} $N_\text{c} = 2\left((m^*_\text{c} kT)/(2\pi\hbar^2)\right)^{3/2}$ and equivalently for $N_\text{v}$.\label{Ftn:EffectiveStates}} For CdTe at 0\textdegree C, $N_\text{c} \approx \unit[7.9\times10^{17}]{cm^{-3}}$ and $N_\text{v} \approx \unit[4.5\times10^{18}]{cm^{-3}}$. With $\E_\text{g}$ = \unit[1.5]{eV} it follows that $n_\text{i}$=$\unit[3.0\times10^4]{cm^{-3}}$. For comparison, the free electron densities in aluminium and platinum are \unit[$1.8 \cdot 10^{23}$]{cm$^{-3}$} and \unit[$1.3 \cdot 10^{23}$]{cm$^{-3}$}, respectively.

For a non-degenerate semiconductor in equilibrium the relation
\begin{equation}
np = n^2_\text{i}
\label{Eq:Product_np}
\end{equation}
holds \cite[(2.22)]{Pier96}, so the Fermi level (\ref{Eq:Fermi-Level}) can also be expressed using $n$ instead of $p$.

The bulk resistivity $\mathcal{R}$ can be calculated from the carrier concentrations and mobilities as
\begin{equation*}
\frac{1}{\mathcal{R}} = e (n \mu_\text{e} + p \mu_\text{h}) = e \left( \frac{n^2_\text{i}}{p} \mu_\text{e} + p \mu_\text{h} \right).
\end{equation*}
The maximum resistivity is obtained for $p =  n_\text{i} \sqrt{\mu_\text{e}/\mu_\text{h}} \approx 3.3 n_\text{i}$, giving $\mathcal{R} = 1/(2 e n_\text{i} \sqrt{\mu_\text{e} \mu_\text{h}}) \approx \unit[3]{G\Omega m}$. In practice, much smaller resistivity values around $\unit[50]{M\Omega m}$ are found for good CdTe crystals, translating to $p=\unit[1.3\times10^{7}]{cm^{-3}}$ for p-type material.

The actual doping concentration is not easy to determine experimentally. The matter is complicated in CdTe by self-compensation processes, the influence of deep traps, and resulting \emph{Fermi level pinning}. In particular, a comparatively high density of deep traps, which by the mechanism of bend bending (described in the next section) might be below the Fermi level in part of the crystal, will contribute to the effective doping density already in equilibrium.

Studies with the STIX sensors indicate that good performance is obtained at bias voltages above about \unit[50]{V}. Assuming that this corresponds to $V_\text{d}$ in (\ref{Eq:Potential-DepletionApproximation}), one finds a density of ionized acceptors $\rho/e = \unit[6\times10^{10}]{cm^{-3}}$, for the sensor thickness of \unit[1]{mm}. Modelling those ionized dopants as effective, shallow acceptors close to the valence band edge, all are nearly fully ionized in equilibrium (each acceptor has captured one electron from the valence band). Accordingly, in Fig.\,\ref{Fig:IsolatedEnergyLevels} these acceptors are drawn as a narrow band close to the valence band. The full ionization state is expressed by the acceptor band lying well below the Fermi level, thus indicating an occupation probability $f(\E)$ close to unity.

Full ionization means that in equilibrium $p = \unit[6\times10^{10}]{cm^{-3}}$ as well, and thus (\ref{Eq:Fermi-Level}) yields $\F_\text{CdTe} = \F_\text{i} - \unit[0.34]{eV}$ (approximately \unit[0.43]{eV} above the valence band edge), as drawn in Fig.\,\ref{Fig:IsolatedEnergyLevels}. The Fermi level below mid-gap indicates the p-type nature of the material.

A \emph{deep acceptor level}, mentioned above, is also indicated in Fig.\,\ref{Fig:IsolatedEnergyLevels} for later usage.

\section{Equilibrium energy levels in depletion approximation}
\label{Levels-Depletion-Approximation}

When the electrodes and the semiconductor are brought into contact, according to the definition of the Fermi level, charges will flow so that in equilibrium the Fermi level throughout the system is constant.\footnote{This is also the origin of the contact or Volta potential if two dissimilar metals are brought into close contact.} This means that the net flow of charges due to electric fields (electric potential) and due to concentration gradients (chemical potential) just balances. In Fig.\,\ref{Fig:IsolatedEnergyLevels} this can be expressed by joining the three individual level diagrams and then offsetting them vertically so that the Fermi levels indicated by the blue lines are at the same height. The semiconductor bands deform smoothly away from the interface.

The band edges in an energy band diagram indicate the \textbf{electrostatic potential energy of an electron} at a given location.\footnote{Increasing the energy of a conduction band electron with respect to the conduction band edge increases its kinetic energy and moves it further up. In analogy, increasing the energy of a valence band hole with respect to the valence band edge increases its kinetic energy, but moves it further down in the energy band diagram. The latter is equivalent to moving an electron further up.} To describe quantitatively how the bands deform when materials are in contact, (\ref{Eq:Poisson-Equation}) has to be solved, thus the charge density has to be known.

\begin{figure}
\centering
\includegraphics[width=0.9\textwidth]{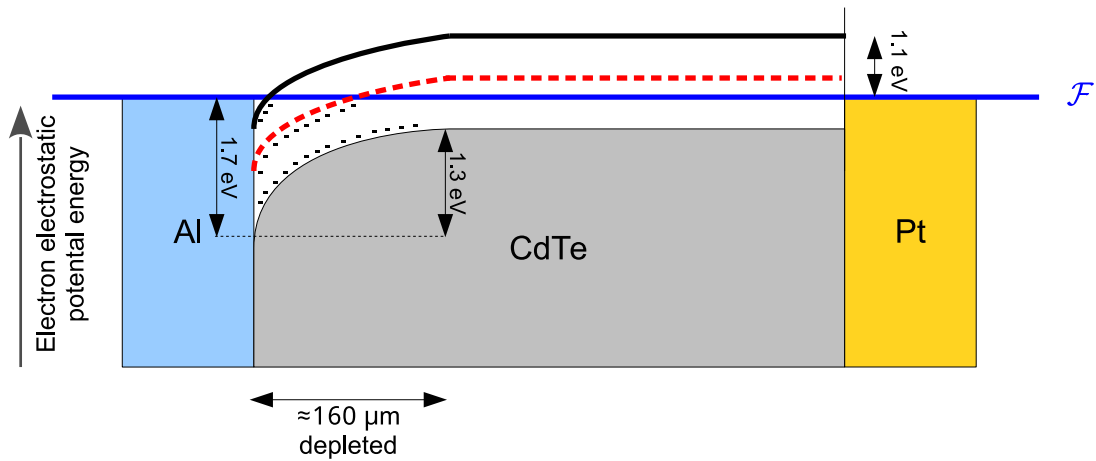}
\caption[Equilibrium energy level diagram]{Equilibrium electronic energy level diagram for aluminium, CdTe, and platinum in contact. Nearly all energy levels below the constant Fermi level are occupied with electrons.}
\label{Fig:BandBending}
\end{figure}

In Section~\ref{Sect:Electric-Field}, the \emph{depletion approximation} has been employed implicitly to arrive at (\ref{Eq:Potential-DepletionApproximation}), (\ref{Eq:ElectricField}) and Fig.\,\ref{Fig:ElectricField}. This approximation amounts to stating that a certain thickness of the semiconductor close to an interface is fully depleted of mobile charges and thus has a constant space-charge density resulting from the ionized dopants. The potential drop across this depleted region can be directly calculated by integrating (\ref{Eq:Poisson-Equation}) for $\rho(x)=Ne$, assuming a constant ionized dopant density $N$, for a certain thickness $x<W$. The change of electric potential over the depleted region is called \emph{built-in voltage}, $V_\text{bi}$, so the depletion width without applied bias voltage is, from (\ref{Eq:Potential-DepletionApproximation}),
\begin{equation}
W=\sqrt{\frac{2\varepsilon_{\text{r}}\varepsilon_{0}V_\text{bi}}{Ne}}, \qquad\qquad e V_\text{bi} = \Phi_\text{CdTe} - \Phi_\text{Al}.
\label{Eq:DepletionWidth}
\end{equation}
The electrostatic potential is constant beyond this distance, and therefore the electric field vanishes, and the crystal is undepleted (conducting). The quadratic dependence (\ref{Eq:Potential-DepletionApproximation}) of the electric potential on distance for $V=0$ then results in a band diagram as shown in Fig.\,\ref{Fig:BandBending}. Numerically, with $V_\text{bi}$ = \unit[1.3]{V} and $N=\unit[6\times10^{10}]{cm^{-3}}$, $W$=\unit[160]{\textmu m}. The relative spacing between the conduction band, valence band, and the deep acceptor level remains constant, as these are intrinsic characteristics of the material.


Since holes are the majority carriers in a p-type semiconductor, hole injection from the aluminium electrode into the valence band (equivalent to removal of an electron from the semiconductor) is inhibited by the potential barrier height of about \unit[1.7]{eV} (the only free electron states in aluminium are near the Fermi level). Holes from the undepleted (conducting) bulk part of the CdTe crystals need to acquire an extra energy of \unit[1.3]{eV} to enter the aluminium electrode (again, a hole moving towards the left is equivalent to an electron moving to the right, which then sees an increasing valence band energy). A Schottky contact, blocking majority carriers, has formed near the aluminium electrode.

No appreciable bending occurs near the platinum interface since the Fermi levels of the isolated materials are nearly identical, so only very little charge has to flow across the boundary to equalize the levels, and consequently the space-charge density is low. The effect of deep acceptor traps on the band structure is treated in the next section.

Special care is needed for the conduction band, which is partly below the Fermi level in Fig.\,\ref{Fig:BandBending} and then violates the assumption of non-degeneracy. More generally, the type of a semiconductor, n or p, is determined by its Fermi level being above or below the intrinsic level $\F_\text{i}$, which is located approximately at mid-gap after (\ref{Eq:Intrinsic-Level}). The mid-gap level falls below the Fermi level at some distance from the interface in Fig.\,\ref{Fig:BandBending}, therefore an inversion occurs - the minority carriers in the conduction band will have a higher density than the majority carriers. The density of states in the conduction band of typically \unit[$10^{17}$]{cm$^{-3}$} is orders of magnitude higher than the effective doping density, therefore the charge density in the vicinity of the interface, where the conduction band is below the Fermi level, is dominated by conduction band electrons. In a very short distance, this large charge will result in an electrostatic potential lifting the conduction band above the Fermi level, when the usual considerations apply again. Section~\ref{Sect:Degeneracy} considers this more in depth.

\section{General equilibrium band diagram}
\label{Sect:GeneralEquilibriumBandDiagram}

To find a generally valid, quantitative description of the band diagram, i.e. of the dependence of the electrostatic potential on position, an expression for $\rho(x)$ in (\ref{Eq:Poisson-Equation}) has to be found. Considering in the following one single, shallow acceptor (modelling the effective result of compensation between donors and acceptors) with concentrations $N_\text{a}$ (assumed fully ionized since shallow) and one deep acceptor trap with energy level at $\E_\text{t}(x)$ and constant concentration $N_\text{t}$, the charge density at position $x$ is
\begin{equation}
\frac{\rho(x)}{e} = p(x) - n(x) - N_\text{a} - N_\text{t}f(\E_\text{t}(x)).
\label{Eq:ChargeDensity_GeneralBandDiagram}
\end{equation}
The charge density contains the mobile charges and the fixed space-charge (here negative since only acceptors are considered).

Inserting (\ref{Eq:ChargeDensity_GeneralBandDiagram}) into (\ref{Eq:Poisson-Equation}) yields
\begin{equation}
\frac{\varepsilon_{\text{r}}\varepsilon_{0}}{e} \frac{\partial^2 U(x)}{\partial x^2} = p(x) - n(x) - N_\text{a} - N_\text{t} f(\E_\text{t}(x)).
\label{Eq:Poisson_GeneralBandDiagram}
\end{equation}
The coordinate is chosen such that $x=0$ is the location of the interface. The semiconductor is assumed infinite in extent, and neutral away from the interface, $\lim_{x\to\infty}\rho(x) = 0$.

For compact notation, the definition $y(x)=(\F_\text{i}(x)-\F)/(kT)$ will be used. Further, $\lim_{x\to\infty}y(x) = y_\text{bulk}$ is defined by the bulk material properties, and $y(0) = y_\text{surface}$ by the contact characteristics, such that $y_\text{surface}-y_\text{bulk} = -eV_\text{bi}/(kT)$. The relation $\partial y/\partial x = -e/(kT) \, \partial U/\partial x$ holds.

The assumption of a position-independent Fermi level $\F$ assures that the currents from carrier drift and from diffusion cancel everywhere for both electrons and holes.

Interface states, due to electron or hole levels within the band gap at the surface, are completely neglected in this derivation. Depending on the nature and quality of the interface, they can have a large impact and result in Fermi level pinning, potentially rendering the Schottky barrier height almost independent of the metal work function.

\subsection{Non-degenerate case without traps}

First, the case without traps ($N_\text{t}=0$) and with non-degenerate location of the Fermi level is considered. The mobile charge densities can then be expressed by
\begin{equation*}
n(x) = n_\text{i} \, e^{-y(x)}, \qquad p(x) = n_\text{i} \, e^{y(x)},
\end{equation*}
and, from the charge neutrality in the bulk, $N_\text{a} = n_\text{i} \, (e^{y_\text{bulk}} - e^{-y_\text{bulk}})$. Inserting all into (\ref{Eq:Poisson_GeneralBandDiagram}),
\begin{equation}
\frac{\partial^2 y(x)}{\partial x^2} = \frac{1}{L_\text{D}^2} \left( e^{y(x)} - e^{-y(x)} - e^{y_\text{bulk}} + e^{-y_\text{bulk}} \right).
\label{Eq:Exact-DifferentialEquation}
\end{equation}
where the \emph{Debye length} $L_\text{D}=\sqrt{\varepsilon_{\text{r}}\varepsilon_{0}kT/(e^2n_\text{i})}$ for intrinsic material has been introduced. Defining normalized derivatives $y^{(n)}=L^n_\text{D} \partial^n y/\partial x^n$, multiplying both sides with $y'$, and integrating formally from the bulk to a distance $x$, this equation becomes
\begin{equation}
\int_{\infty}^{x} y'' y' \,\partial x = \int_{\infty}^{x} \left( e^{y(x)} - e^{-y(x)} - e^{y_\text{bulk}} + e^{-y_\text{bulk}} \right) y' \,\partial x.
\label{Eq:Exact-FormalIntegral}
\end{equation}
Carrying out the integration yields
\begin{equation*}
\frac{1}{2} \left. \left(y'\right)^2 \right|_{\infty}^{x} = \left.\left( e^{y(x)} + e^{-y(x)} + \left( e^{-y_\text{bulk}} - e^{y_\text{bulk}} \right) y(x) \right) \right|_{\infty}^{x}.
\end{equation*}
$y'$ is proportional to the electric field, which vanishes in the bulk away from the interface, so
\begin{equation*}
\frac{1}{\sqrt{2}} y' = \sqrt{F(y)}, \qquad F(y) = e^{y} + e^{-y} + e^{y_\text{bulk}} (y_\text{bulk} - y - 1) + e^{-y_\text{bulk}} (y - y_\text{bulk} - 1).
\end{equation*}
Separating the variables and integrating from the interface to position $x$,
\begin{equation}
\boxed{\int_0^x \partial x = x = \frac{L_\text{D}}{\sqrt{2}} \int_{y_\text{surface}}^{y(x)} \frac{\partial y}{\sqrt{F(y)}}.}
\label{Eq:Exact-Solution}
\end{equation}
There is no analytic result for the integral on the right hand side. But since $y_\text{surface}$ is given, this equation is an implicit equations for $y(x)$, and thus for the intrinsic level $\E_\text{i}(x)$ relative to the Fermi level. The equation can be evaluated and inverted numerically. Conduction and valence bands have a fixed offset to $\E_\text{i}$.

\begin{figure}
\centering
\begin{subfigure}{0.8\textwidth}
	\includegraphics[width=\textwidth]{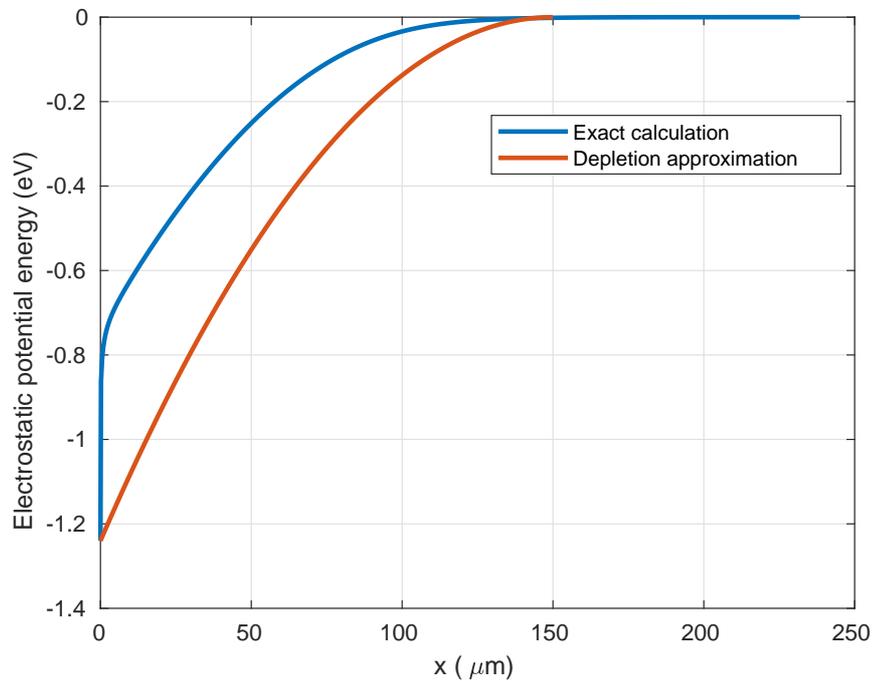}
    \caption{Electron electrostatic potential energies}
	\label{Fig:NumericalBandStructure-a}
\end{subfigure}
\begin{subfigure}{0.8\textwidth}
	\includegraphics[width=\textwidth]{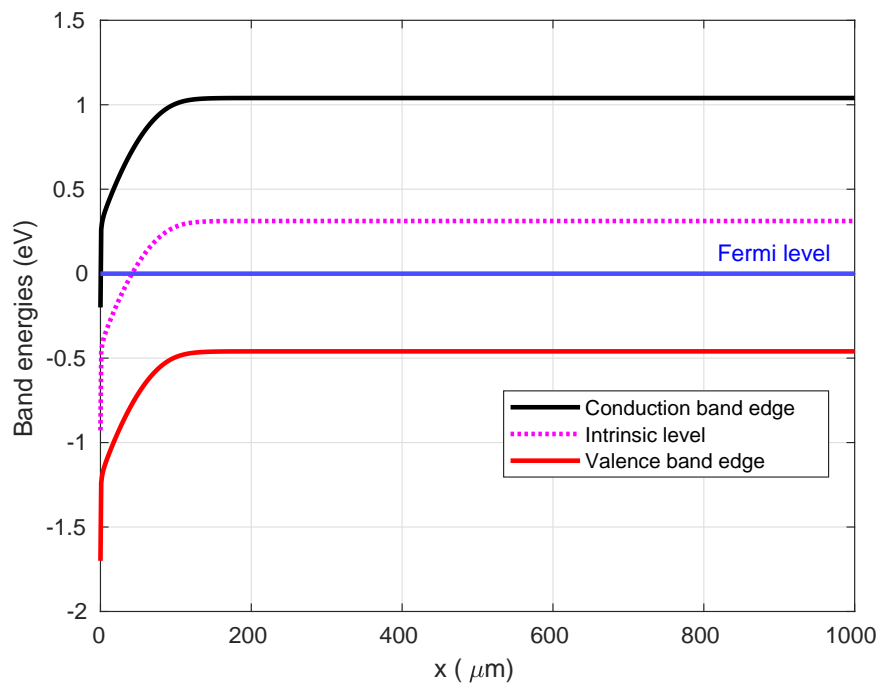}
    \caption{Band diagram for the exact solution}
   	\label{Fig:NumericalBandStructure-b}
\end{subfigure}
\caption[Band diagram comparison between depletion approximation and exact solution]{Illustrations of the depletion approximation and the exact solution. In the immediate vicinity of the interface, the conduction band edge drops below the Fermi level.}
\label{Fig:NumericalBandStructure}
\end{figure}

For the parameters in Section~\ref{Levels-Depletion-Approximation}, $y_\text{bulk} = \ln (p/n_\text{i}) = 12$, $y_\text{surface} = y_\text{bulk} - eV_\text{bi}/(kT) = -37$, and $L_\text{D} = \unit[7.3]{mm}$. A comparison of the exact result (\ref{Eq:Exact-Solution}) and the depletion approximation is shown in Fig.\,\ref{Fig:NumericalBandStructure}. The steep rise in potential near the interface seen for the exact solution is due to conduction band falling below the Fermi level very near the interface. The conduction band has a very high density of states, compared to the effective acceptor concentration, thus a narrow, dense electron sheet forms.

The sign of $y(x)$ indicates if the Fermi level is above or below the intrinsic level, thus defines the type of the semiconductor. In the present case, $y(x)=0$ for $x \approx$ \unit[42]{\textmu m}, therefore a \emph{type inversion} occurs at this distance from the interface. The bulk behaves p-type, with valence band holes being the more numerous carriers, whereas close to the interface, conduction band electrons dominate.

The electric field $E(x)$ as function of position can be determined from (\ref{Eq:Exact-Solution}) by
\begin{equation}
E(x) = \frac{kT}{e} \frac{\partial y}{\partial x} = \frac{kT}{e} \left( \frac{\partial x}{\partial y} \right)^{-1} = \frac{k T}{e L_\text{D}} \sqrt{2 F(y)}.
\label{Eq:Exact-ElectricField}
\end{equation}
Since $F(y_\text{bulk}) = 0$, the electric field vanishes in the bulk, as expected. At the interface, the electric field is very high due to the high charge density in the conduction band. Correct evaluation will need to consider the degeneracy (see next section).

\vspace{2ex}\noindent{\bf Comparison to the depletion approximation}\nopagebreak[4]\\[0.5ex]
The depletion approximation states that $n(x)$ and $p(x)$ are zero up to as distance $W$ from the interface, beyond which the semiconductor is neutral. The charge density is therefore $\rho(x) = -e N_\text{a}$ for $x \le W$ and $\rho(x) = 0$ for $x>W$, with $W$ given by (\ref{Eq:DepletionWidth}) (recall that in this section $N_\text{t} = 0$, thus $N=N_\text{a}$ in that formula). Therefore, $U(x)$ in (\ref{Eq:Poisson_GeneralBandDiagram}), or, equivalently, $y(x)$, must have a quadratic dependence on $x$. Using the boundary values $y_\text{surface}$ at $x = 0$ and  $y_\text{bulk}$ at $x = W$, one finds
\begin{equation*}
y(x) = (y_\text{surface} - y_\text{bulk}) \left( \frac{x^2}{W^2} - \frac{2 x}{W} \right) + y_\text{surface} \qquad \text{for}\; x \le W.
\end{equation*}
Noting that $W = \sqrt{2\varepsilon_{\text{r}}\varepsilon_{0}kT/(e^2 N_\text{a})} \sqrt{y_\text{surface} - y_\text{bulk}}$, and defining the electric potential $U(x)$ such that it vanishes in the bulk, this translates to
\begin{equation*}
U(x) = -V_\text{bi} \left( \frac{x}{W} - 1 \right)^2 \quad\text{for}\; x \le W, \qquad\qquad U(x) = 0 \quad\text{otherwise}.
\end{equation*}
The resulting electric field is
\begin{equation*}
E(x) = -\frac{\partial U}{\partial x} = \frac{2 V_\text{bi}}{W} \left( \frac{x}{W} - 1 \right).
\end{equation*}
At the interface, $E(0) = -2V_\text{bi}/W$.

\subsection{Consideration of conduction band degeneracy}
\label{Sect:Degeneracy}

The considerations in Section~\ref{Levels-Depletion-Approximation} and the calculations in the previous section have shown that, close to the interface, the conduction band lies below the Fermi level. This actually invalidates the approximation of non-degeneracy that was made in the previous section, in particular the expression for the conduction band electron density $n(x) = n_\text{i} \, e^{-y(x)}$. The exact expression follows from integrating over the product of the conduction band density of states and the Fermi-Dirac distribution, and is
\begin{equation*}
n(x) = \frac{2 N_\text{c}}{\sqrt{\pi}} \int_{0}^{\infty} \frac{\sqrt{\eta}}{1 + \exp(\eta - \eta_\text{c})} \text{d}\eta = \frac{2 N_\text{c}}{\sqrt{\pi}} F_{1/2}(\eta_\text{c}),
\end{equation*}
where $\eta_\text{c} = (\F - E_\text{c}(x))/(k T)$ and $F_{1/2}$ is the Fermi-Dirac integral of order 1/2. The condition of non-degeneracy is $\eta_\text{c} \ll 0$ (the conduction band edge is at least several $kT$ above the Fermi level), in which case the integral can be evaluated easily, as $F_{1/2}(\eta_\text{c}) \approx \exp \eta_\text{c}$.

\begin{figure}
\centering
\includegraphics[width=0.8\textwidth]{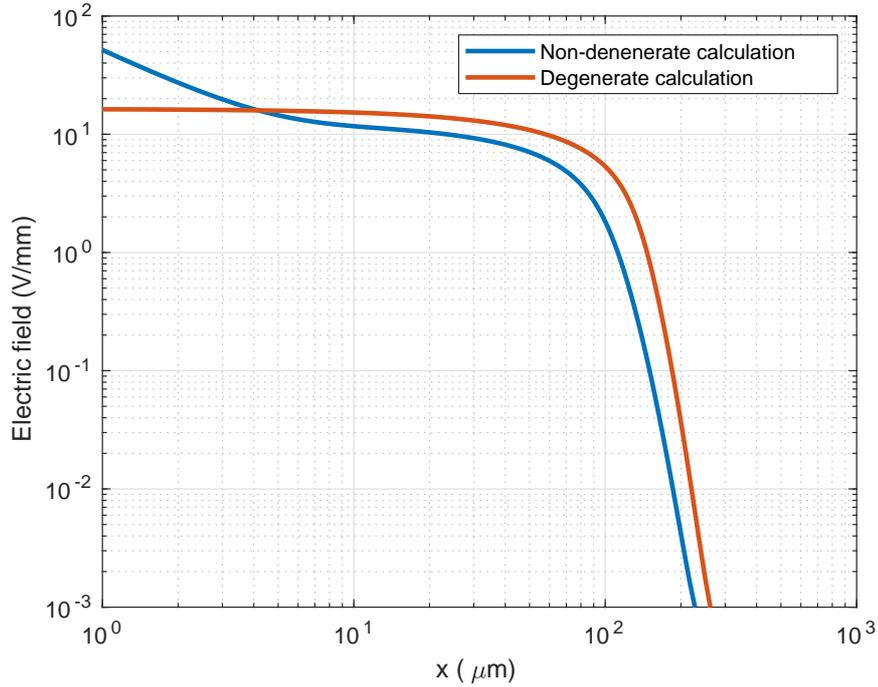}
\caption[Degenerate and non-degenerate calculation of the electric field]{Degenerate and non-degenerate calculation of the electric field near the interface}
\label{Fig:DegenerateElectricField}
\end{figure}

Since degeneracy has to be considered only for the conduction band near the interface, not for the valence band or the bulk behaviour, replacing $\exp(-y(x))$ by $F_{1/2}(-y(x))$ in (\ref{Eq:Exact-DifferentialEquation}) is the only required modification for correct handling of degeneracy. Since furthermore $\int\!F_{1/2}(y) \,\text{d}y = F_{3/2}(y)$, the integration in (\ref{Eq:Exact-FormalIntegral}) can still be carried out. Finally, (\ref{Eq:Exact-Solution}) yields the correct solution including degeneracy if $F(y)$ is modified to
\begin{equation*}
F(y) = e^{y} + e^{y_\text{bulk}} (y_\text{bulk} - y - 1) + e^{-y_\text{bulk}} (y - y_\text{bulk}) + F_{3/2}(-y) - F_{3/2}(-y_\text{bulk}).
\end{equation*}
A comparison of the electric field obtained by (\ref{Eq:Exact-ElectricField}) with the degenerate and non-degenerate expressions for $F(y)$ is shown in Fig.\,\ref{Fig:DegenerateElectricField}. Close to the interface, the degenerate expression yields a field constant with distance. Inserting $y_\text{surface}$ yields a very high value of \unit[460]{kV/mm} with the non-degenerate expression, but more realistic \unit[16]{V/mm} with the degenerate expression.

For the case at hand, the conduction band drops by about \unit[200]{mV} or $8kT$ below the Fermi level. The non-degenerate expression for $n(x)$ overestimates the electron density right at the interface by about a factor of 200 ($n(0) = \unit[1.2 \cdot 10^{19}]{cm^{-3}}$ using the correct expression $F_{1/2}$), but the discrepancy quickly reduces with rising conduction band energy. Therefore, the details of the charge layer immediately at the interface are not correctly captured by (\ref{Eq:Exact-Solution}) with the non-degenerate expression for $F(y)$, but the band diagram Fig.\,\ref{Fig:NumericalBandStructure-a} would be changed only imperceptibly if the exact expression for $n(x)$ would be used.

There are also other effects of a very high carrier concentration that are not considered even when using $F_{1/2}$, among them many-body effects that can change the band gap and result in tailing of the band edges. Since, therefore, correctly handling the immediate interface region is complicated, but also not important for the remaining conclusions to be drawn, the non-degenerate expressions will be used in the following.

\subsection{Non-degenerate case with traps}
\label{Sect:Exact-with-traps}

Now deep traps are included, $N_\text{t} > 0$. Their energy level $\E_\text{t}$ is expressed by a parameter $Z$,
\begin{equation*}
Z = \exp\frac{\E_\text{t} - \F_\text{i}}{k T},
\end{equation*}
which is independent of position, so that the Fermi-Dirac distribution can be written as
\begin{equation*}
f(\E_\text{t}(x)) = \frac{1}{1 + Z \, e^{y(x)}}.
\end{equation*}

\begin{figure}
\centering
\includegraphics[width=0.8\textwidth]{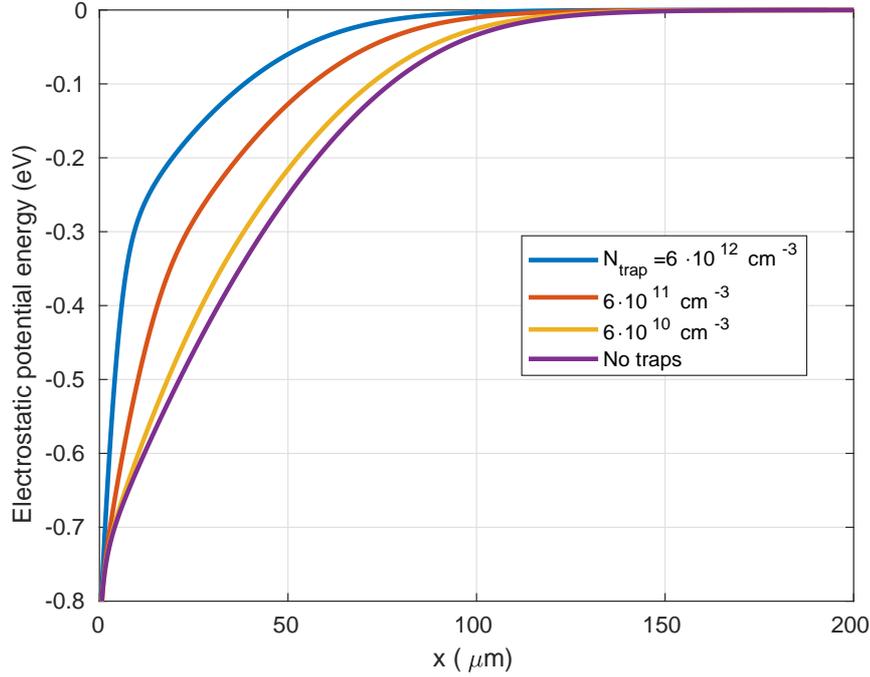}
\caption[Potential energies for a mid-gap trap]{Electrostatic potential energies for a mid-gap trap at \unit[0.77]{eV} above the valence band with different trap densities. The bulk hole concentration is $\unit[6 \cdot 10^{10}]{cm^{-3}}$, the bulk Fermi level \unit[0.46]{eV} above the valence band.}
\label{Fig:DeepTrap-Potentials}
\end{figure}

Maintaining the non-degenerate expressions for $n(x)$ and $p(x)$ and the definitions used before, the equation
\begin{equation*}
y'' = e^{y(x)} - e^{-y(x)} - e^{y_\text{bulk}} + e^{-y_\text{bulk}} - \frac{N_\text{t}}{n_\text{i}}  \left( \frac{1}{1 + Z \, e^{y(x)}} - \frac{1}{1 + Z \, e^{y_\text{bulk}}} \right)
\end{equation*}
is obtained instead of (\ref{Eq:Exact-DifferentialEquation}). Except for the last term, this is identical to the formulas leading to (\ref{Eq:Exact-Solution}). The extra term is handled similar to (\ref{Eq:Exact-FormalIntegral}) by defining a new function $G(y)$,
\begin{eqnarray*}
G(y) & = & \int_{\infty}^{x} \left( \frac{1}{1 + Z \, e^{y(x)}} - \frac{1}{1 + Z \, e^{y_\text{bulk}}} \right) y' \;\partial x \\
 & = & \int_{y_\text{bulk}}^{y} \left( \frac{1}{1 + Z \, e^{\bar{y}}} - \frac{1}{1 + Z \, e^{y_\text{bulk}}} \right) \partial \bar{y} \\
 & = & \left. \left( \bar{y} - \ln (Z \, e^{\bar{y}} + 1) - \frac{\bar{y}}{1 + Z \, e^{y_\text{bulk}}} \right) \right|_{y_\text{bulk}}^{y}\\
 & = & y - y_\text{bulk} + \ln \frac{Z \, e^{y_\text{bulk}} + 1}{Z \, e^{y} + 1} - \frac{y - y_\text{bulk}}{1 + Z \, e^{y_\text{bulk}}}.
\end{eqnarray*}
The electrostatic potential energy is again defined implicitly, now by the relation
\begin{equation}
\boxed{x = \frac{L_\text{D}}{\sqrt{2}} \int_{y_\text{surface}}^{y(x)} \frac{\partial y}{\displaystyle \sqrt{F(y) + \frac{N_\text{t}}{n_\text{i}} G(y)}}.}
\label{Eq:Exact-Solution-With-Traps}
\end{equation}
Resulting electrostatic potential energies for different trap densities are shown in Fig.\,\ref{Fig:DeepTrap-Potentials}. As expected, the ionized traps add space-charge and shorten the depletion region.

\section{Polarization effect}

A possibly performance-limiting characteristic of Schottky-type CdTe detectors is the \emph{polarization effect}. This term refers to a time-dependent change, typically a degradation, of the detector behaviour when bias is applied for a longer time. The leakage current usually increases over time, resulting in higher read-out noise. It has also been found that parts of a detector become practically insensitive.

The consensus is that deep acceptor traps are at the origin of this behaviour. They get progressively ionized when the detector is biased and increase the space-charge. If this acceptor density is large compared to the effective shallow doping density (equivalently, large compared to the equilibrium carrier density), after some time the traps will dominate the space-charge, and a bias voltage initially large enough to fully deplete the sensor volume might not be sufficient anymore.

\subsection{Band levels with external bias}

With a sufficiently large reverse external bias $V$, the details of the electrostatic potential near the Schottky contact, including degeneracy effects of the conduction band, are of no relevance to the discussion of polarization. Also, the exact calculations in Section~\ref{Sect:GeneralEquilibriumBandDiagram} assume an infinite semiconductor and would need to be adapted to be applicable to the actual finite sensors. Instead, the depletion approximation will be employed, neglecting at first acceptor traps ($N_\text{t} = 0$).

\begin{figure}
\centering
\includegraphics[width=0.8\textwidth]{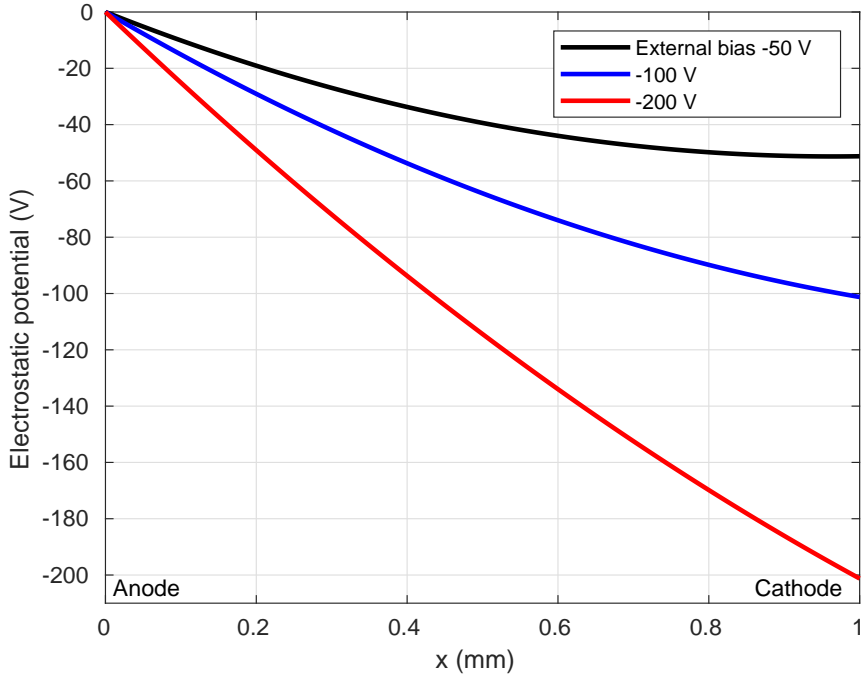}
\caption[Electrostatic potential with external bias]{Electrostatic potential (\ref{Eq:PotentialWithBias}) for three applied values of the external bias}
\label{Fig:Potential-WithBias}
\end{figure}

The depletion width is in this approximation, by extension from (\ref{Eq:DepletionWidth}),
\begin{equation*}
W=\sqrt{\frac{2\varepsilon_{\text{r}}\varepsilon_{0} (V_\text{bi} - V)}{N_\text{a} e}}.
\end{equation*}
Full depletion of a sensor with thickness $d$ is obtained at $V < V_\text{d}$ with
\begin{equation*}
V_\text{d} = V_\text{bi} - \frac{N_\text{a} e d^2}{2\varepsilon_{\text{r}}\varepsilon_{0}}.
\end{equation*}
Note that by convention $V_\text{bi}>0$. For $d$=\unit[1]{mm} and $N_\text{a}=\unit[6\times10^{10}]{cm^{-3}}$ as assumed before, $V_\text{d}$=\unit[-48]{V}. The electric potential $U(x)$ is quadratic in $x$ from (\ref{Eq:Poisson_GeneralBandDiagram}), and can be written as
\begin{equation}
U(x) = \frac{e N_\text{a}}{2\varepsilon_{\text{r}}\varepsilon_{0}} (x^2 - d x) + \frac{V - V_\text{bi}}{d}x, \qquad\text{for}\quad V < V_\text{d}.
\label{Eq:PotentialWithBias}
\end{equation}
The electrostatic potential for three bias values is plotted in Fig.\,\ref{Fig:Potential-WithBias}. The electric field was shown earlier in Fig.\,\ref{Fig:ElectricField}.

\subsection{Basic description of trap ionization}

Assume there is a deep acceptor concentration $N_\text{t}$, constant throughout the material, and possibly much higher than the effective shallow acceptor doping density. A single trap energy level $\E_\text{t}(x)$ is considered, where the position dependence is included to account for band bending. The concentration of ionized traps that contribute negative space-charge is labelled $N^-_\text{t}(x,t)$, which can be time and position dependent. $N_\text{t}-N^-_\text{t}(x,t)$ is the concentration of neutral (unoccupied) traps.

Starting at $t=0$ from equilibrium conditions,
\begin{equation}
N^-_\text{t}(x,0) = f(\E_\text{t}(x)) \cdot N_\text{t}.
\label{Eq:InitialPolarization}
\end{equation}
As indicated in Fig.\,\ref{Fig:BandBending} and calculated in Section~\ref{Sect:Exact-with-traps}, those traps that lie below the Fermi level will be ionized already in thermal equilibrium, and effectively increase the doping concentration close to the anode.

If now a sufficiently large reverse bias voltage is applied, nearly all mobile charges are swept out by the resulting electric field. Only a very small density of mobile carriers, forming the leakage current, remains. Equilibrium conditions do not apply anymore. In particular, an ionized trap can release its electron only to the conduction band, but not to the valence band since the hole concentration is nearly zero, and thus no free energy levels for the electron are available. If the energy gap to the conduction band is sufficiently large, the trap will remain ionized indefinitely while bias is applied.

Since the thermal ionization process of the deep traps will continue as long as there are empty traps, the space-charge will gradually increase. This will increase the electric field near the Schottky anode\footnote{This can be seen by (\ref{Eq:ElectricField}): increasing the space-charge density increases $V_\text{d}$, and thus also increases the magnitude of the electric field at the anode, $E(d)$.}, and therefore, through image force lowering of the Schottky barrier, increase the leakage current by thermionic emission\footnote{Thermionic emission refers to thermal excitations over the barrier, the most probable process to overcome the barrier for the parameters relevant here. The rate depends exponentially on the barrier height.}. A higher electric field also increases the probability of Frenkel-Poole charge transport through the depleted region. More leakage current decreases the energy resolution by its shot noise contribution in the amplification stage. If the trap density is large enough, the bias voltage will eventually not be sufficient to deplete the full detector volume and a low-field region near the cathode can develop, affecting primarily low energy X-rays that convert close to the surface. Increasing the applied bias voltage can recover full depletion, but results in an even higher leakage current.

To address these dynamic effects quantitatively, first the deep acceptor trapping and de-trapping rates in equilibrium are calculated, and then the assumption is made that under bias the trapping rate is unchanged while the de-trapping rate becomes zero. This approach closely follows \cite{Shock52}.

\subsection{Equilibrium rate considerations}

The rate of trap ionization per volume, $R_\text{ion}$, i.e. of an acceptor acquiring an electron from the valence band, can be written as
\begin{equation*}
R_\text{ion}(x) = N_\text{v}\,f(\E_\text{v}(x)) <\!\!v \, \sigma_\text{ion}\!\!> N_\text{t}\left( 1-f(\E_\text{t}(x)) \right).
\end{equation*}
The term left of the angled brackets is the number density of electrons in the valence band, the term on the right side the number density of unoccupied traps. The brackets indicate a suitable average of the electron thermal velocity $v$ times the velocity-dependent electron capture cross-section $\sigma_\text{ion}$ of the trap. The equation expresses the notion that an electron that passes within an area $\sigma_\text{ion}$ around an unoccupied trap is captured. Averaging is required to account for the thermal velocity distribution. Since only electrons that have sufficient energy can overcome the energy barrier to the trap, a Boltzmann factor will appear.

A similar expression is applicable for the inverse process, i.e. release of a electron from a deep acceptor into the valence band, where however no energy barrier is present. The release rate per volume $R_\text{rel}$ is
\begin{equation*}
R_\text{rel}(x) = N_\text{v}\,(1-f(\E_\text{v}(x))) <\!\!v \, \sigma_\text{rel}\!\!> N_\text{t}\,f(\E_\text{t}(x)).
\end{equation*}
In thermal equilibrium $R_\text{ion}(x)=R_\text{rel}(x)$ holds everywhere, thus\footnote{The identity $1-f(\E)=f(\E)\exp((\E-\F)/kT)$ is useful for evaluating hole occupation probabilities.}
\begin{equation}
\frac{<\!\!v \, \sigma_\text{rel}\!\!>}{<\!\!v \, \sigma_\text{ion}\!\!>} = \frac{f(\E_\text{v})(1-f(\E_\text{t}))}{f(\E_\text{t})(1-f(\E_\text{v}))} = \exp\left(\frac{\E_\text{t}-\E_\text{v}}{kT}\right).
\label{Eq:RatioCrossSections}
\end{equation}
Note that although $\E_\text{t}$ and $\E_\text{v}$ might be position dependent, their difference is not, and therefore $x$ has been omitted. Capture from the conduction band is ignored since this band is almost empty in equilibrium. Release to the conduction band is suppressed by the additional energy barrier, but might become significant if the hole density becomes very low and the density of ionized traps high.

\subsection{Ionization rate with applied bias}

For the non-equilibrium case with applied bias, it is now assumed that a similar formalism applies, but with separate Fermi levels, the \emph{quasi Fermi levels} $\F_\text{v}$ and $\F_\text{t}$, for the valence band and the traps. In analogy to (\ref{Eq:FermiDiracDistribution}), a \emph{quasi Fermi distribution}\footnote{In the terminology of the chemical potential, the electrons in the valence band, the conduction band, and the traps are each considered a separate particle species. The term \emph{quasi} expresses the expectation that the occupation probabilities of the energy levels within each species still follows the Fermi-Dirac distribution for the same temperature.}
\begin{equation}
f_\text{q}(\E,\F_\text{v,t}) = \frac{1}{1+\exp\left(\frac{\E-\F_\text{v,t}}{kT}\right)},
\label{Eq:QuasiFermiDistribution}
\end{equation}
is defined. Assuming that (\ref{Eq:RatioCrossSections}) still holds, the net ionization rate $R$ can be written as
\begin{eqnarray}
R  & = & R_\text{ion} - R_\text{rel} \nonumber\\
   & = & N_\text{v} N_\text{t} \,f_\text{q}(\E_\text{v},\F_\text{v}) f_\text{q}(\E_\text{t},\F_\text{t}) \left( \exp\left( \frac{\E_\text{t}-\F_\text{t}}{kT} \right) - \exp\left( \frac{\E_\text{t}-\F_\text{v}}{kT} \right) \right) \, \!<\!\!v \, \sigma_\text{ion}\!\!> \!\!.
\label{Eq:NetIonizationRate}
\end{eqnarray}
In equilibrium $\F_\text{v} = \F_\text{t}$, and thus $R=0$ as expected.

\begin{figure}
\centering
\includegraphics[width=0.9\textwidth]{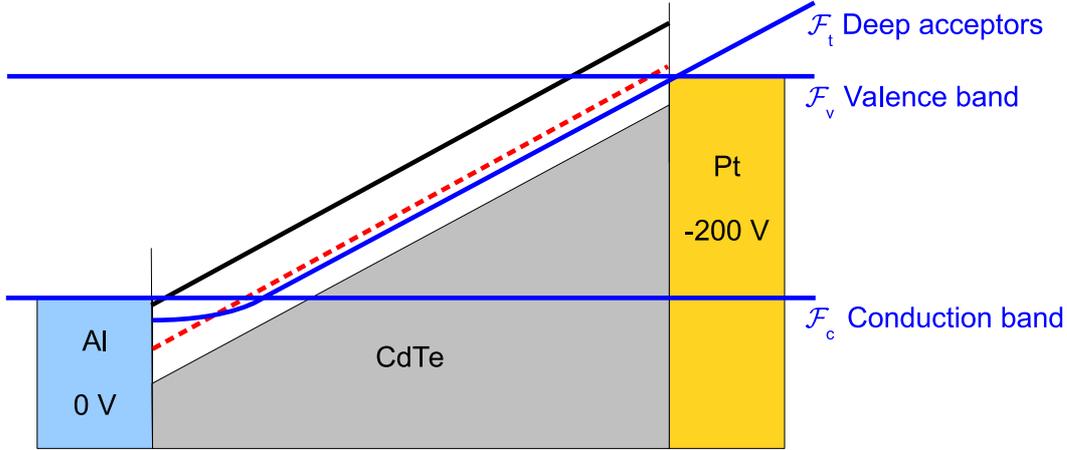}
\caption[Energy level diagram with reverse bias]{Energy level diagram after \unit[-200]{V} reverse bias has just been applied (not to scale). As seen in Fig.\,\ref{Fig:Potential-WithBias}, the band edges are nearly straight for larger bias.}
\label{Fig:BandsWithBias}
\end{figure}

With external bias, the quasi Fermi levels won't be equal, and can also become position dependent. Making the assumption of zero leakage current, however, will render the valence band and conduction band quasi Fermi levels position independent (any effective current is the result of non-constant electrochemical potential). The occupation of the deep trap acceptors will initially be the same as prior to the application of bias. Therefore, $\E_\text{t}-\F_\text{t}$ in (\ref{Eq:QuasiFermiDistribution}) has to remain constant. The resulting level diagram is shown in Fig.\,\ref{Fig:BandsWithBias}.

$\F_\text{t}$ is introduced mainly for formal reasons. Since the traps are not mobile, changes in the trap occupation can only occur via the conduction or valence bands, not by drift or diffusion of the 'trap species'. However, the inclined trap Fermi level in the figure illustrates that the depicted situation is not a steady state. With the above assumption that traps are interacting only with the valence band, the trap ionization will increase until $\F_\text{t}$ and $\F_\text{v}$ are equal.

A time constant $\tau_\text{ion}$ for the net ionization process can be defined by dividing the density of unoccupied traps by the net ionization rate, giving
\begin{equation*}
\frac{1}{\tau_\text{ion}(x)} = \frac{R}{(1-f_\text{q}(\E_\text{t},\F_\text{t})) N_\text{t}} = N_\text{v}\,f_\text{q}(\E_\text{v},\F_\text{v}) <\!\!v \, \sigma_\text{ion}\!\!> \left(1-\exp\left(\frac{\F_\text{t}(x)-\F_\text{v}}{kT}\right)\right).
\end{equation*}
Just after applying a large reverse bias, $\F_\text{v} \gg \F_\text{t}(x)$ (except close to the cathode, see Fig.\,\ref{Fig:BandsWithBias}), so the exponential vanishes. There are also very few holes, $f_\text{q}(\E_\text{v}(x),\F_\text{v}) \approx 1$, therefore the time constant at the beginning of polarization becomes
\begin{equation*}
\tau^\text{init}_\text{ion} = \frac{1}{N_\text{v} <\!\!v \, \sigma_\text{ion}\!\!>}.
\end{equation*}

A similar reasoning leads to the time constant $\tau_\text{rel}$ for the release process,
\begin{equation*}
\frac{1}{\tau_\text{rel}(x)} = \frac{-R}{f_\text{q}(\E_\text{t},\F_\text{t}) N_\text{t}} = N_\text{v}\,f_\text{q}(\E_\text{v},\F_\text{v}) <\!\!v \, \sigma_\text{ion}\!\!> \left(\exp\left(\frac{\E_\text{t}(x)-\F_\text{v}}{kT}\right) - \exp\left(\frac{\E_\text{t}(x)-\F_\text{t}}{kT}\right)\right).
\end{equation*}
The minus sign is introduced since $R$ is negative in case the release process dominates. For a fully polarized detector $\F_\text{v} = \F_\text{t}(x)$. If the applied bias is then reduced to \unit[0]{V}, initially $\F_\text{v} \ll \F_\text{t}(x)$. $f_\text{q}(\E_\text{v}(x),\F_\text{v}) \approx 1$ still approximately holds, so that the initial time constant
\begin{equation*}
\tau^\text{init}_\text{rel} = \frac{1}{N_\text{v} <\!\!v \, \sigma_\text{ion}\!\!>} \exp\left(\frac{\F_\text{v} - \E_\text{t}(x)}{kT}\right).
\end{equation*}
Since $\E_\text{t} > \F_\text{v}(x)$ (see Fig.\,\ref{Fig:BandBending}), the exponential can be significantly smaller than unity, and then $\tau^\text{init}_\text{rel} \ll \tau^\text{init}_\text{ion}$. This demonstrates that the release process is significantly faster than the ionization process, reflecting the fact that there is no energy barrier to overcome, if only the electrons in the trap level find holes in the valence band below to recombine with. Still, this \emph{depolarization} process is not instantaneous because the hole density is finite.

\subsection{Further considerations}

Further practical progress is hampered by the need to know $<\!\!v \, \sigma_\text{ion}\!\!>$ for performing numerical evaluations. Qualitative data indicates that the polarization time scale is in excess of a week at temperatures of -20\textdegree C. The long-term leakage current measurements shown in Fig.\,\ref{Fig:LongTermCurrent} indicate that the current doubles in 5.5 days at -6\textdegree C and in 12 hours at +4\textdegree C. At +25\textdegree C, a time scale of minutes is found.

\begin{figure}
\centering
\includegraphics[width=0.8\textwidth]{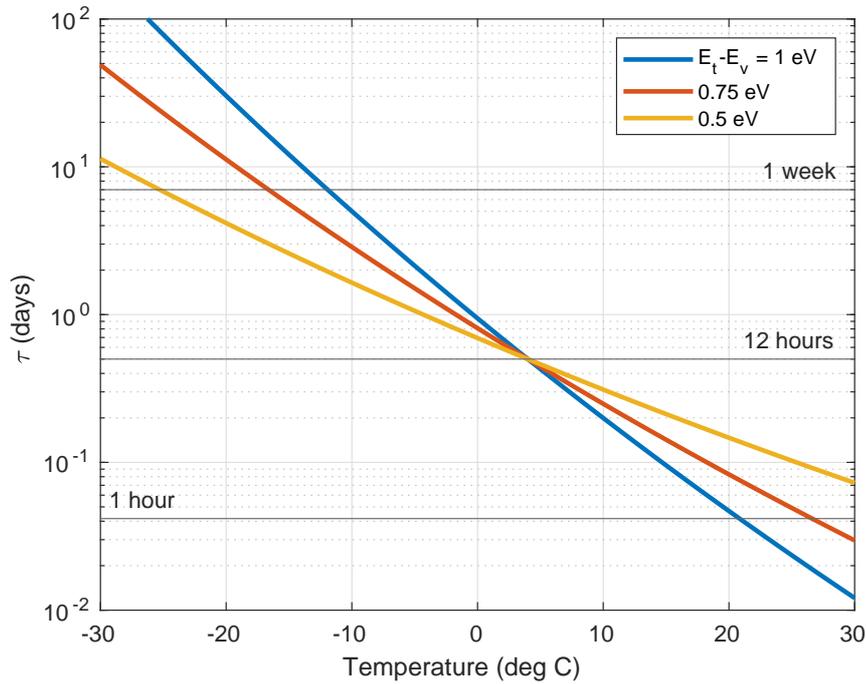}
\caption[Initial ionization timescale]{Initial ionization timescale (\ref{Eq:IonizationTimescale}) for three values of $\E_\text{t}-\E_\text{v}$}
\label{Fig:IonizationTimescale}
\end{figure}

Assuming for the purpose of a crude estimate that all temperature dependence in (\ref{Eq:RatioCrossSections}) belongs to the ionization term, and using that $N_\text{v} \sim T^{3/2}$, the temperature dependence of the initial ionization time scale becomes
\begin{equation}
\tau^\text{init}_\text{ion} \sim \left( T^{3/2} \; \exp\left(\frac{\E_\text{v}-\E_\text{t}}{kT}\right) \right)^{-1}.
\label{Eq:IonizationTimescale}
\end{equation}
This relation is plotted in Fig.\,\ref{Fig:IonizationTimescale} for three values of $\E_\text{t}-\E_\text{v}$ and assuming that at +4\textdegree C the time scale is 12 hours. Qualitatively, the experimentally observed temperature dependence is reproduced, but there are too few data points, and no estimates of uncertainties, that would allow a quantitative analysis. In principle, an estimate of the relevant trap energy level could be done with this equation.

\chapter{Signal induction}
\label{Sect:SignalInduction}

The carriers induce mirror charges on the detector electrodes. Initially, with an equal number of electrons and holes being generated at one location, the induced charge is zero. This changes as the carriers are moving away from each other due to the applied electric field. Carrier motion thus induces a current, until the motion stops due to trapping, recombination loss, or due to arrival at the electrodes. Depending on the electric circuit connected to the electrodes, this induced current results in an electrical signal.

The derivation of the induced signal given in the following closely follows \cite{He01}.

\section{Shockley-Ramo theorem}

Consider a number of electrodes, indexed by $i$, held at constant potentials $V_i$. These are the pixels, guard ring and cathode of the CdTe sensors as shown in Fig.\,\ref{Fig:PixelPattern}.\footnote{The pixels and the guard ring are held near ground potential by the ASIC, the cathode at bias potential by the high-voltage power supply.} This arrangement is enclosed at some distance by an equipotential surface, for example by a metallic box. This surface provides the reference electrical potential which is set to zero.

The induced charge on an electrode can be determined by considering the energy balance within some given volume. To this end, separate the electric field at position $\mathbf{r}$ into three components: $\mathbf{E}(\mathbf{r},\mathbf{r}_\text{q}) = \mathbf{E}_\text{e}(\mathbf{r}) + \mathbf{E}_\text{sc}(\mathbf{r}) + \mathbf{E}_\text{q}(\mathbf{r},\mathbf{r}_\text{q})$. $\mathbf{E}_\text{e}(\mathbf{r})$ is the field resulting from the constant electrode potentials in the absence of any space or moving charge, $\mathbf{E}_\text{sc}(\mathbf{r})$ is the field from the space-charge alone when all electrodes are grounded, and $\mathbf{E}_\text{q}(\mathbf{r},\mathbf{r}_\text{q})$ is the field from the moving charge at position $\mathbf{r}_\text{q}$ alone when all electrodes are grounded.

The change $\Delta Q$ of the charge enclosed in an arbitrary volume with surface S by the movement of the mobile charge $q$ from some initial position $\mathbf{r}_\text{start}$ to a final position $\mathbf{r}_\text{end}$ can be calculated by Gauss's law,
\begin{equation}
	\Delta Q = \oint_{\text{S}} \varepsilon(\mathbf{r}) \left( \mathbf{E}(\mathbf{r},\mathbf{r}_\text{start}) - \mathbf{E}(\mathbf{r},\mathbf{r}_\text{end}) \right) \text{d}\mathbf{S}
	= \oint_{\text{S}} \varepsilon(\mathbf{r}) \left( \mathbf{E}_\text{q}(\mathbf{r},\mathbf{r}_\text{start}) - \mathbf{E}_\text{q}(\mathbf{r},\mathbf{r}_\text{end}) \right) \text{d}\mathbf{S}.
	\label{Eq:GaussLaw}
\end{equation}

The permittivity $\varepsilon(\mathbf{r}) = \varepsilon_{\text{r}}(\mathbf{r})\varepsilon_{0}$ might be position dependent because of the chosen integration surface. If the surface encloses an electrode, the change of induced charge on this electrode, thus the measurable signal, is found. This equation demonstrates the intuitively clear fact that the fixed space-charge has no influence on the resulting signal. In the following, $\mathbf{E}_\text{sc}(\mathbf{r}) = 0$ can therefore be used without loss of generality to calculate the signal. However, all electric field contributions must be taken into account for determining the charge trajectory.

If the mobile charge $q$ moves from $\mathbf{r}_\text{start}$ to $\mathbf{r}_\text{end}$, the induced charge on the electrodes changes by $\Delta Q_i$. Keeping the potentials constant requires work from the the power supplies of magnitude $V_i \Delta Q_i$. The work done on the charge by moving from start to end position by the electric field generated by the electrodes is $q \int \mathbf{E}_\text{e}(\mathbf{r}) \text{d}\mathbf{r}$. The difference between the power supply work and this energy given to the charge must be equal to the change in electric field energy because of energy conservation, thus
\begin{equation}
\begin{split}
	\sum_i & \left( V_i \Delta Q_i \right) - q \!\!\! \int_{\mathbf{r}_\text{start}}^{\mathbf{r}_\text{end}} \!\!\! \mathbf{E}_\text{e}(\mathbf{r}) \,\text{d}\mathbf{r} \\
	& = \frac{1}{2} \!\!\! \int_\text{volume} \!\!\!\! \varepsilon(\mathbf{r}) \left( (\mathbf{E}_\text{e}(\mathbf{r}) + \mathbf{E}_\text{q}(\mathbf{r},\mathbf{r}_\text{start}))^2 - (\mathbf{E}_\text{e}(\mathbf{r}) + \mathbf{E}_\text{q}(\mathbf{r},\mathbf{r}_\text{end}))^2 \right) \text{d}V \\
	& = \!\!\! \underbrace{\int_\text{volume} \!\!\!\! \varepsilon(\mathbf{r}) \, \mathbf{E}_\text{e}(\mathbf{r}) \left( \mathbf{E}_\text{q}(\mathbf{r},\mathbf{r}_\text{start}) - \mathbf{E}_\text{q}(\mathbf{r},\mathbf{r}_\text{end}) \right) \text{d}V}_{\displaystyle A} + \frac{1}{2} \!\!\! \underbrace{\int_\text{volume} \!\!\!\! \varepsilon(\mathbf{r}) \left( \mathbf{E}^2_\text{q}(\mathbf{r},\mathbf{r}_\text{start}) - \mathbf{E}^2_\text{q}(\mathbf{r},\mathbf{r}_\text{end}) \right) \text{d}V.}_{\displaystyle B}
\end{split}
\label{Eq:EnergyBalance}
\end{equation}
The volume integral is taken over the the full extend of the electric fields, and is shown in the following to vanish.

Expressing the electric displacement and the electric field by their respective potential gradients, $\varepsilon(\mathbf{r}) \mathbf{E}_\text{e}(\mathbf{r}) = -\nabla U_\text{e}(\mathbf{r})$ and $\mathbf{E}_\text{q}(\mathbf{r},\mathbf{r}_\text{q}) = -\nabla U_\text{q}(\mathbf{r},\mathbf{r}_\text{q})$, and applying Green's first identity\footnote{Green's first identity is $\int \psi \nabla^{2} \varphi + \nabla \varphi \cdot \nabla \psi \,\text{d}V = \oint \psi \nabla \varphi \cdot \,\text{d}\mathbf{S}$. $\psi$ and $\varphi$ are scaler fields, the integrals are evaluated over a volume and its surface.} yields
\begin{equation*}
\begin{split}
	\int_\text{volume} \!\!\!\! & \varepsilon(\mathbf{r}) \, \mathbf{E}_\text{e}(\mathbf{r}) \cdot \mathbf{E}_\text{q}(\mathbf{r},\mathbf{r}_\text{q}) \,\text{d}V = \!\!\!\! \int_\text{volume} \!\!\!\! \nabla U_\text{e}(\mathbf{r}) \cdot \nabla U_\text{q}(\mathbf{r},\mathbf{r}_\text{q}) \,\text{d}V \\
	& = \oint_\text{surface} \!\!\!\! U_\text{q}(\mathbf{r},\mathbf{r}_\text{q}) \nabla U_\text{e}(\mathbf{r}) \cdot \text{d}\mathbf{S} - \!\!\!\! \int_\text{volume} \!\!\!\! U_\text{q}(\mathbf{r},\mathbf{r}_\text{Q}) \nabla^2 U_\text{e}(\mathbf{r}) \,\text{d}V.
\end{split}
\end{equation*}
The surface integral on the right hand side vanishes because the potential $U_\text{q}(\mathbf{r},\mathbf{r}_\text{q})$ is zero on the surface bounding the volume as per construction (enclosing metallic box at zero potential).\footnote{An alternative reasoning to explain the vanishing of the surface integral without assuming an enclosing box is that potentials at large distance from the sources fall off inversely to the distance, and the gradient of a potential to the inverse square. The integrand of the surface integral thus falls of as inverse cube, whereas the surface area increases only proportional to the distance squared.} From Poisson's equation $\nabla^2 U_\text{e}(\mathbf{r}) = 0$ because all charges inside the volume are removed for defining $\mathbf{E}_\text{e}(\mathbf{r})$, so the volume integral on the right hand side vanishes as well, and therefore both terms of the volume integral $A$ in (\ref{Eq:EnergyBalance}) are zero.

Volume integral $B$ is evaluated using Green's identity again.
\begin{equation}
\begin{split}
	\int_\text{volume} & \!\!\! \varepsilon(\mathbf{r}) \left( \mathbf{E}^2_\text{q}(\mathbf{r},\mathbf{r}_\text{start}) - \mathbf{E}^2_\text{q}(\mathbf{r},\mathbf{r}_\text{end}) \right) \text{d}V \\
	& = \oint_\text{surface} \!\!\! \varepsilon(\mathbf{r}) \left( U_\text{q}(\mathbf{r},\mathbf{r}_\text{start}) \nabla U_\text{q}(\mathbf{r},\mathbf{r}_\text{start}) - U_\text{q}(\mathbf{r},\mathbf{r}_\text{end}) \nabla U_\text{q}(\mathbf{r},\mathbf{r}_\text{end}) \right)\cdot \text{d}\mathbf{S} \\
	& \quad + \int_\text{volume} \!\!\! \varepsilon(\mathbf{r}) \left( U_\text{q}(\mathbf{r},\mathbf{r}_\text{end}) \nabla^2 U_\text{q}(\mathbf{r},\mathbf{r}_\text{end}) - U_\text{q}(\mathbf{r},\mathbf{r}_\text{start}) \nabla^2 U_\text{q}(\mathbf{r},\mathbf{r}_\text{start}) \right) \text{d}V \\
	& = q \left( U_\text{q}(\mathbf{r}_\text{start},\mathbf{r}_\text{start}) - U_\text{q}(\mathbf{r}_\text{end},\mathbf{r}_\text{end}) \right).
\end{split}
\end{equation}
The surface integral vanishes for the same reason as above. The volume integral is evaluated by using Poisson's equation for a point charge at location $\mathbf{r}_\text{q}$, $\nabla^2 U_\text{q}(\mathbf{r},\mathbf{r}_\text{q}) = -q\,\delta(\mathbf{r}-\mathbf{r}_\text{q})/\varepsilon(\mathbf{r})$, with $\delta()$ the Dirac delta function. The electric potential of a point charge in free space is $U_\text{q}(\mathbf{r},\mathbf{r}_\text{q}) = q/(4\pi\epsilon(\mathbf{r}) |\mathbf{r}-\mathbf{r}_\text{q}|)$. It diverges for $\mathbf{r}=\mathbf{r}_\text{q}$, but the divergence is independent of position. The electric field very near to a point charge is also independent of the boundary conditions further away, so the difference of the two diverging potentials in the last equation vanishes.

Since both volume integrals in (\ref{Eq:EnergyBalance}) vanish, it follows that
\begin{equation}
	\sum_i V_i \Delta Q_i = q \!\!\! \int_{\mathbf{r}_\text{start}}^{\mathbf{r}_\text{end}} \!\!\! \mathbf{E}_\text{e}(\mathbf{r}) \text{d}\mathbf{r} = q \cdot \left( U_\text{e}(\mathbf{r}_\text{start}) - U_\text{e}(\mathbf{r}_\text{end}) \right).
	\label{Eq:GeneralRamoRelation}
\end{equation}
The energy extracted from the power supplies to keep the electrode potentials fixed is transferred only to the moving charge, none to the electric fields. In vacuum the charge would increase its kinetic energy by accelerating, in a semiconductor the energy is lost again to the crystal lattice.

The actual values of the potential on the electrodes are irrelevant for calculating the change in charge, as $\mathbf{E}_\text{e}(\mathbf{r})$ does not appear in (\ref{Eq:GaussLaw}). The $V_i$ can therefore be chosen at will.  To finally calculate the signal on one particular electrode $k$, all $V_i$ in (\ref{Eq:GeneralRamoRelation}) are set to zero, except $V_k$ which is set to an arbitrary non-zero value. Then,
\begin{equation}
	\Delta Q_k = \frac{q}{V_k} \left( U_\text{e}^k(\mathbf{r}_\text{start}) - U_\text{e}^k(\mathbf{r}_\text{end}) \right),
\end{equation}
where the electric potential $U_\text{e}^k(\mathbf{r})$ has to be determined for the chosen values of the electrode potentials and with all charges removed. Defining the dimensionless \emph{weighting potential} $\psi_k(\mathbf{r}) = U_\text{e}^k(\mathbf{r}) / V_k$, the Shockley-Ramo theorem follows:
\begin{equation}
	\boxed{\Delta Q_k = q \cdot \left( \psi_k(\mathbf{r}_\text{start}) - \psi_k(\mathbf{r}_\text{end}) \right).}
	\label{Eq:ShockleyRamo}
\end{equation}
Note that the weighting field is distinctively different to the electric field, except if there are only two electrode.

The application of the theorem for calculating the signal shape in an actual semiconductor device proceeds as follows:
\begin{enumerate}
	\item Calculate the trajectory of the electrons and holes using the electric field $\mathbf{E}_\text{e}(\mathbf{r}) + \mathbf{E}_\text{sc}(\mathbf{r})$ within which the charge is moving. In general, this field has to be determined numerically, though an approximation as (\ref{Eq:ElectricField}) might be sufficient.
	\item Calculate the weighting fields $\psi_k(\mathbf{r})$, using a similar approach to the first step, but setting all except one electrode to zero potential. This nearly always has to be done numerically since the weighting field will have a more complicated distribution than the actual electric field.
	\item The fractional induced charge on an electrode when the electron or hole moves from one location to another is given by the difference in the weighting potential of this electrode.
	\item The signal as function of position can be converted into a time-dependent signal by using an equation as (\ref{Eq:TimeDistanceRelation}), for example, or an equivalent for a more accurate electric field representation.
\end{enumerate}
If only the total induced charge is required, but not the signal shape, then step 3 can be performed using only the location of charge creation and the end point on one of the electrodes.

\begin{figure}
\centering
\includegraphics[width=0.8\textwidth]{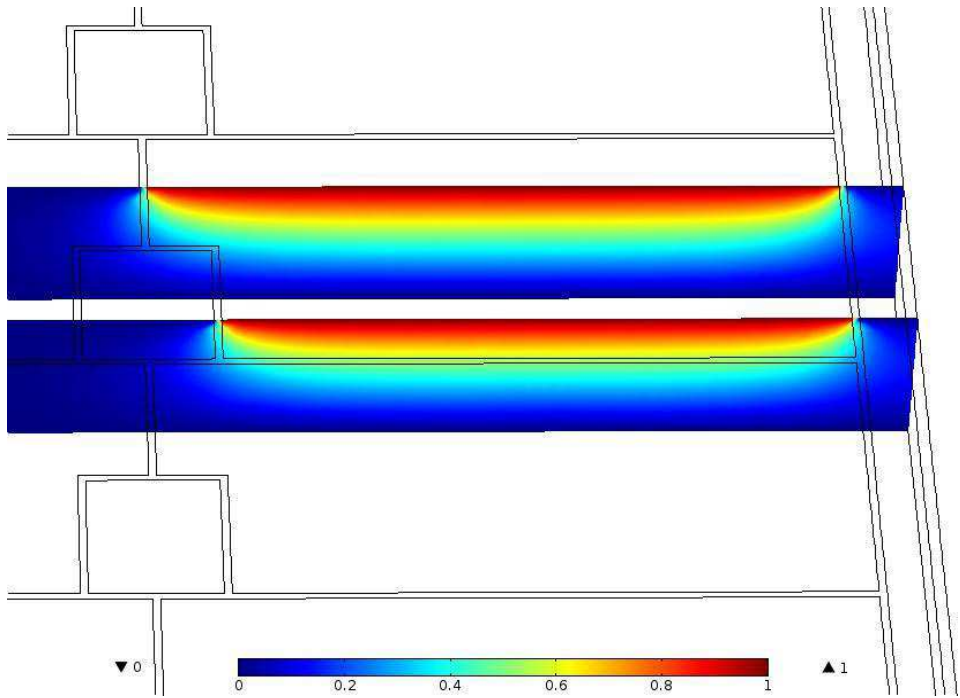}
\caption[Weighing potential slices for a large pixel]{Two slices of the weighting potential for a large pixel of the pattern shown in Fig.\,\ref{Fig:PixelPattern}. The calculation is done with Comsol.}
\label{Fig:WeightingPotential}
\end{figure}

The weighting potential for a large pixel of the STIX CdTe sensors is shown in Fig.\,\ref{Fig:WeightingPotential}. It extends beyond the pixel boundaries, thus transient signals from charges moving close to the boundary occur.

\section{Example signal shapes}

\begin{figure}
\centering
\begin{subfigure}{0.8\textwidth}
	\includegraphics[width=\textwidth]{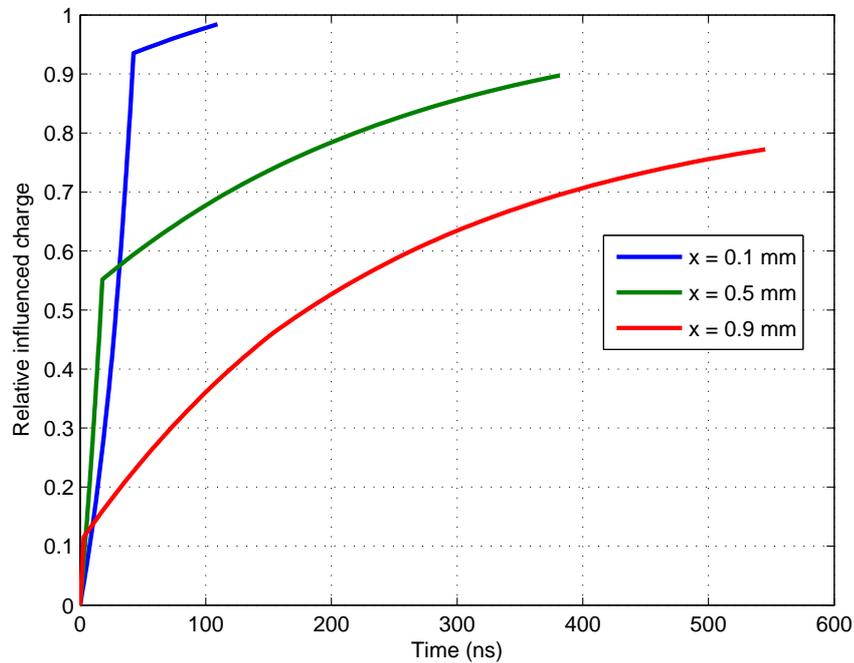}
    \caption{Signal on L5 for an interaction at $y$=\unit[1.1]{mm}, $z$=\unit[3]{mm}, near the centre of L5.}
    \label{Fig:SignalShapes_a}
\end{subfigure}\quad
\begin{subfigure}{0.8\textwidth}
	\includegraphics[width=\textwidth]{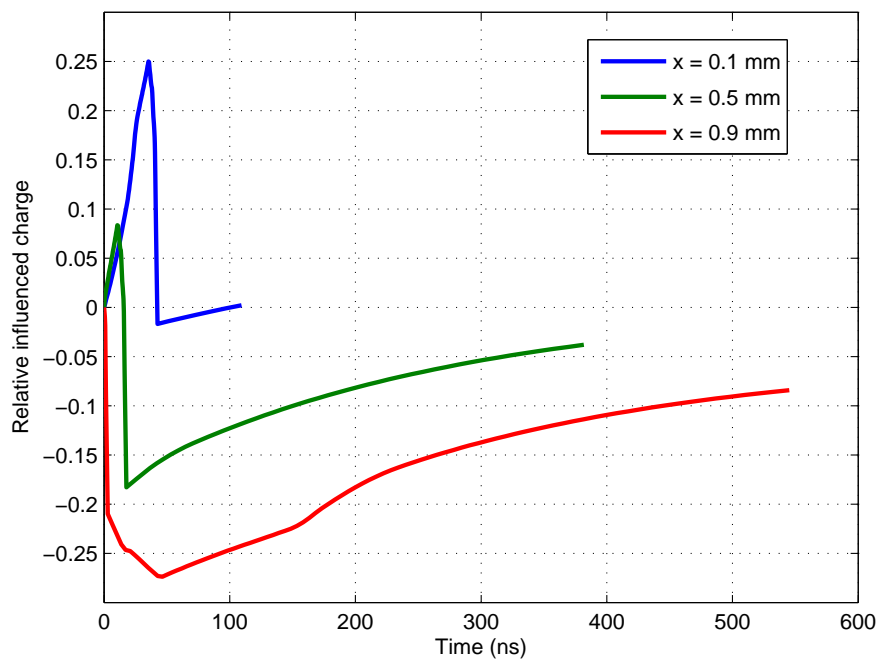}
    \caption{Signal on L4 for an interaction at $y$=\unit[2.1]{mm}, $z$=\unit[3]{mm}, \unit[100]{\textmu m} from the geometric boundary between large pixels L4 and L5. Signal on L5 is almost identical to (a).}
    \label{Fig:SignalShapes_b}
\end{subfigure}
\caption[Signal shapes for different interaction depths]{Signal shapes for different depths $x$ of interaction. Pixel locations are indicated on Fig.\,\ref{Fig:PixelPattern}.}
\label{Fig:SignalShapes}
\end{figure}

In Fig.\,\ref{Fig:SignalShapes} signal shapes as function of time are shown for several interaction positions in the STIX CdTe sensors. This is for point-like charges, thus diffusion of the charge cloud is not considered. The signal is given relative to the total signal if all charges that were initially created would be collected on a single electrode. The calculation includes charge loss (\ref{Eq:CarrierDecay}) with parameters as given in Section~\ref{Sect:Bulk-Properties}.

An interaction that occurs transversely near the centre of a large pixel is plotted in Fig.\,\ref{Fig:SignalShapes_a}. Depending of the depth of interaction, the fraction of the fast-rising electron signal (due to the high electron mobility) to the slow rising hole signal varies. For an interaction at \unit[0.5]{mm} depth, half-way between cathode and anode, the electron signal contributes more than 50\% due to the shape of the weighting potential. Even at the centre of a large pixel this differs from the shape of an idealized parallel-plate capacitor potential. With increasing interaction depth the contribution from the holes to the total signal increases. The signal takes longer to rise and, due to increasing hole loss with drift time, attains smaller final values.

In Fig.\,\ref{Fig:SignalShapes_b} an interaction at a distance of \unit[100]{\textmu m} from the geometric boundary between two large pixels is considered. This is still large compared to the diffusion size. The interaction occurs over pixel L5. The signal of L5 is nearly unchanged compared to Fig.\,\ref{Fig:SignalShapes_a}. On the neighbouring pixel L4 transient signals are observed. Initially, the electron signal rises, but then drops sharply because the electrons are all collected on L5. The sharp drop reflects the shape of the weighting potential close to the pixel boundaries. The hole signal has a negative transient only for interactions close to the anode. If no charge loss would occur, the transient signals would cancel exactly, but because not all holes reach the cathode a small, negative signal remains.

Note that these are the signals influenced on the electrodes and normally not directly observable. Signal amplification and shaping by the read-out electronic that results in the measurable signal is described in Section~\ref{Sect:ASIC}.

\section{Inclusion of diffusion, charge sharing}

The weighting field of an electrode usually extends into the geometric area of a neighbouring pixel, as seen in Fig.\,\ref{Fig:WeightingPotential}. In addition, charges may also move horizontally due to diffusion, even if the electric field, and thus their drift, is assumed to be perfectly perpendicular to the pixel surface. A pixel adjacent to the initial energy deposition may therefore see not only transient, but also permanent signals.

The carriers move along the direction of the field as an expanding cloud. Neglecting diffusion in the drift direction, as this would only modify very slightly the temporal behaviour of the signal\footnote{This can be seen from (\ref{Eq:MaximumSpread}): while the charges drift over the full crystal thickness, they spread by diffusion only to a size small compared to the thickness. $\sigma(t_0)/d$ is only a few percent.}, the charge distribution $\rho(x,y,z)$ is modelled as a flat, circular packet with a transversely Gaussian distribution, and with spread increasing with drift distance according to (\ref{Eq:DiffusionSigma}),
\begin{equation*}
	\rho(x,y,z) = \frac{q(x)}{2\pi\sigma^2(x)} \exp\left( -\frac{y^2 + z^2}{2 \sigma^2(x)} \right).
\end{equation*}
The coordinate system origin is at interaction point for this equation, where the (initially point-like) charge cloud is generated. $\sigma(x)$ was found to be the same for electrons and holes.

In the time domain, $q$ decreases exponentially as described in Section~\ref{Sect:CarrierLoss}. Using the inversion of (\ref{Eq:TimeDistanceRelation}), the loss can be expressed as a function of drift distance, and so $q(x)$ determined. $q(x)$ will be distinctly different for electrons and holes due to their different lifetimes.

Multiplying this with the weighting field of an electrode $k$ under consideration, and integrating transversely, defines a function $I_k(x)$,
\begin{equation*}
	I_k(x) = \int\!\!\!\int \rho(x,y,z) \, \psi_k(x,y,z) \; \text{d}y\,\text{d}z.
\end{equation*}
Now (\ref{Eq:ShockleyRamo}) can be re-formulated as
\begin{equation*}
	\Delta Q_k = I_k(x) - I_k(x + \Delta x),
\end{equation*}
for a movement of the flat charge packet from $x$ to $x + \Delta x$.

If charge loss can be neglected, the total induced signal on pixel $k$ is given simply by $I_k(x)$, evaluate on pixel surface ($x$=\unit[1]{mm} for the present set-up).

\section{Note on anode-side electrode gaps}

Carriers arriving at the surface within an electrode gap are possibly quickly lost by surface recombination, without reaching the electrode. Those not lost immediately might accumulate and result in local charging, modifying the electric field. Such charging would repel subsequent charges and effectively guide them more directly to an electrode.

The exact transport behaviour near an electrode gap is difficult to model. However, only a very small fraction of the signal charge is induced by motion in this vicinity, as the major part of the weighting field has already been traversed. Here the electrode gaps are ignored and charges are assumed to drift exactly perpendicular to the electrodes.

\section{Two electrode configuration, Hecht relation}
\label{Sect:HechtRelation}

For a thin parallel-plate geometry with two electrodes, the weighting field as function of $x$ perpendicular to the electrodes is $\psi(x)=x/d$. Assuming a constant electric field $E$, an exponential decay of the carrier numbers with time as in (\ref{Eq:CarrierDecay}), and the initial creation of $n_0$ carrier pairs at location $x_0$, the observed charge signal after electrons and holes drifted to their electrodes is
\begin{equation}
	\frac{\Delta Q}{e} =  \int_0^{t_\text{e}} n_\text{e}(t) \,\text{d}\psi + \int_0^{t_\text{h}} n_\text{h}(t) \,\text{d}\psi = \frac{n_0}{d} \int_{x_0}^d \exp\left( -\frac{x-x_0}{\lambda_\text{e}} \right) \text{d}x + \frac{n_0}{d} \int_0^{x_0} \exp\left( -\frac{x_0-x}{\lambda_\text{h}} \right) \text{d}x.
	\label{Eq:Integral_Hecht}
\end{equation}
The integrals are written in terms of the drift lengths for electrons and holes, defined by $\lambda_\text{e}=\mu_\text{e} \tau_\text{e} E$ and $\lambda_\text{h}=\mu_\text{h} \tau_\text{h} E$. Electrons drift towards $x=d$, holes towards $x=0$. This equation evaluates to the \emph{Hecht relation}
\begin{equation}
	\frac{\Delta Q}{n_0\,e} = \frac{\lambda_\text{e}}{d}\left(1 - \exp\left(-\frac{d-x_0}{\lambda_\text{e}} \right) \right) + \frac{\lambda_\text{h}}{d}\left(1 - \exp\left(-\frac{x_0}{\lambda_\text{h}} \right) \right).
	\label{Eq:HechtFormula}
\end{equation}

By differentiating (\ref{Eq:HechtFormula}), the maximum signal $\Delta Q^\text{max}$ is found to result from interactions at depth $x_0^\text{max}$,
\begin{equation*}
	x_0^\text{max} = \frac{d}{\lambda_\text{e}/\lambda_\text{h} + 1},
\end{equation*}
and has the value
\begin{equation*}
	\frac{\Delta Q^\text{max}}{n_0\,e} = \frac{\lambda_\text{e} + \lambda_\text{h}}{d} \left(1 - \exp\left(-\frac{d}{\lambda_\text{e} + \lambda_\text{h}} \right) \right).
\end{equation*}

If the average electric field is used, these relations give an estimate of the observed signals in the STIX detectors for events occurring not too close to a pixel boundary and for the large pixels. Since the transverse dimension of the small pixels is comparable to the sensor thickness, the Hecht relation is less accurate for them. The weighting field is noticeably different from the simplified form given above. In particular, the contribution of the electron signal is enhanced by the so-called \emph{small pixel effect}.

Computationally, the Hecht relation allows much faster calculations than the weighting field approach if both diffusion and charge loss are included in the latter.

\chapter{ASIC signal read-out}
\label{Sect:ASIC}

The pixel electrodes are connected to channels of the read-out ASIC \cite{Mich10}. Its response has to be considered to determine the final measurement result. An ASIC channel contains essentially a charge sensitive preamplifier with settable gain, a pole-zero cancellation stage and an CR-RC2 filter with a baseline holder. The output of the filter is connected to a peak detector for energy measurement and also to the trigger circuit. The details of the gain, shaping and baseline restoring circuitry are unknown. Only the idealized CR-RC2 filter response is considered in the following.

The ASIC has a single, differential analogue output. The signals stored by the peak detectors are multiplexed to this output under control of a digital communication sequence initiated by the controller after the ASIC generated a trigger. The controller then also digitizes the signals with an ADC. In STIX, the controller is implemented in an FPGA.

\section{Charge sensitive amplifier and shaping}

\begin{figure}
\centering
\includegraphics[width=0.8\textwidth]{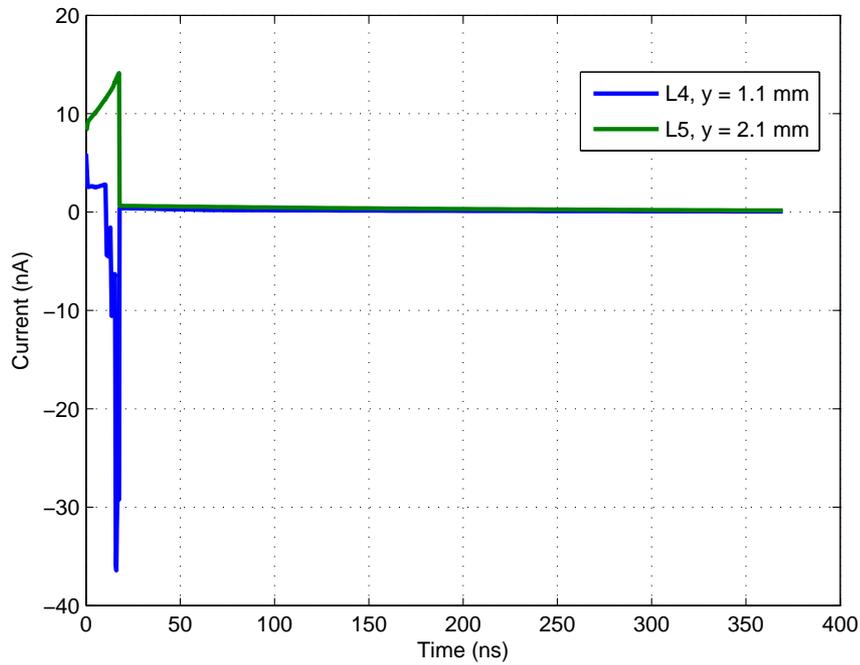}
\caption[Signal current]{Signal current for the interactions at $x=\unit[0.5]{mm}$ in Fig.\,\ref{Fig:SignalShapes_a} and Fig.\,\ref{Fig:SignalShapes_b}. The deposited energy is \unit[10]{keV}.}
\label{Fig:SignalCurrent}
\end{figure}

The peak signal currents, given by the change of the influenced charge with time, can be three orders of magnitude larger than the typical large-pixel leakage current of \unit[50]{pA}, as seen in Fig.\,\ref{Fig:SignalCurrent} for events that correspond to \unit[10]{keV} deposited energy. Transient currents in adjacent pixels can also flow briefly in reverse to the leakage current. The charge-sensitive amplifier at the input of the ASIC channels is assumed in the following to behave ideal, that is it instantaneously integrates the current\footnote{It is known that a minimum amount of leakage current is needed for correct functioning of the ASIC analogue channel, but not if a short reverse transient might cause problems.}, and therefore its output, which is fed into the shaper, is proportional to the signals shown in Fig.\,\ref{Fig:SignalShapes}.

\begin{figure}
\centering
\includegraphics[width=0.8\textwidth]{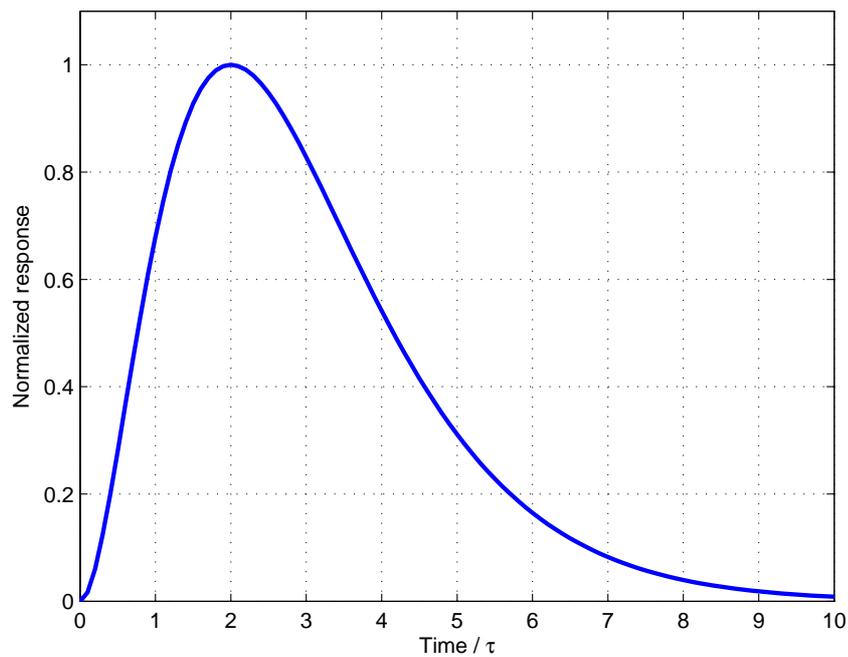}
\caption[ASIC filter response]{Normalized ASIC filter response (\ref{Eq:Filter})}
\label{Fig:ASICFilter}
\end{figure}

The shaping circuit of the ASIC implements a CR-RC2 filter. Its normalized response $R(t)$ to a step input is described by
\begin{equation}
	R(t) = \frac{e^2}{4} \left(\frac{t}{\tau}\right)^2 \exp\left(\frac{-t}{\tau}\right).
	\label{Eq:Filter}
\end{equation}
The response, shown in Fig.\,\ref{Fig:ASICFilter}, peaks at $t=2\tau$. The peak value of the filtered response is held by the ASIC peak detector. The output amplitude is related to the integrated charge by the ASIC gain, which in STIX nominally is set to \unit[200]{mV/fC}. The peaking time $2\tau$ is adjustable between \unit[0.98]{\textmu s} and \unit[12.92]{\textmu s}.

The fast transients within the first \unit[100]{ns} seen in Fig.\,\ref{Fig:SignalShapes} are strongly attenuated by the filtering even for the shortest peaking time, so that the signal shape seen by the peak detector is very similar to the filter response, with an amplitude proportional to the total influence charge. Incomplete charge collection (ballistic deficit) might occur at the shortest shaping time for the hole signal, as the carrier drift time might not be much shorter than the shaping time in this case. Due to the strong, lifetime-induced tailing that is anyway present on hole-dominated spectral lines, this however is likely difficult to observe in practice.

Positive transients, as seen for the pixel L4 signal in Fig.\,\ref{Fig:SignalShapes_b}, are not expected to generate an erroneous trigger even for high energy depositions and a low threshold.

This conclusion relies on the stated ideal integration behaviour of the charge-sensitive amplifier. If it would cut off reverse currents, the total signal would remain positive after filtering and triggering might occur.

\section{Analogue pile-up}

\begin{figure}
\centering
\begin{subfigure}{0.8\textwidth}
	\includegraphics[width=\textwidth]{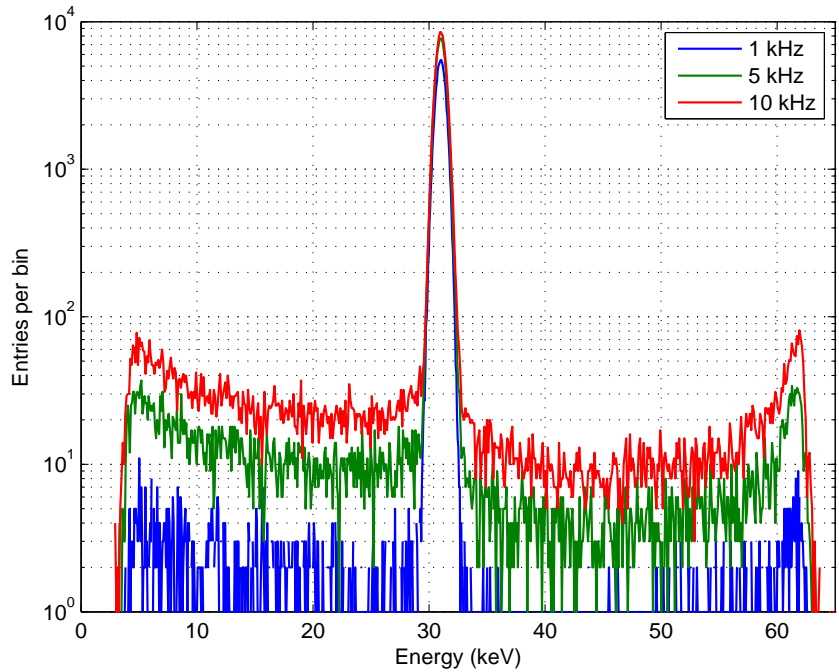}
    \caption{Different photon rates, peaking time \unit[4]{\textmu s}}
\end{subfigure}\quad
\begin{subfigure}{0.8\textwidth}
	\includegraphics[width=\textwidth]{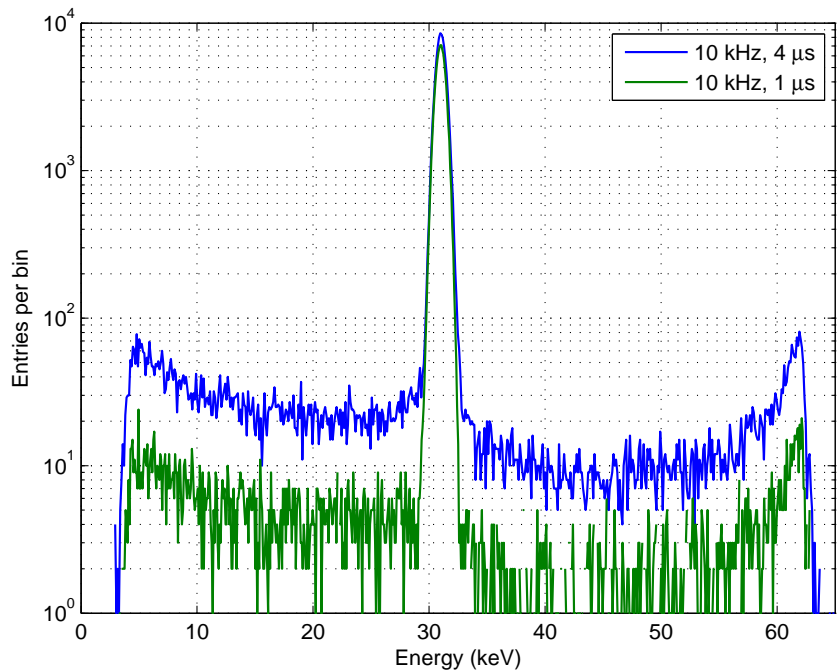}
    \caption{Photon rate \unit[10]{kHz}, different peaking times}
\end{subfigure}
\caption[Pile-up effect]{Effect of pile-up for \unit[31]{keV} monochromatic photons. The energy resolution is \unit[1]{keV} FWHM, the trigger threshold \unit[4]{keV}. A photon rate of \unit[10]{kHz} for a large pixel corresponds to about \unit[84]{kHz} for one Caliste-SO unit.}
\label{Fig:PileUp}
\end{figure}

During the time from the first photon interaction until the peak detector is frozen by the control unit, pile-up from a subsequent photon interaction can possibly disturb the energy measurement. This will occur if the peak amplitude after filtering exceeds the peak in case there would have been only the first interaction. The resulting effect for monochromatic photons is illustrated in Fig.\,\ref{Fig:PileUp}. The plot results from
\begin{itemize}
\itemsep0em
\item generating a time-domain event sequence using the filter response from Fig.\,\ref{Fig:ASICFilter} with exponentially distributed random delay times and given average delay,
\item modelling the trigger generation by determining the times when the signal level crosses \unit[4]{keV},
\item enforcing a \unit[12]{\textmu s} dead time after each trigger to account for the read-out time,
\item determining the maximum value of the signal within the peaking time after the trigger generation,
\item adding \unit[1]{keV} FWHM noise to the obtained peak value.
\end{itemize}
No fluorescence is considered, so in the absence of pile-up a single, Gaussian line is expected.

The pile-up results not only in a low-level, distributed background, but also in a peak at twice the photon energy. To minimize this disturbance, the peaking time needs to be short at high photon rates and peak detector freezing should occur as soon as possible after the shaped signal reached its maximum. A limitation to the precision with which this can be achieved results from the uncertainty in time between the actual photon interaction and the trigger generation. In Fig.\,\ref{Fig:ASICFilter} this is the delay between $t=0$ until the signal crosses a given level. This \emph{time walk} depends on the amplitude of the event and is near zero for a large events and small thresholds and equal to the peaking time for an event just large enough to reach the trigger threshold. As the event amplitude is unknown before read-out, the minimum delay before peak detector freezing after trigger generation is equal to the peaking time.

The energy resolution of the main spectral line is unchanged for all the cases shown in Fig.\,\ref{Fig:PileUp}. An effect on the resolution becomes significant only when the photon rate approaches the inverse of the peaking time. For example, a photon rate of \unit[200]{kHz} at peaking time \unit[1]{\textmu s} results in a 7\% increase of the spectral width.

However, a significantly more important degradation of resolution with rate is observed experimentally with the Caliste-SO. At \unit[5]{kHz} trigger rate the FWHM resolution at \unit[31]{keV} is found to be \unit[1.6]{keV} versus \unit[1.1]{keV} at \unit[50]{Hz}. This has been traced to an effect of the ASIC baseline holder circuitry, see document STIX-TN-0175-FHNW \emph{Caliste-SO rate dependencies}.

\addtocontents{toc}{\protect\newpage}
\chapter{Signal generation simulation}
\label{Sect:Implementation}

The considerations presented in the previous chapters on the signal generation mechanisms were implemented into a software tool to allow simulations of spectra and comparisons with experimental data. The tool allows to calculate, for a given interaction location in the STIX CdTe sensors, the response for each pixel. Using a suitable grid, three-dimensional response maps are constructed that are independent of the deposited energy. These maps can be coupled with a Monte-Carlo code that generates a list of interactions (location and energy) from a given source description to achieve an \emph{end-to-end simulation} of the detector front-end.

\section{Summary of the simulation steps}

The software implementation to simulate the various processes that generate the final measurement signal for given interaction locations proceeds along the following steps:
\begin{enumerate}
\itemsep0ex
\item Calculation of the three dimensional weighting fields for the pixel electrodes. This step needs to be performed only once.
\item Generating an x coordinate grid for the electron and hole diffusion cones depending on the interaction location.
\item Determination of the time-position relation $t(x)$ from (\ref{Eq:TimeDistanceRelation}) for the given electric field parameters and carrier mobilities.
\item Determination of the diffusion width $\sigma(x)$ as function of position from (\ref{Eq:DiffusionSigma}).
\item Generating an yz grid for each x position depending on $\sigma(x)$.
\item Integrating the product of weighting field and charge distribution over the yz grid separately for electrons and holes.  Subtracting the weighting field at the initial interaction location yields one point in the position-dependent signal $S(x)$ for both carriers.
\item Applying charge loss for electrons and holes according to their lifetimes by using (\ref{Eq:CarrierDecay}) and $t(x)$. The entrance layer effect (Section~\ref{Sect:CarrierLoss}) is applied by modifying the carrier lifetimes as function of distance from the surface as described in \cite{Gri19}.
\item Summing the contributions from electrons and holes to obtain the time-domain signal on an electrode $S(t)$.
\item Folding of the time-dependent signal with the CR-RC2 shaper response function and determination of the signal amplitude as the maximum within a given time.
\item Multiplication with the deposited energy to get the measured signal amplitude.
\item Applying carrier statistics including the Fano factor and Gaussian noise of given amplitude to the signal (Section\,\ref{Sect:Bulk-Properties}).
\item Applying a given trigger threshold to each pixel and, optionally, rejecting events with more than one pixel above threshold.
\item Converting the signals to voltage or ADC units as required, using given calibration parameters.
\end{enumerate}

Charge transport by diffusion along the x axis is neglected, therefore the assumption is that the dominant transport in this direction is drift imposed by the electric field. Partially depleted sensors cannot be simulated.

The simulation is coded in Matlab, except for the weighting field calculation, and can be obtained from the author. It includes a simple Monte-Carlo tracking code that models photon interactions and the main fluorescence processes in CdTe.

\section{Meshing}

The weighting fields are calculated in Comsol, using meshing as determined by the program from the physical geometry, which results in varying distances between mesh points in all three dimensions. The weighting fields are then read into Matlab, where the \lstinline{scatteredInterpolant()} function is used to linearly interpolate the field for the required coordinates. Because of the decoupling of the physical edge of the CdTe sensor from the pixels by the guard ring, it is sufficient to calculate one response map for a large pixel and one for a small pixel to handle all pixels.

The x coordinate grid for the signal determination has a logarithmic density near the surfaces and is linear in the bulk, to sufficiently sample the near surface lifetime reduction. About 60 grid points are used in total in the x direction.

Transversely, the Gaussian charge distribution from diffusion is subdivided in a grid of 10$\times$10 points out to a distance of $5\sigma$ from the charge centre-of-gravity.

The meshing granularity was checked empirically by refining the mesh to the point where further refinement did not change the results appreciably. Computing a spectrum with good statistics for all pixels of a CdTe sensor takes of the order of ten hours with the meshing as described and on a medium-recent four-core CPU.

\section{Results}
\label{Sect:Results}

\begin{figure}
\centering
\includegraphics[width=0.8\textwidth]{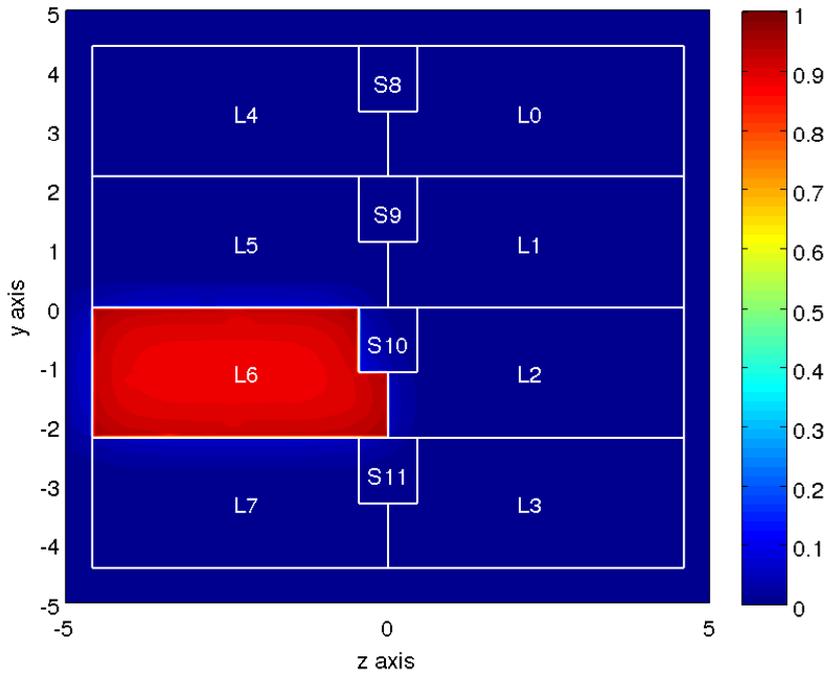}
\caption[Pixel response for given interaction depth]{Response for an interaction depth of \unit[0.5]{mm} for pixel L6. The units of the y and z axis are millimetres.}
\label{Fig:ResponseFull}
\end{figure}

The response of pixel L6 for an interaction depth of \unit[0.5]{mm} is shown in Fig.\,\ref{Fig:ResponseFull}. The response extends beyond the pixel boundaries mainly because of charge diffusion. The low hole lifetime is the reason why the response within the pixel boundaries is significantly less than unity. The charge loss also indirectly results in the rise of the response towards the edges of the pixel for a given interaction depth: the electron signal contribution becomes relatively larger, and the electron loss is less than for holes. The concentration of the weighting field closer to the pixel at its edge can be seen in Fig.\,\ref{Fig:WeightingPotential}. This is the origin of the small-pixel effect, the preferential sensitivity to the electron signal of a finely pixilated sensor.

\begin{figure}
\centering
\includegraphics[width=0.8\textwidth]{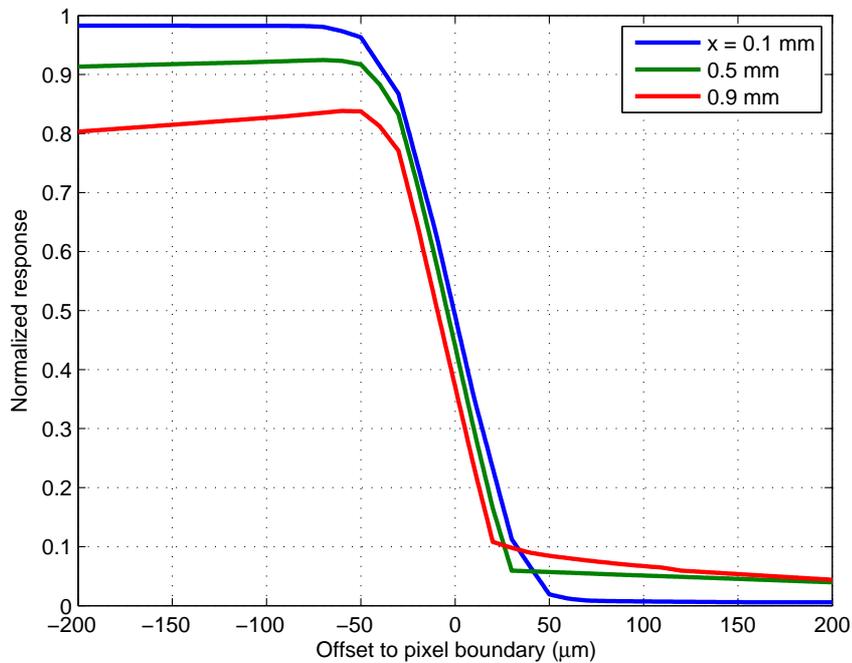}
\caption[Pixel response close to the boundary]{Response of pixel L6 for three interaction depths close to the boundary. The monolithic cathode is located at $x$=0, the pixelized anode at $x$=\unit[1]{mm}. The metallized surface of L6 starts at offset \unit[-25]{\textmu m}.}
\label{Fig:EdgeResponse}
\end{figure}

The response close to a pixel boundary is shown for three interaction depths in Fig.\,\ref{Fig:EdgeResponse}. Again the response is largest close to the pixel boundary, and this effect becomes more pronounced for interactions at larger $x$ which are closer to the anode.

The parameters for these calculation are $V_\text{d}=\unit[-120]{V}$, $V=\unit[-200]{V}$, $\mu_\text{e}=\unit[1100]{cm^2/(V s)}$, $\mu_\text{h}=\unit[100]{cm^2/(V s)}$, $\tau_\text{e}=\unit[3]{\text{\textmu}s}$, $\tau_\text{h}=\unit[0.67]{\text{\textmu}s}$, $T=-20$\textdegree C, peaking time $2\tau=\unit[1]{\text{\textmu}s}$.

\section{Sensitivity to input parameters}

This section illustrates how the results depend on some of the input parameters. The \emph{nominal} parameter set referred to is given in Section~\ref{Sect:Results}.

There is no discernible effect on Fig.\,\ref{Fig:EdgeResponse} when changing the temperature by some 100\textdegree C. That is because the diffusion scale still remains small compared to the pixel gap.\footnote{A slight smoothing of the response over the pixel boundary becomes visible only at $\sim$300\textdegree C.} However, the leakage current rises exponentially with temperature and thus, through amplification noise, the spectral energy resolution will degrade significantly for temperatures above about -10\textdegree C.

\begin{figure}
\centering
\includegraphics[width=0.8\textwidth]{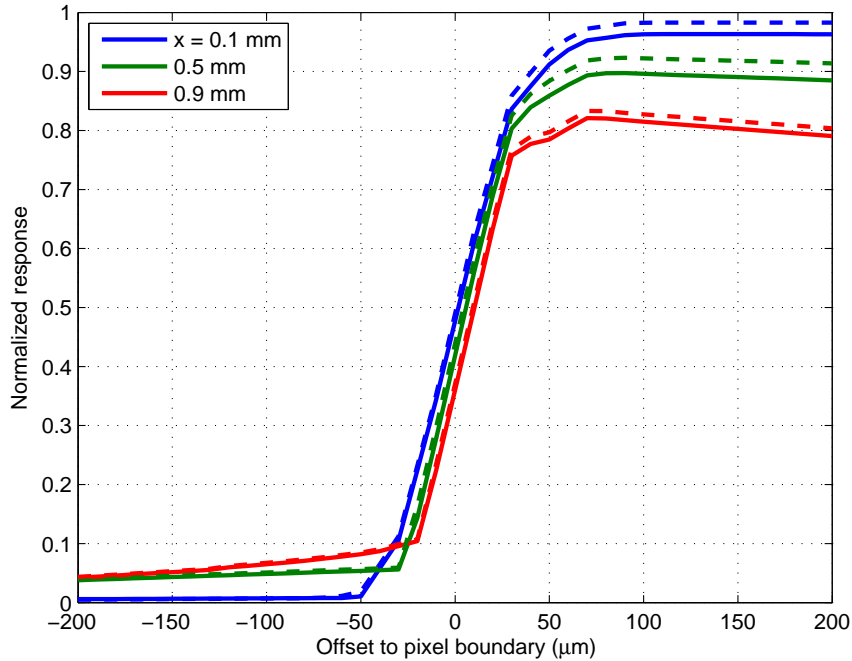}
\caption[Pixel response close to the boundary at low voltage]{Response close to the boundary of pixel L5 for nominal parameters (dashed) and for $V_\text{d} = -\unit[190]{V}$ instead of \unit[-120]{V}, that is for only \unit[10]{V} above full depletion (solid).}
\label{Fig:ResponseVd}
\end{figure}

There is little effect if the applied voltage is only slightly exceeding the depletion voltage, as shown in Fig.\,\ref{Fig:ResponseVd}. The electric field is now very low close the the cathode, which affects the drift time for all charge carriers, and thus especially the loss of holes. The effect for all three curves is mainly due to this hole loss.

\begin{figure}
\centering
\includegraphics[width=0.8\textwidth]{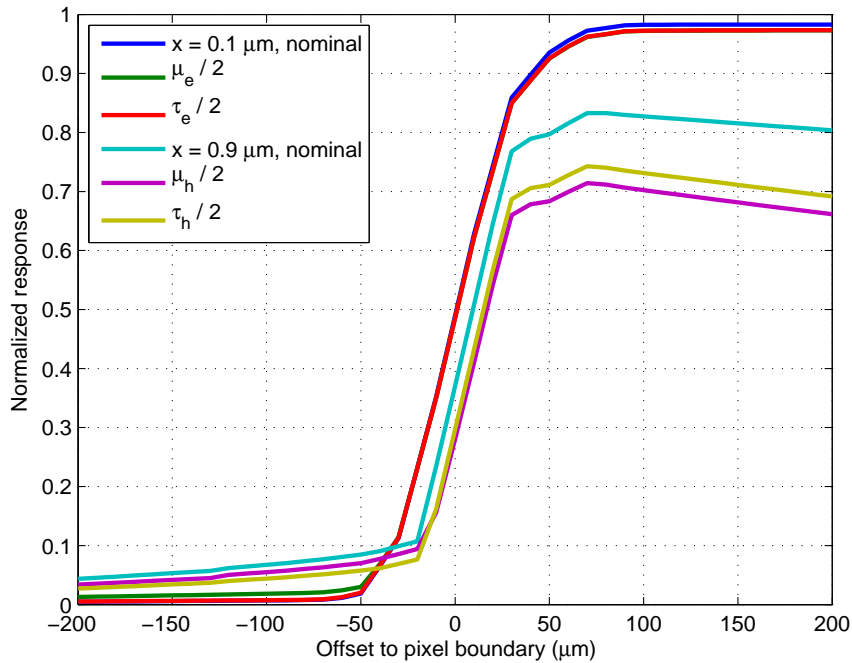}
\caption[Electron and hole pixel response]{Response close to a pixel boundary for nominal parameters at two interaction depths where the signal is dominated by either electrons (0.1 mm) or holes (0.9 mm), for mobility and lifetime values changed as indicated in the legend.}
\label{Fig:ResponseMuTau}
\end{figure}

The Hecht formula (\ref{Eq:HechtFormula}) indicates that to first order the signal depends only on the product of mobility and lifetime. The influence seen in the simulation is shown in Fig.\,\ref{Fig:ResponseMuTau} for two interaction depths, \unit[0.1]{mm} (signal electron dominated) and \unit[0.9]{mm} (hole dominated). For the former, the electron mobility and lifetime where individually changed by a factor 2, for the latter the same has been done for the holes. The signal does indeed depend mainly on the product of the two parameters, though the signal reduction due to reduced mobility is a bit larger than due to reduced lifetime. This is especially clear for holes. The reason is that the ballistic deficit from signal shaping is not completely negligible for a peaking time of $2\tau=\unit[1]{\text{\textmu}s}$. The lower mobility results not only in more charge loss for a given lifetime, but also in an additional ballistic deficit.

Outside of the pixel boundary, the relative influence of reduced mobility and lifetime is reversed. Here, a part of the signal is due to a fraction of carriers not reaching the electrode, or reaching it late, so that the transient signals from electrons and holes do not exactly cancel. A ballistic deficit is this case results in a larger signal.

\begin{figure}
\centering
\includegraphics[width=0.8\textwidth]{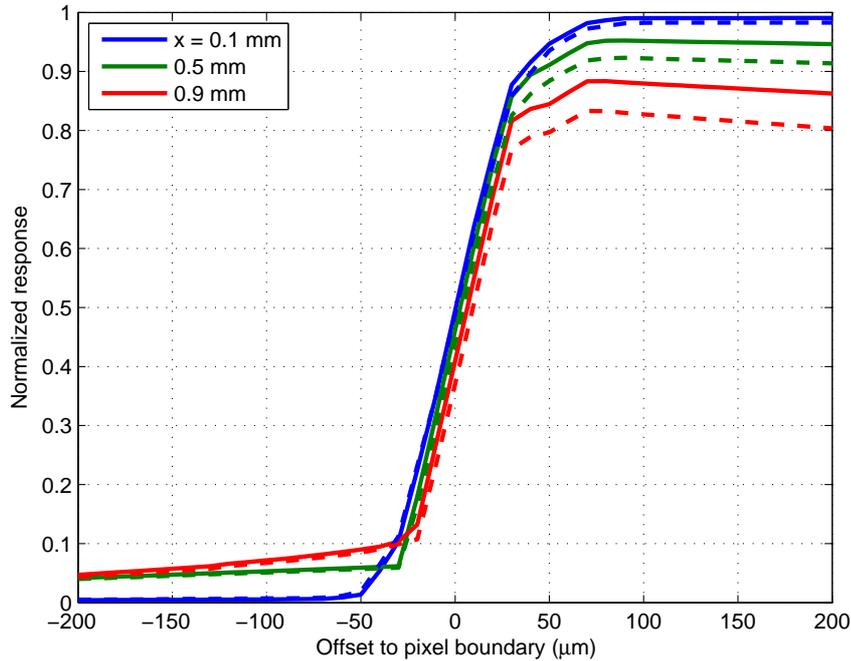}
\caption[Reponse at pixel boundary for larger bias voltage]{Response close to a pixel boundary for nominal parameters (dashed) and for the applied voltage increased to \unit[300]{V} (solid).}
\label{Fig:ResponseVoltage}
\end{figure}

Finally, the effect of increasing the applied voltage from \unit[200]{V} to \unit[300]{V} is shown in Fig.\,\ref{Fig:ResponseVoltage}. The loss of holes is significantly reduced by the decreased drift time. This is the reason why the performance of crystals with a larger defect density, for example after proton irradiation and resulting shorter carrier life time, can be partly recovered by higher voltage. However, a larger leakage current will also ensue.

\section{Comparison with measurements}

The theoretical sensitivity to the input parameters and to the meshing resolution can be demonstrated with a plot like Fig.\,\ref{Fig:EdgeResponse}. Such a plot is, however, not directly reproducible experimentally, as interactions at a defined depth cannot be enforced (except for low energy X-rays that all interact near the cathode), and also because generating a pencil beam of the required small width is difficult. Verification of the simulation against measurements is best performed by acquiring precision spectra with a well-defined X-ray source and comparing to simulations that reproduce the same geometry.

\begin{figure}
\centering
\begin{subfigure}{0.82\textwidth}
	\includegraphics[width=\textwidth]{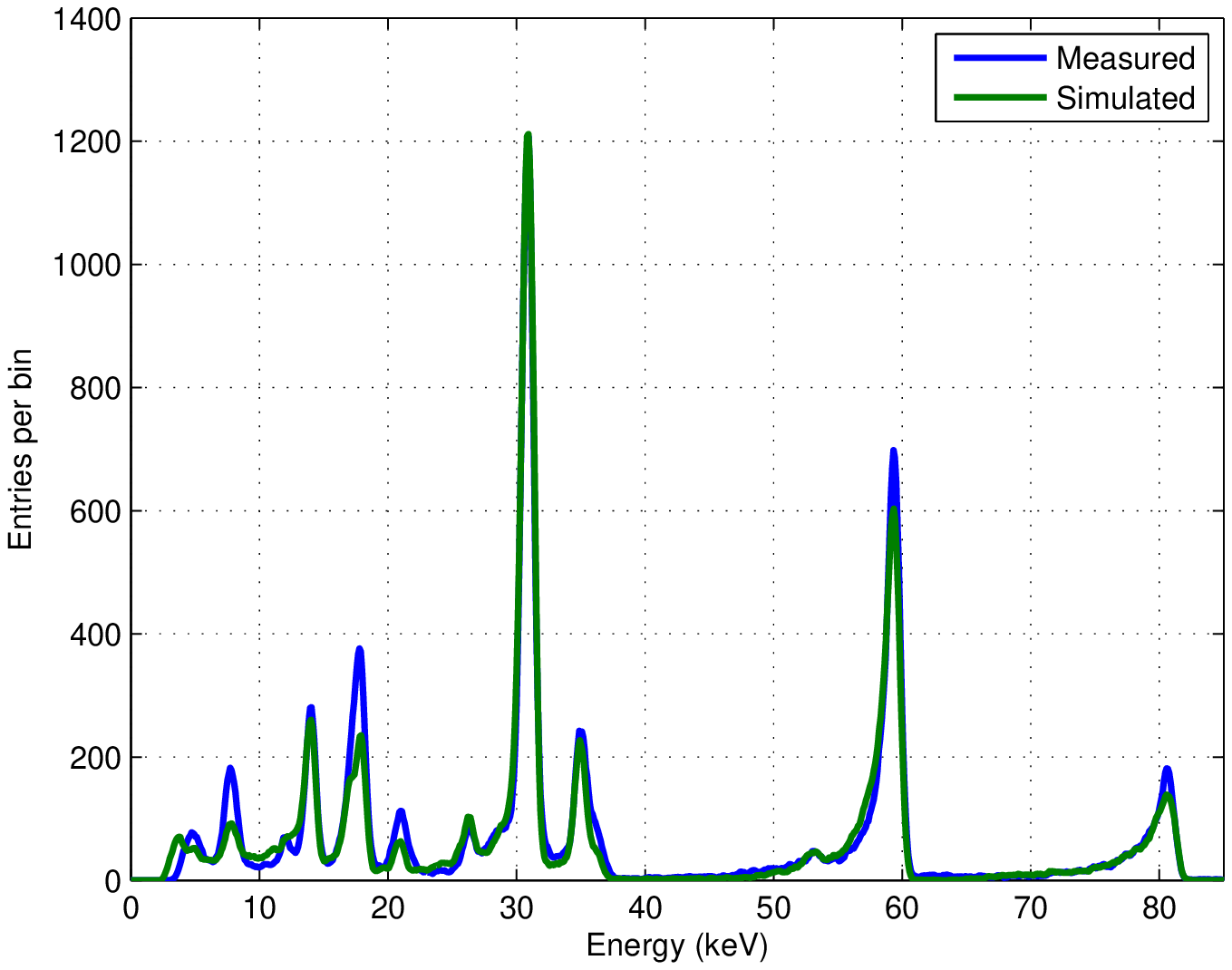}
    \caption{Sum spectrum of large pixels}
\end{subfigure}
\begin{subfigure}{0.82\textwidth}
	\includegraphics[width=\textwidth]{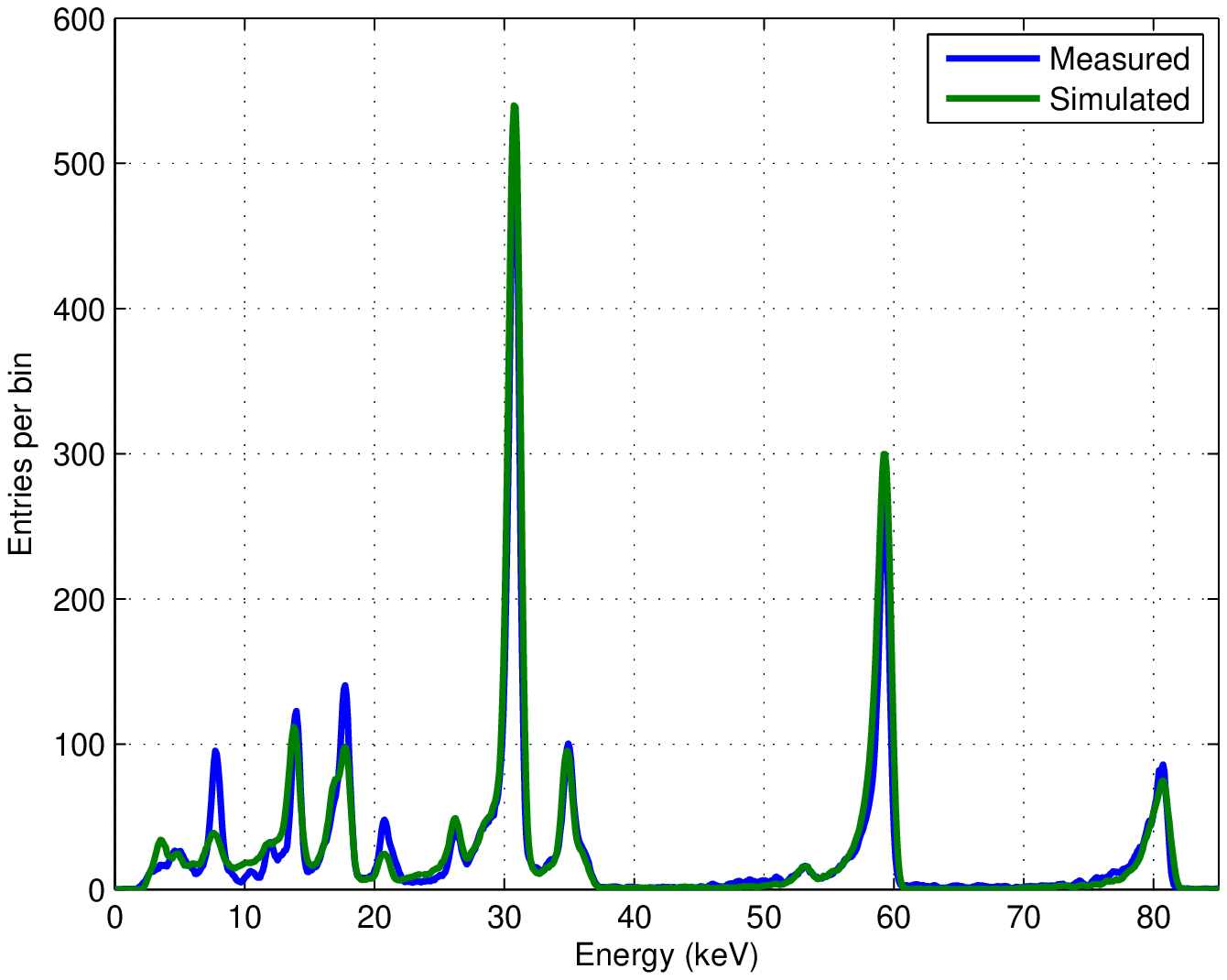}
    \caption{Sum spectrum of small pixels}
\end{subfigure}
\caption[Comparison of measured and simulated spectra]{Comparison between measurement and simulation for spectra obtained with a Caliste-SO at \unit[-200]{V} bias voltage and -20\textdegree C. The sum spectra for the 8 large and 4 small pixels are shown. Barium-133 and americium-241 point sources were used simultaneously, located at a distance of a few centimetres from the CdTe sensor.}
\label{Fig:Comparison}
\end{figure}

The accuracy for reproducing measured spectra is shown in Fig.\,\ref{Fig:Comparison}. The measurement was performed with a Caliste-SO cooled to -20\textdegree C and at \unit[-200]{V} bias. It was irradiated by using simultaneously a barium-133 and an americium-241 radioactive point source, both located at several centimetres distance.

The nominal parameters from Section~\ref{Sect:Results} are used for the simulation, except that the hole lifetime has been increased from \unit[0.67]{\textmu s} to \unit[1]{\textmu s} to obtain a better agreement of the \unit[81]{keV} line profile. The electronic noise is set to \unit[700]{eV} FWHM. The bin population is normalized at \unit[31]{keV} between measurement and simulation.

The relevant source emission lines are\\[1.5ex]
\begin{tabular}{l|cccccccccc}
    Barium-133 energy (keV) & 4.3 & 4.6 & 4.9 & 30.6 & 31.0 & 34.9 & 35.8 & 52.2 & 79.6 & 81.0 \\\hline
    Emission probability (\%) & 6.0 & 3.8 & 1.2 & 34.9 & 64.5 & 17.6 & 3.6 & 2.2 & 2.6 & 34.1
\end{tabular}\\[1.5ex]
\begin{tabular}{l|cccccc}
    Americium-241 energy (keV) & 13.8 & 16.9 & 17.8 & 20.8 & 26.3 & 59.5 \\\hline
    Emission probability (\%) & 10.7 & 4.0 & 7.1 & 1.4 & 2.4 & 35.9
\end{tabular}\\[1.5ex]
In addition, escape lines are located at \unit[7.5]{keV}, \unit[7.9]{keV}, \unit[11.9]{keV}, \unit[32.0]{keV}, \unit[36.4]{keV}, \unit[53.5]{keV} and \unit[57.9]{keV}.

Most of these lines are clearly visible in both measured and simulated spectra, and for cases where the energies are too close for separation still a line-shoulder deformation is seen. A measured spectrum annotated with line energies is shown on the title page of this report.

A discrepancy is found in the amplitudes of escape lines at low energies. The fluorescence probabilities were taken as fixed values from \cite{Iwan79} for this simulation. Better agreement will be attainable by a more detailed simulation of the X-ray absorption and fluorescence processes, e.g. with Geant4.

\chapter[Experimental results]{Experimental results}
\label{Sect:Experimental-Results}

These sections provide a collection of measurement data that was obtained with the STIX CdTe sensors during instrument development.

\section{Voltage scans}
\label{Sect:Voltage-Scans}

A Caliste-SO (serial number SN4) was mounted with a clamping socket to a STIX 3B test board and read out via a commercial FPGA evaluation board at \unit[12.5]{MHz} strobe frequency. A \unit[370]{kBq} barium-133 radioactive source illuminated the CdTe crystal from some distance. The ASIC registers were set as follows:\\[2ex]
\begin{tabular}{llccllccllc}
    \multicolumn{2}{l}{Register} & Value & \hspace{5ex} & \multicolumn{2}{l}{Register} & Value & \hspace{5ex} & \multicolumn{2}{l}{Register} & Value \\\cline{1-3}\cline{5-7}\cline{9-11} \\[-1em]
     1 & ICOMP   & 1      & & 5 & TPEAK & 1   & & 9  & SPY    & 0 \\
     2 & IREQ    & 4      & & 6 & I0 & 1      & & 10 & VREF2P & 3 \\
     3 & TH      & 62/63  & & 7 & RDELAY & 0  & & 11 & TUNE   & 2 \\
     4 & SELTEST & all 0  & & 8 & GAIN & 3    & & 12 & ALIMON & all 1
\end{tabular}\\[2ex]
In particular, the baseline holder (BLH) was switched off via register 11. The trigger threshold was 62 ($\approx$\unit[16]{keV}) for the twelve pixels, and 63 (off) for all other channels.

The set-up is mounted in a vacuum chamber, evacuated to below \unit[$10^{-5}$]{mbar}, and cooled by a Peltier element attached to the mounting plate in the vacuum chamber. Heat is removed from the warm side of the Peltier element with a liquid cooler. The electronics is covered with multi-layer insulation.

After applying the bias voltage and waiting for two minutes, a spectrum was accumulated for ten minutes, followed by two minutes of depolarization at zero voltage. Spectra were taken in steps of \unit[20]{V} between \unit[-20]{V} and \unit[-200]{V}, and in steps of \unit[40]{V} down to \unit[-400]{V}. The cooling was adjusted to obtain temperatures, as measured by the ASIC, of -17\textdegree C,  -5\textdegree C, and +5\textdegree C. Spectra were analysed as described in \cite{Gri19}. The main results are shown in Fig.\,\ref{Fig:Voltage-Scan}.

\begin{figure}
\centering
\begin{subfigure}{0.65\textwidth}
	\includegraphics[width=\textwidth]{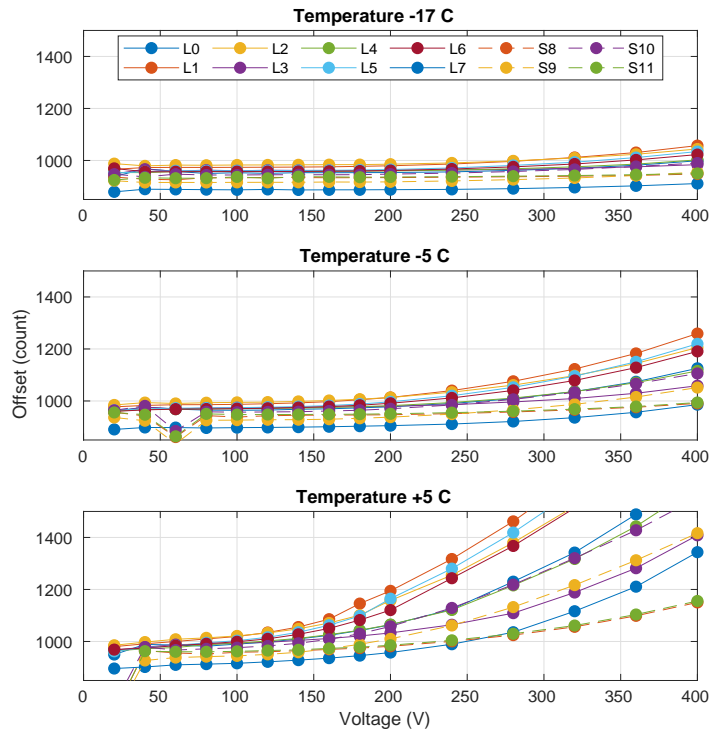}
    \caption{Calibration offset}
\end{subfigure}
\begin{subfigure}{0.65\textwidth}
	\includegraphics[width=\textwidth]{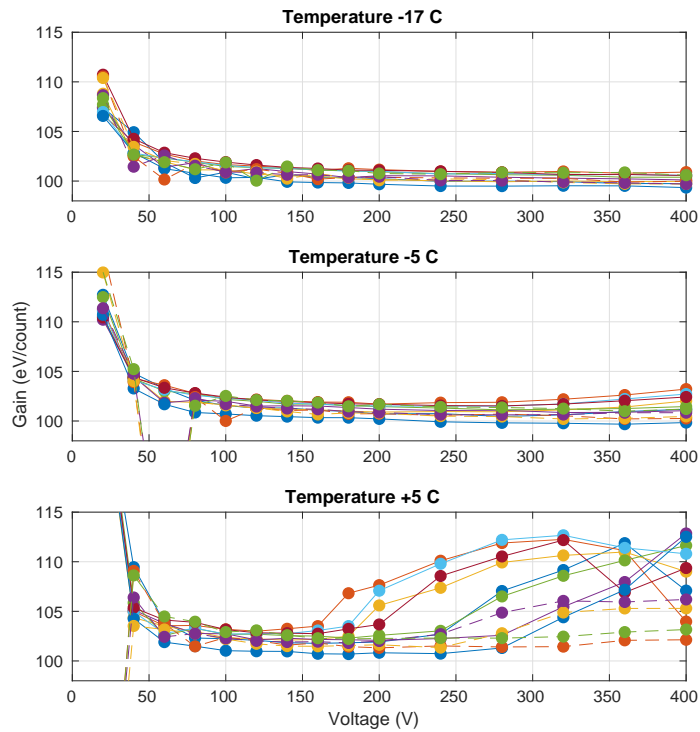}
    \caption{Calibration gain}
\end{subfigure}
\caption[Voltage scan]{Results from the voltage scan measurements. The pixel legend is shown on the first plot.}
\label{Fig:Voltage-Scan}
\end{figure}

\begin{figure}
\ContinuedFloat
\captionsetup{list=off}
\centering
\begin{subfigure}{0.65\textwidth}
	\includegraphics[width=\textwidth]{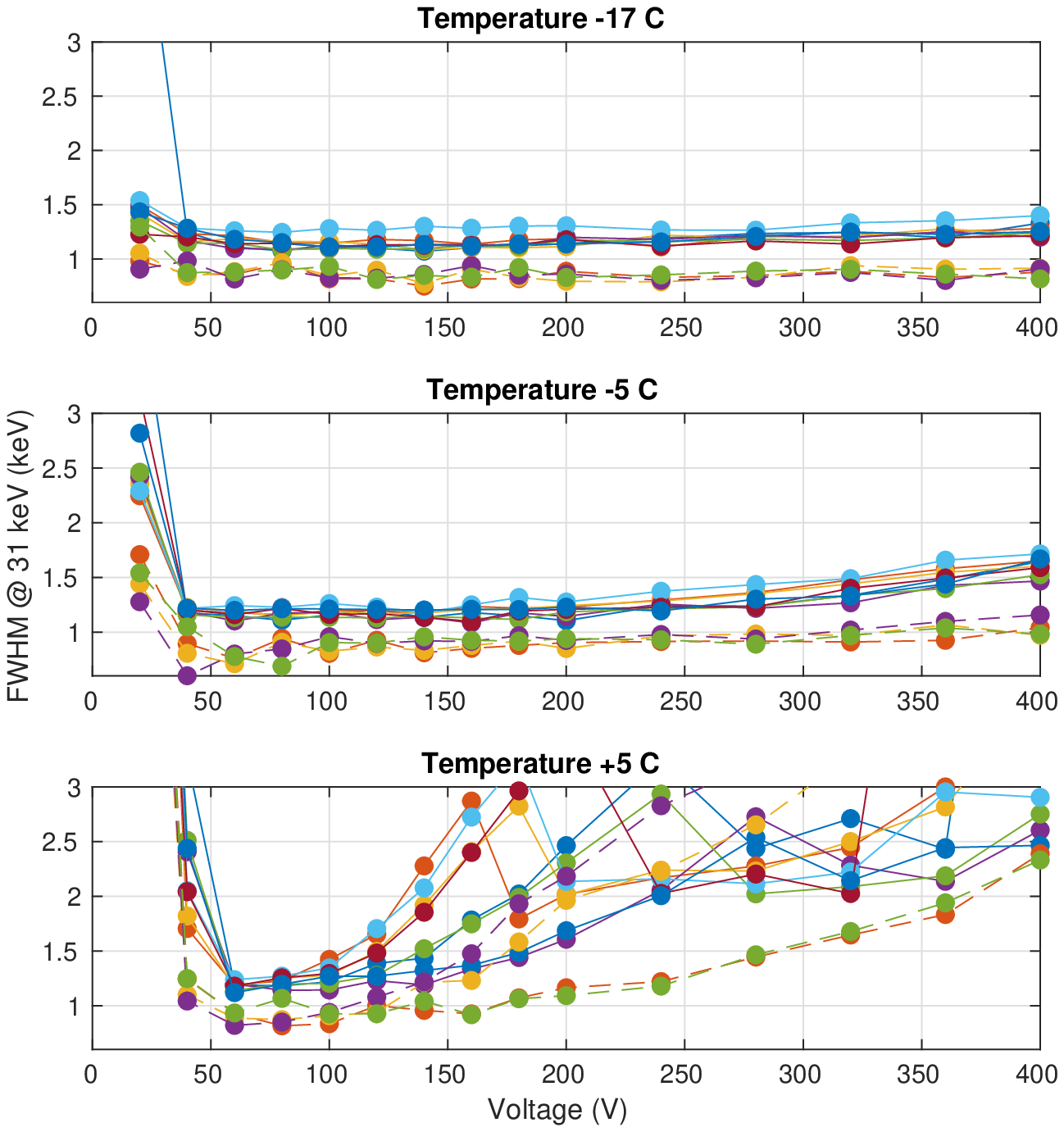}
    \caption{FWHM at \unit[31]{keV}}
\end{subfigure}
\begin{subfigure}{0.65\textwidth}
	\includegraphics[width=\textwidth]{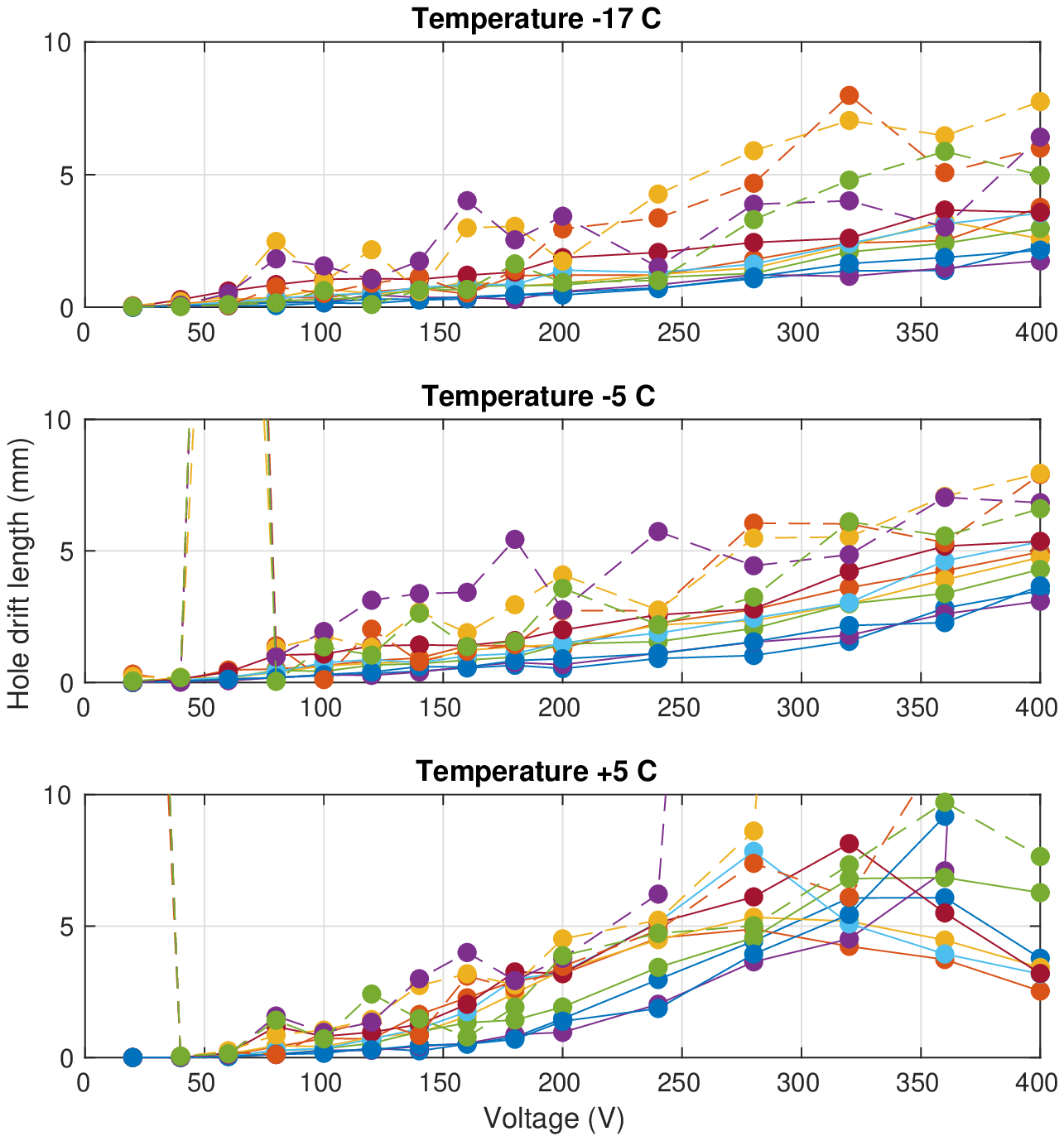}
    \caption{Hole drift length}
\end{subfigure}
\caption{(continued)}
\end{figure}

A linear dependence of the drift length on applied bias is to be expected (see Section~\ref{Sect:HechtRelation}). The significant bias dependency of the offset is a consequence of the disabled BLH, which makes the ASIC susceptible to leakage current variations. Similar measurements with enabled BLH showed no offset dependence on bias beyond \unit[60]{V} at +4\textdegree C.

\section{Current-voltage relationship}

\begin{figure}
\centering{}
\begin{subfigure}[t]{0.45\textwidth}
	\includegraphics[width=\textwidth]{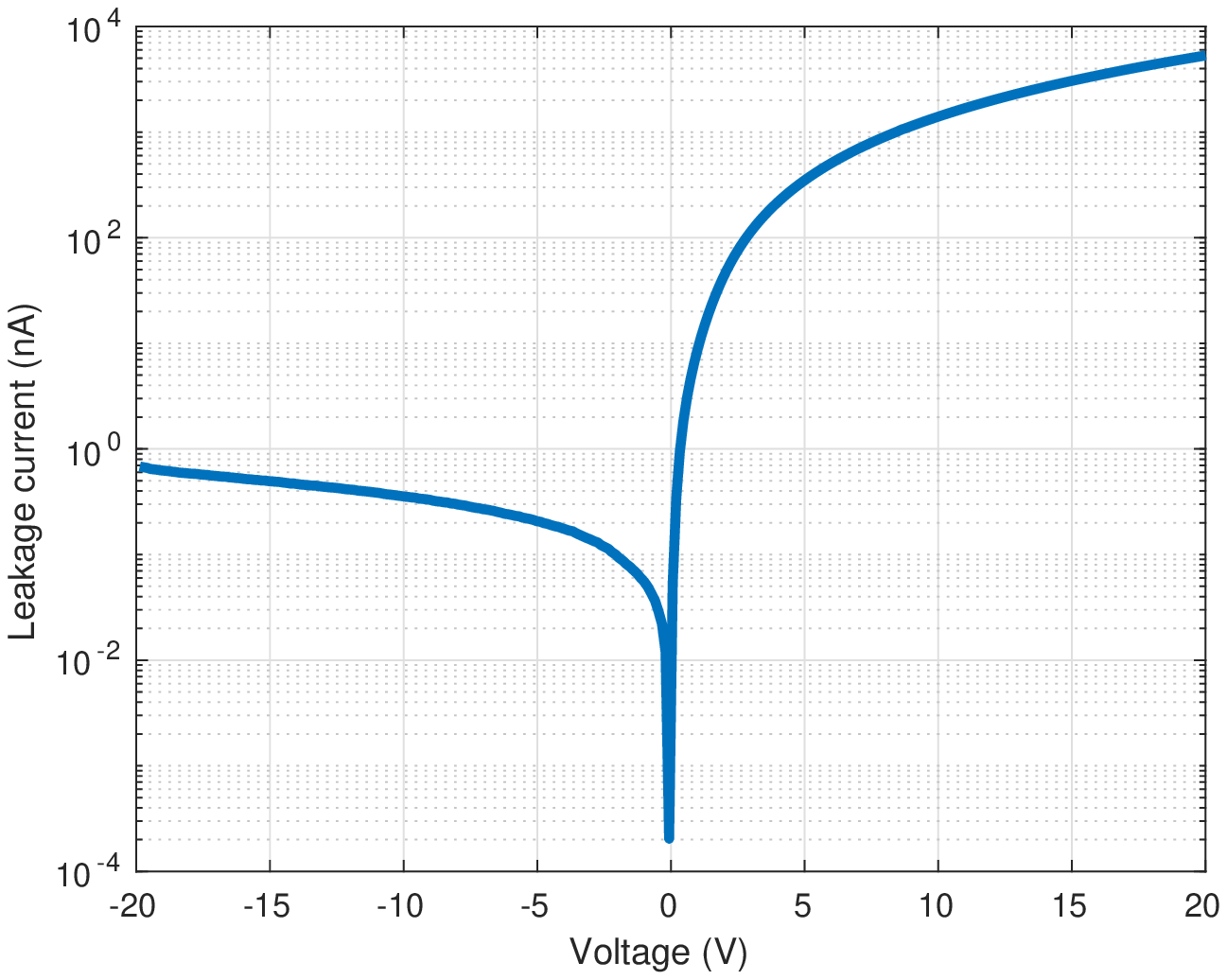}
    \caption{Absolute current on logarithmic ordinate}
\end{subfigure}
\begin{subfigure}[t]{0.53\textwidth}
	\includegraphics[width=\textwidth]{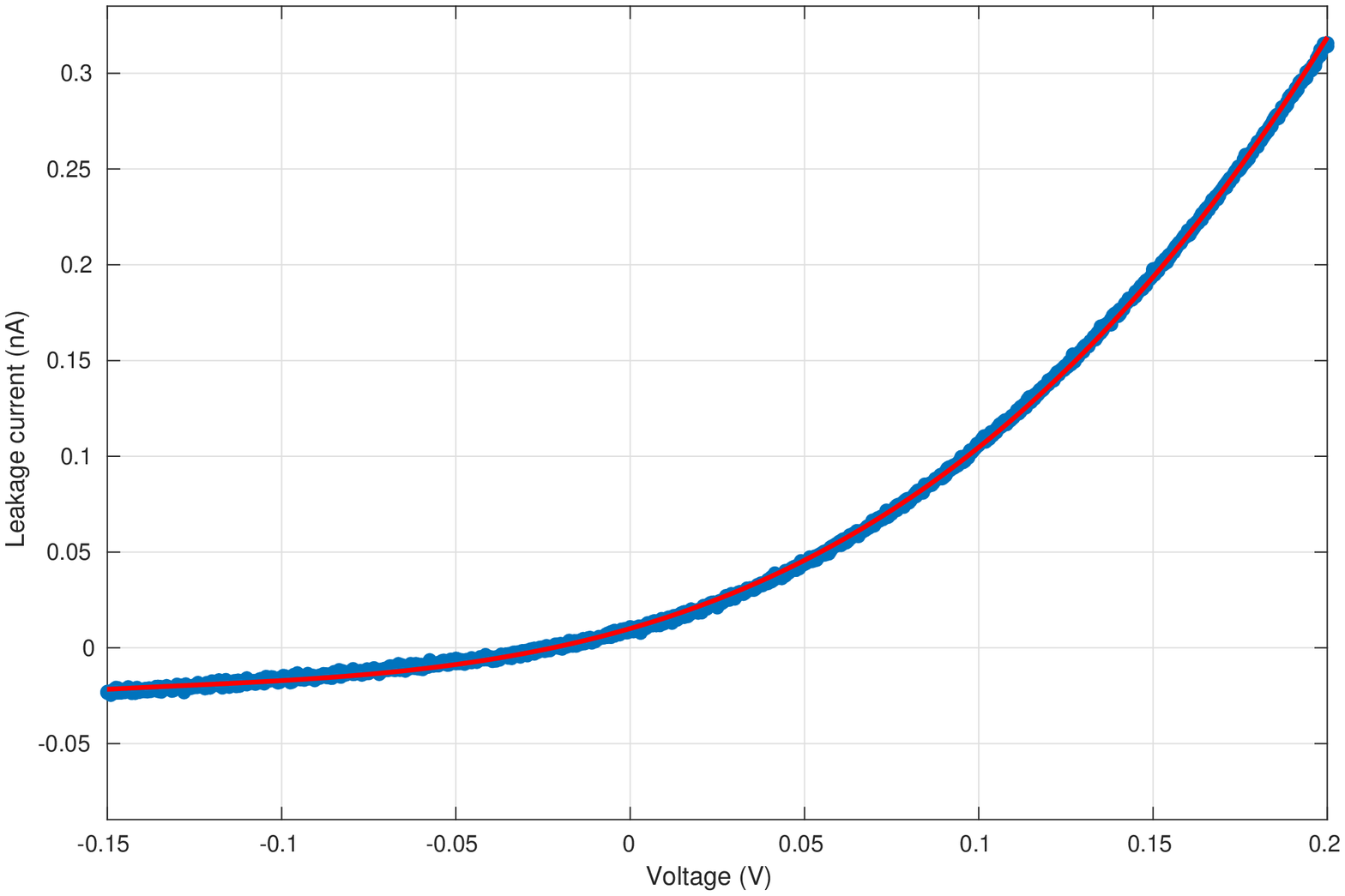}
    \caption{Nero zero voltage with linear ordinate and indicating a third-order polynominal fit}
	\label{Fig:CurrentVoltage-NearZero}
\end{subfigure}
\caption[Current-voltage curves]{CdTe current voltage curves at room temperature}
\label{Fig:CurrentVoltage}
\end{figure}

A Keithley sourcemeter was directly connected to all anode pixels and the guard ring of a CdTe crystal and to the rear monolithic cathode. The current was measured 15 seconds after bias application, followed by 10 seconds at zero volt before the next measurement. The result of the measurement, done at room temperature, is shown in Fig.\,\ref{Fig:CurrentVoltage}. The rectifying characteristic of the Schottky contact is evident.

When thermionic emission of majority carriers over the Schottky barrier dominates, the current as function of applied voltage can be written as \cite[Section 3.3.1]{Sze07}
\begin{equation}
	I(V) = A A^* T^2 \exp\frac{e \Phi}{k T} \left( \exp\frac{eV}{kT} - 1 \right), \qquad\text{with}\quad A^* = \frac{4 \pi e m^* k^2}{h^3}.
\label{Eq:IV-Relation}
\end{equation}
$A$ is the area of the contact of approximately \unit[1]{cm$^2$}, $A^*$ the effective Richardson constant, and $m^*$ the effective mass of the relevant carrier. Here, since holes are the majority carriers, $m^* = m^*_\text{v} \approx 0.35\,m_\text{e}$. The slope at zero bias of this formula,
\begin{equation*}
	\left. \frac{\text{d}I}{\text{d}V} \right|_{V=0} =  \frac{e A A^* T}{k} \exp\frac{-e \Phi}{k T},
\end{equation*}
can be compared to the slope of the data at zero bias in Fig.\,\ref{Fig:CurrentVoltage-NearZero}. To this end, a third-order polynomial is fit, $I(V) = p_3 V^3 + p_2 V^2 + p_1 V + p_0$, and $p_1$ is equated to the previous formula. One finds $p_1=\unit[5.2 \cdot 10^{-10}]{1/\Omega}$, and  thus $\Phi \approx \unit[1]{V}$.

A similar result for $\Phi$ is also obtained when fitting (\ref{Eq:IV-Relation}) directly to the data, but only when dividing $V$ by a factor of about 4. Such a factor appears in semiconductor junction current-voltage relations as \emph{ideality factor}, but is usually in the range between 1 and 2.

The barrier would be expected to be about \unit[1.7]{V} from Fig.\,\ref{Fig:BandBending}. It will be lowered due to the electric field from the depletion zone, but the effect is on the millivolt scale.\footnote{The barrier lowering $\Delta\Phi$ as function of the electric field $E$ at the interface is given by $\Delta\Phi = \sqrt{e E / (4 \pi \varepsilon_{\text{r}}\varepsilon_{0})}$. For the field of \unit[16]{V/mm}, estimated in Section~\ref{Sect:Degeneracy}, $\Delta\Phi$ is \unit[1.4]{mV}}. The substantially lower value that was measured might indicate an effect of Fermi level pinning at the interface, that is a modification of the barrier characteristics by surface defect states.

There is a small current at zero voltage. Although the sensor was carefully shielded from light, some small residual illumination cannot be excluded.

\section{Long-term and transient leakage current}

\subsection{Long-term polarization effect}

\begin{figure}
\centering
\begin{subfigure}[t]{0.49\textwidth}
	\includegraphics[width=\textwidth]{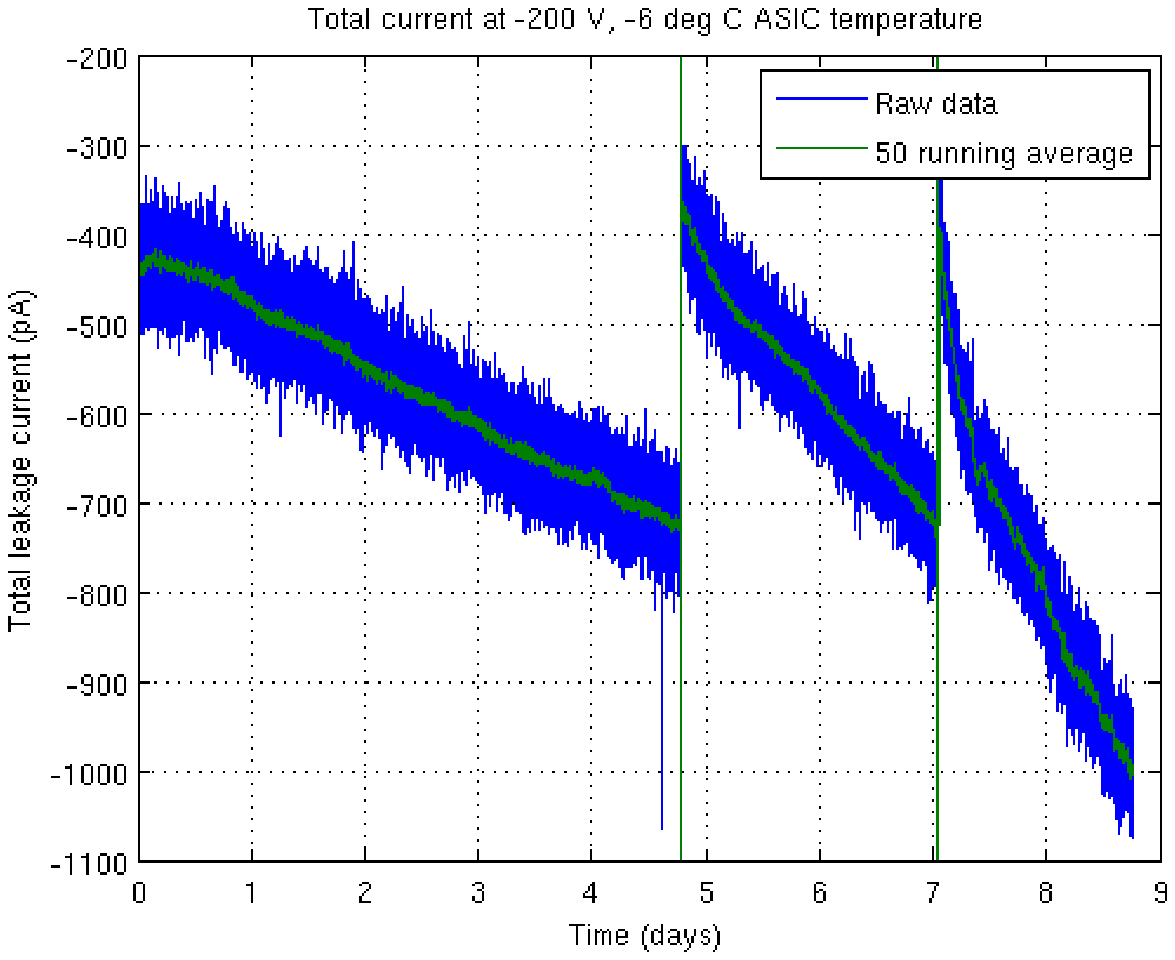}
    \caption{-6\textdegree C, bias reset twice for one minute}
\end{subfigure}
\begin{subfigure}[t]{0.49\textwidth}
	\includegraphics[width=\textwidth]{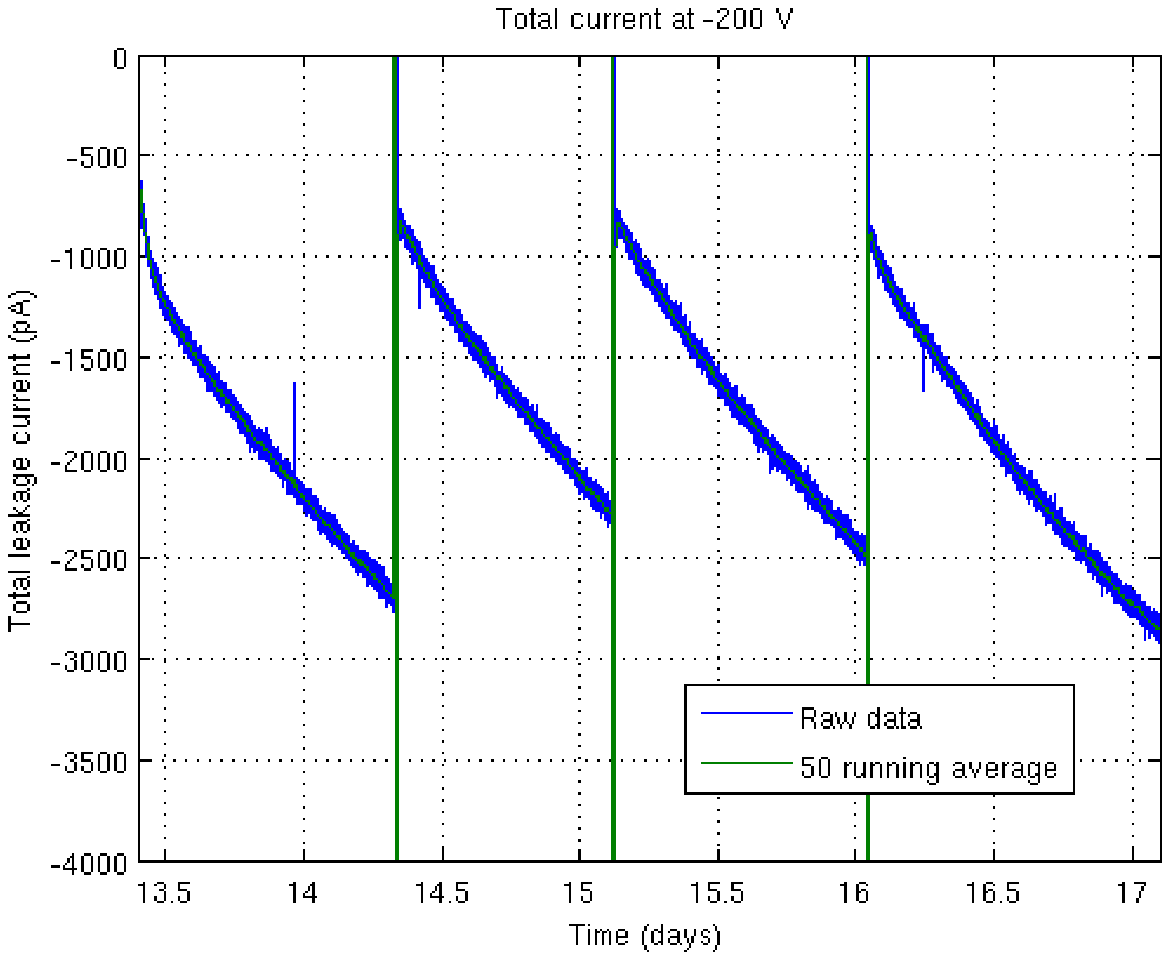}
    \caption{+4\textdegree C, bias reset for ten/ten/two minutes}
\end{subfigure}
\begin{subfigure}[t]{0.49\textwidth}
	\includegraphics[width=\textwidth]{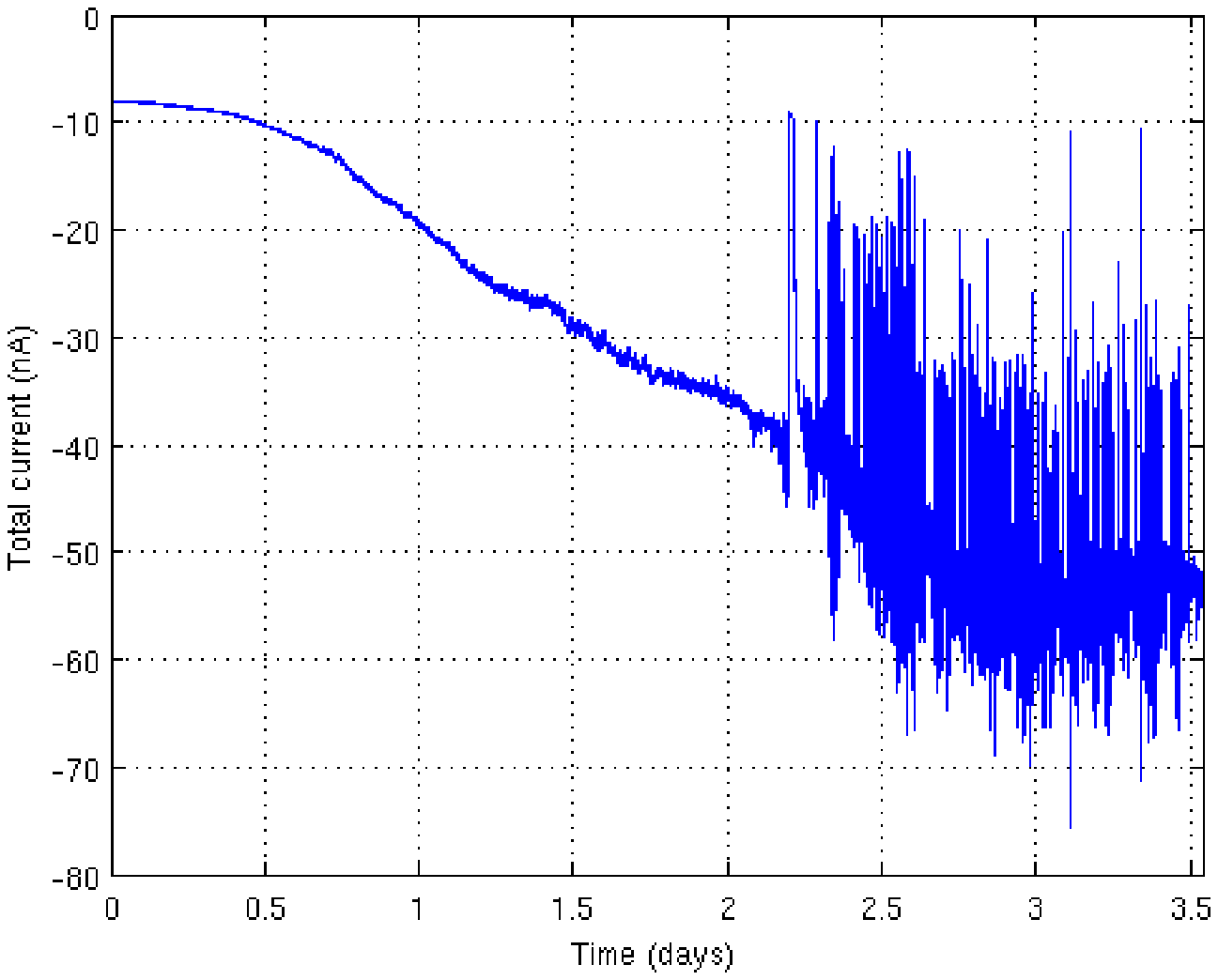}
    \caption{+9\textdegree C, no bias reset}
\end{subfigure}
\caption[Leakage current measurement]{Total leakage current as function of time for three temperatures and for \unit[-200]{V} bias. Incomplete removal of polarization is seen at -6\textdegree C due to too short bias resets. The instability at +9\textdegree C starts at a large current of \unit[40]{nA}.}
\label{Fig:LongTermCurrent}
\end{figure}

A set-up similar to the one in Section~\ref{Sect:Voltage-Scans} was used for long-term studies of the leakage current, except that no Caliste-SO was used, but a CdTe sensor was connected to a separate read-out ASIC with an interface board using low-force spring contacts \cite{Gri15}. The bias voltage and temperature of the CdTe sensor were kept constant, except for short bias resets. The total leakage current over several days, measured with a Keithley sourcemeter at three temperatures, is shown in Fig.\,\ref{Fig:LongTermCurrent}.

Also barium-133 calibration spectra were taken repeatedly, showing the expected widening of the lines at \unit[31]{keV} due to the increased leakage current. At -17\textdegree C, all sensor parameters were found to be completely stable for at least a week.

\subsection{Transient temperature effect}

\begin{figure}
\centering
\begin{subfigure}{0.80\textwidth}
	\includegraphics[width=\textwidth]{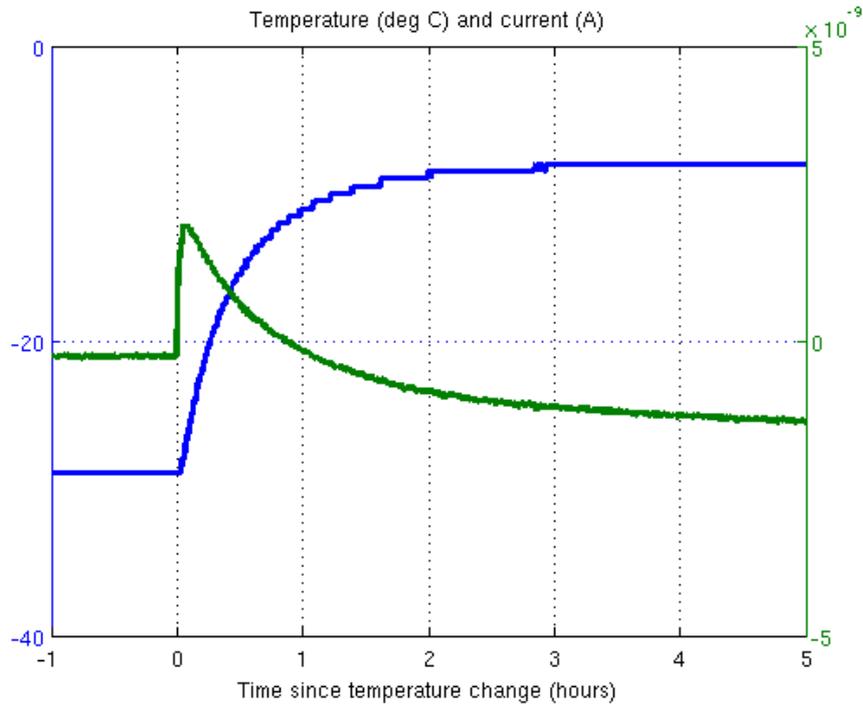}
    \caption{Temperature of the cold plate (blue) and total leakage current (green)}
\end{subfigure}
\begin{subfigure}{0.80\textwidth}
	\includegraphics[width=\textwidth]{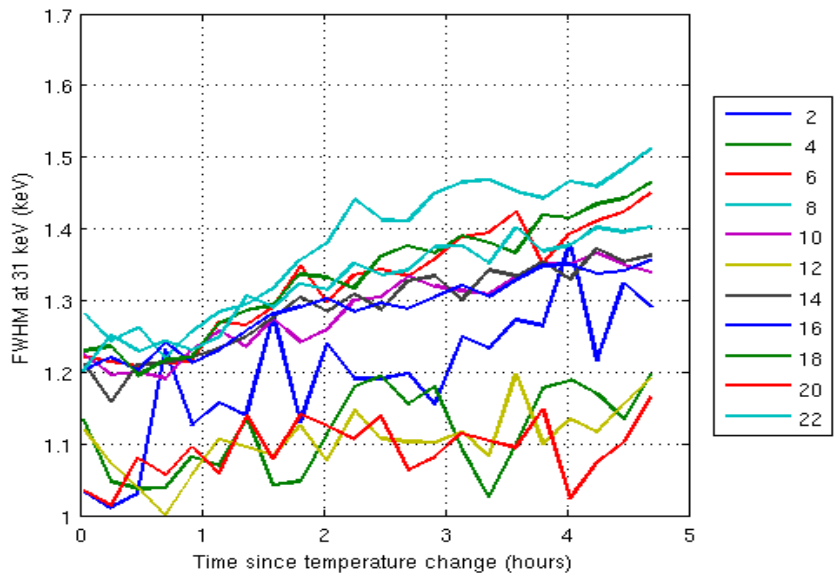}
    \caption{FWHM at \unit[31]{keV}}
\end{subfigure}
\caption[Transient temperature effect]{Effect on total leakage current and FWHM at \unit[31]{keV} for a fast change of temperature at \unit[-200]{V} bias. The indicated temperature is that of the cold plate. The temperature of the ASIC changed from -17\textdegree C to +4\textdegree C during this time.}
\label{Fig:TemperatureTransient}
\end{figure}

The set-up is as in the previous section. The temperature was changed as fast as the thermal inertia allowed. The measured temperature, total leakage current, and FWHM at \unit[31]{keV} are shown in Fig.\,\ref{Fig:TemperatureTransient}. The initial, positive current peak is likely due to the large piezoelectric effect in CdTe. A temperature gradient in the material gives rise to mechanical stress, and, via the piezoelectric effect, to charge separation and to an induced current.

The positive peak current is about \unit[2]{nA}, or \unit[200]{pA} per large pixel. This reverse current appears to have no detrimental effect on the ASIC performance, as seen in the FWHM plot.

\section{Spatial response scan}

\begin{figure}
\centering
\begin{subfigure}[t]{0.44\textwidth}
	\includegraphics[width=\textwidth]{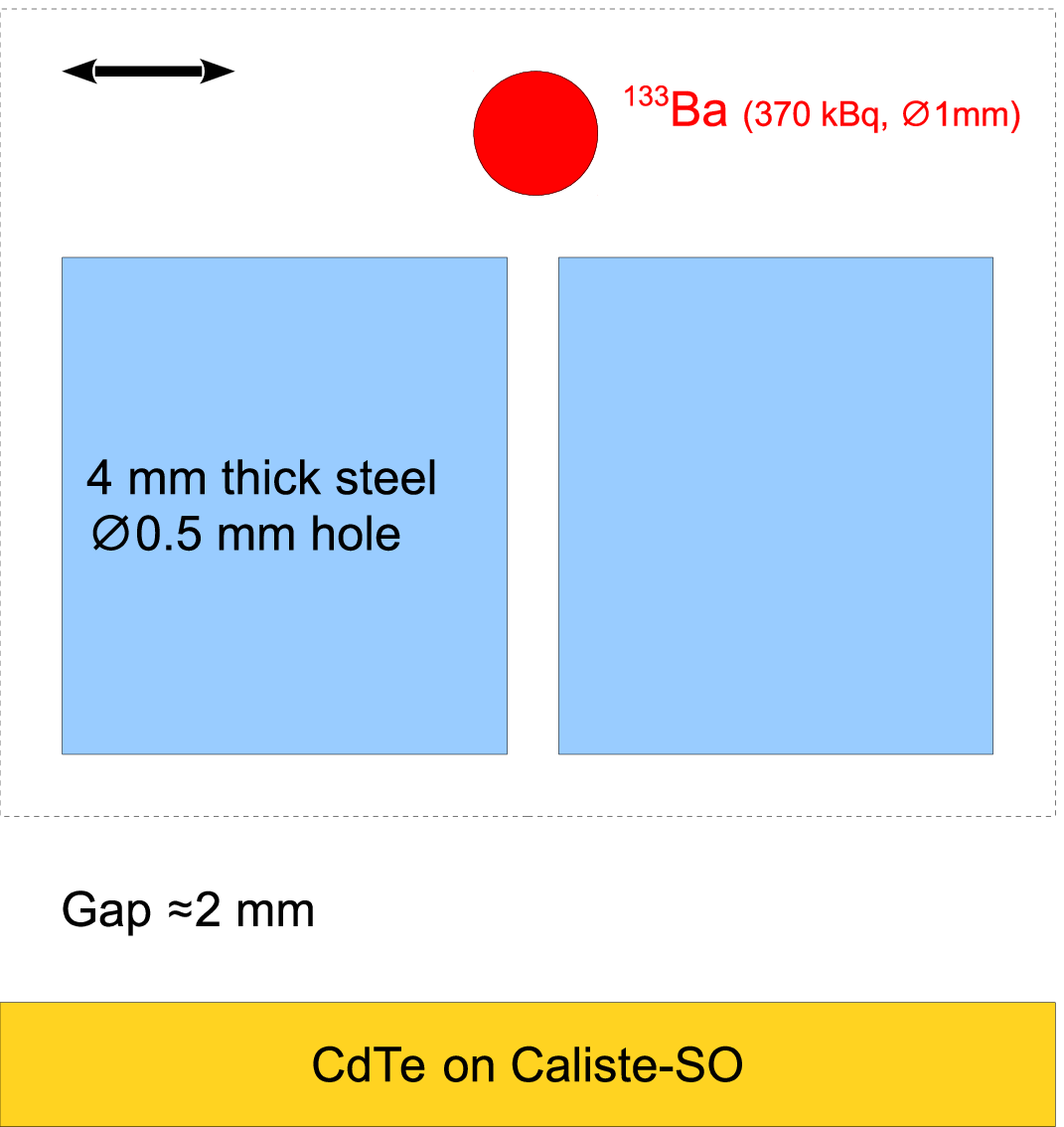}
    \caption{Collimation geometry. The red source bead is sealed inside a \unit[2]{mm} thick plastic plate (not shown).}
    \label{Fig:Scan-Setup-a}
\end{subfigure}
\hspace{2em}
\begin{subfigure}[t]{0.34\textwidth}
	\raisebox{2em}{\includegraphics[width=\textwidth,angle=90]{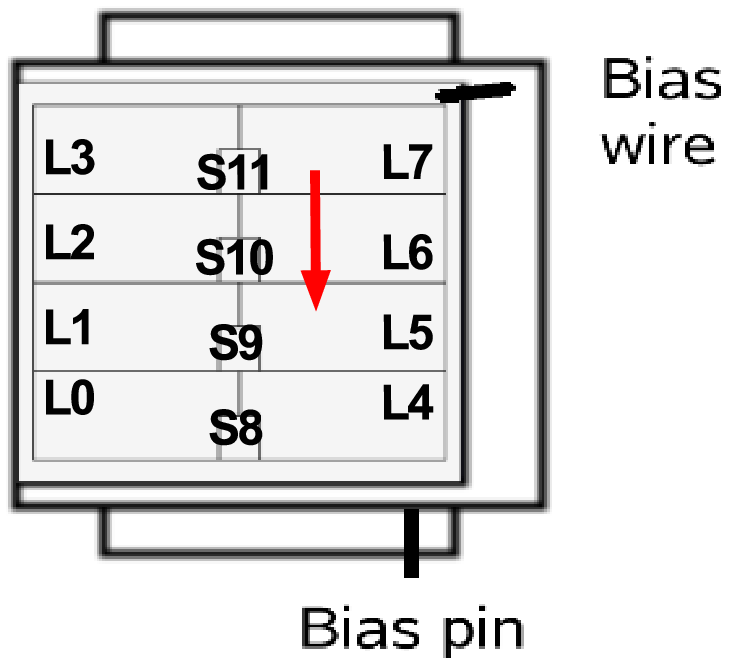}}
    \caption{The collimated X-ray beam moved from the beginning to the end of the red arrow.}
    \label{Fig:Scan-Setup-b}
\end{subfigure}
\caption[Spatial response scan geometry]{Geometries of the spatial response scan}
\end{figure}

The spatial pixel response was studied using a collimated barium-133 radioactive source. The collimator, a \unit[4]{mm} thick steel disk with a \unit[0.5]{mm} diameter hole, and the geometry is shown in Fig.\,\ref{Fig:Scan-Setup-a}. Measurements were performed in vacuum at -16\textdegree C (temperature measured by the ASIC) and bias \unit[-200]{V}. The scan consisted of 500 steps with size \unit[6]{\textmu m}, thus covered \unit[3]{mm}, along the path indicated in Fig.\,\ref{Fig:Scan-Setup-b}.

\begin{figure}
\centering
\begin{subfigure}{0.48\textwidth}
	\includegraphics[width=\textwidth]{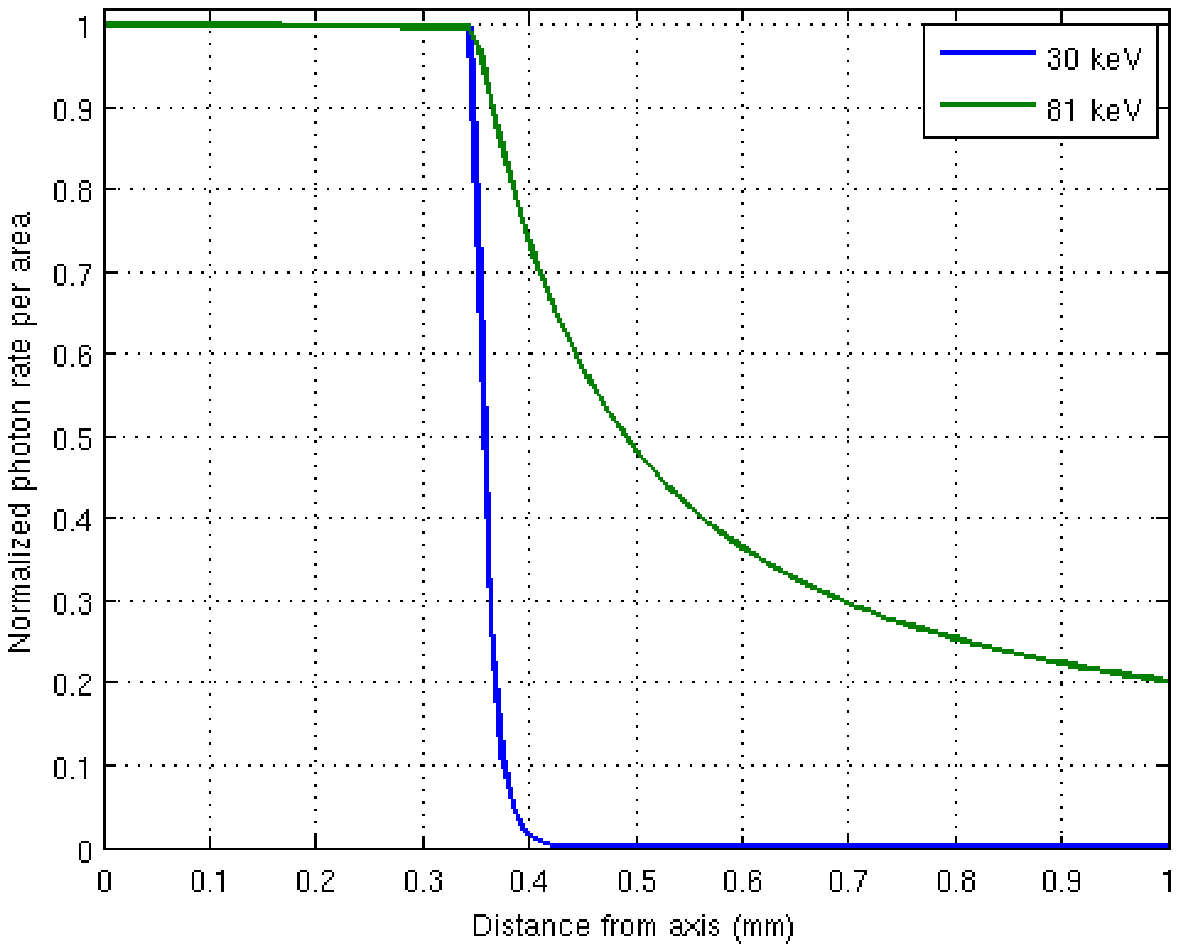}
    \caption{Calculated photon distribution on the CdTe surface at \unit[31]{keV} and \unit[81]{keV}.}
    \label{Fig:Scan-SpotSize}
\end{subfigure}
\begin{subfigure}{0.48\textwidth}
	\includegraphics[width=\textwidth]{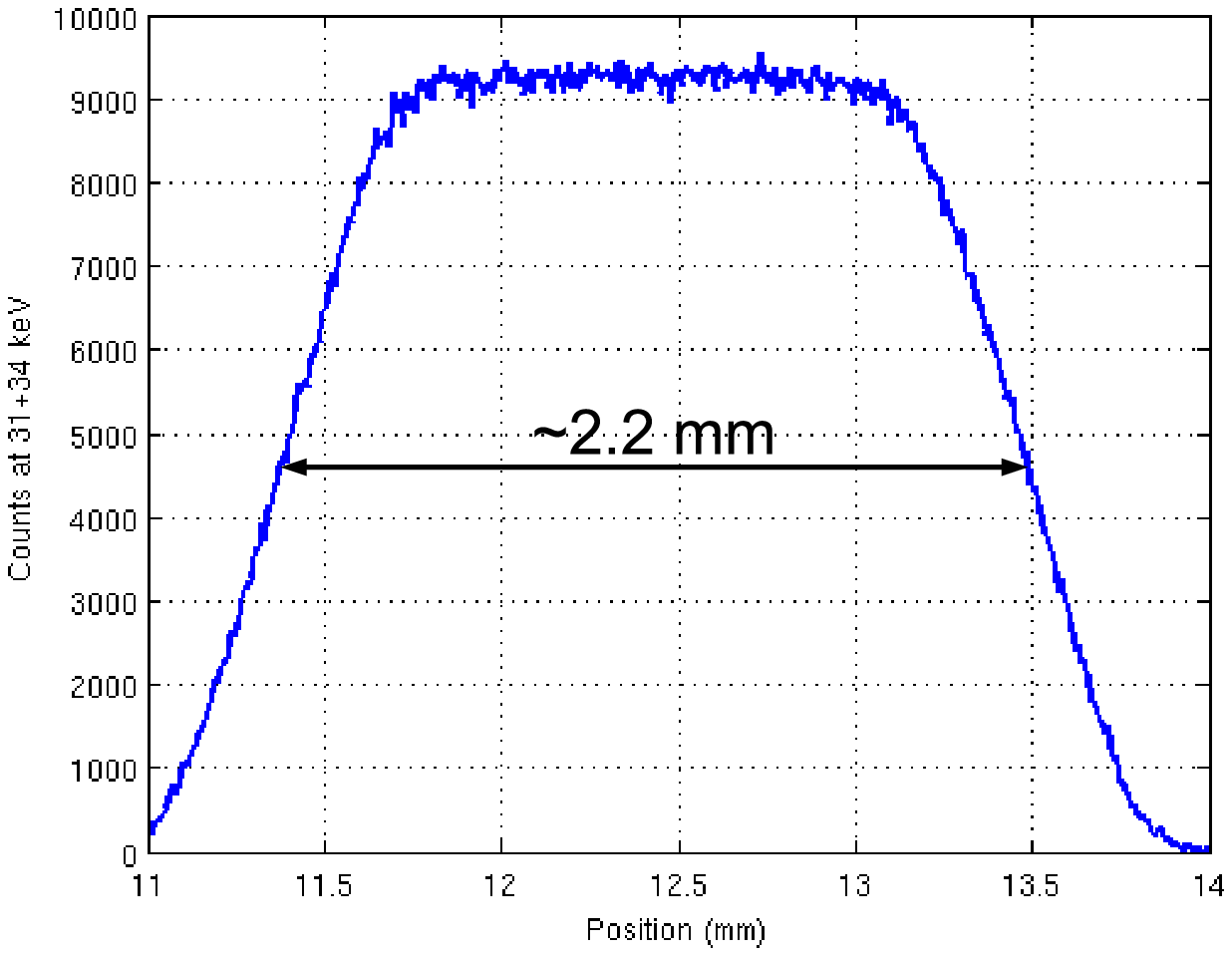}
    \caption{Count rate between \unit[31]{keV} and \unit[35]{keV} as function of scan position.}
    \label{Fig:Scan-Result}
\end{subfigure}
\caption[Spatial response scan result]{Spatial response scan photon spot size and count rate. The width of a large pixel in the scan direction is \unit[2.2]{mm} when measured from the centre of the electrode-free gap between the pixels. The metallized width is \unit[2.15]{mm}.}
\end{figure}

The collimator is much more effective at \unit[31]{keV} than at \unit[81]{keV}, which is illustrated in Fig.\,\ref{Fig:Scan-SpotSize}, showing the photon distribution on the sensor. The distribution was calculated using the actual collimator geometry and the photon absorption lengths of steel (\unit[0.15]{mm} and \unit[2.1]{mm} at the two energies), but modelling the source as a point, not as a finite-sized bead.

At each step, spectra with \unit[100]{s} integration time were taken (14 hours total measurement time). The counts accumulated in the lines around \unit[31]{keV} as function of position are shown in Fig.\,\ref{Fig:Scan-Result}.

\chapter*{Further reading}
\addtocontents{toc}{\protect\vspace{6ex}}
\addcontentsline{toc}{section}{Further reading}

This section lists texts that, beyond the references given in this report, can aid in further understanding of the principles underlying semiconductor detectors.

A good, accessible overview of semiconductor physics is given in the first part of \cite{Pier96}, and of Schottky contacts in part three. Some topics, for example the physical basics of band structure, are explained in greater depth in the partially overlapping \cite{Pier02}. A full account on band structures can be found in \cite{Boer02} and \cite{Ash76}.

A complete, detailed account of semiconductor devices is given in \cite{Sze07}, of many aspects of semiconductor detector systems in \cite{Spi05}, and of general detector physics in \cite{Kno10}. Comprehensive coverage of interface effects is given in \cite{Mon01}.

Problems specific to compound semiconductors are treated in \cite{Owens12}, which also gives an extensive discussion of many specific materials.

Discussions of the bias polarization in CdTe can be found in \cite{Malm74}. Note that this must not be confused with rate polarization, as described for example in \cite{Var88}. Both effects are a result of electric field changes in the sensor, the former coming from changes in the space-charge, the latter from the high signal currents.

Characteristics of materials used in radioactive sources can be found in \cite{Fire06}.

\listoffigures
\addtocontents{toc}{\protect\vspace{3ex}}

\end{document}